\documentclass[12pt]{article}
\usepackage{mathtools,amssymb,mathrsfs,microtype,slashed,amsmath}
\pdfoutput=1 %to solve the issue as in: https://tex.stackexchange.com/questions/719215/opacity-not-working-when-uploading-to-arxiv

\usepackage{array}
\usepackage{comment}
\usepackage{placeins}
\usepackage{standalone}
\usepackage{amsmath}
\usepackage{mathtools}
\usepackage{graphicx}
\usepackage{tikz, subfigure}
\usetikzlibrary{arrows.meta, positioning, quotes,decorations.markings}
\usetikzlibrary{external}
\usetikzlibrary{knots}
\usetikzlibrary{shapes.geometric}
\usepackage{physics}
\usetikzlibrary{calc}
\usepackage[linktocpage]{hyperref}
\hypersetup{colorlinks=true,citecolor=blue,linkcolor=blue, urlcolor=blue, breaklinks=true}
\usepackage{cite}
\usepackage{moresize}
\usepackage{url}
\usepackage[numbers,sort&compress]{natbib}

\definecolor{darkgreen}{RGB}{34, 140, 55}
\definecolor{darkyelloworange}{RGB}{219, 149, 29}

\numberwithin{equation}{section}
\newcommand{\lcm}{\text{lcm}}

\newcommand{\be}{\begin{equation}}
\newcommand{\ee}{\end{equation}}

\renewcommand{\title}[1]{\vbox{\center\LARGE{#1}}\vspace{5mm}}
\renewcommand{\author}[1]{\vbox{\center\large#1}\vspace{5mm}}
\newcommand{\address}[1]{\vbox{\center\em#1}}

\usepackage{ytableau}

\begin{document}

\begin{titlepage}

\begin{center}

\hfill \\
\hfill \\
\vskip 1cm

\title{Non-Invertible Symmetries as Condensation Defects\\ in Finite-Group Gauge Theories}

\author{Clay C{\'o}rdova$^1$, Davi B.~Costa$^1$ and Po-Shen Hsin$^{2}$
}

\address{${}^1$ Enrico Fermi Institute \& Kadanoff Center for Theoretical Physics, University of Chicago
}

% \address{${}^2$Mani L. Bhaumik Institute for Theoretical Physics,\\
% Department of Physics and Astronomy,\\
% University of California, Los Angeles, CA 90095, USA}

\address{${}^2$ Department of Mathematics, King’s College London, Strand, London WC2R 2LS, UK.}

\end{center}

\abstract{
In recent work, we developed a method to construct invertible and non-invertible symmetries of finite-group gauge theories as topological domain walls on the lattice. In the present work, we consider abelian and non-abelian finite-group gauge theories in general spacetime dimension, and demonstrate how to realize these symmetries as condensation defects, i.e., as suitable insertions of lower dimensional topological operators. We then compute the fusion rules and action of these symmetries using their condensation expression and the algebraic properties of the lower dimensional objects that make them. We illustrate the discussion in $\mathbb{Z}_N$ gauge theory, where we derive the correspondence between domain walls, labeled by subgroups and actions for the doubled gauge group, and higher gauging condensation defects, labeled by subalgebras of the global symmetry. As a primary application, we obtain the condensation expression for the invertible symmetries of abelian gauge theories defined by outer automorphisms of the gauge group. We also show how to use these ideas to derive the action for certain non-abelian groups.  For instance, one can obtain the action for the Dihedral group $\mathbb{D}_4$ by gauging a swap symmetry of $\mathbb{Z}_2\times\mathbb{Z}_2$ gauge theory.
}

\vfill

\today

\vfill

\end{titlepage}

\eject

\setcounter{tocdepth}{3}
\tableofcontents
\newpage

\section{Introduction}

Symmetries are often the first line of attack for constraining the dynamics of physical systems.  As such, new symmetries can give powerful new insights into strongly-coupled theories.  Motivated by these considerations, in this work we explore non-invertible symmetries in simple topological field theories: gauge theories with finite gauge group \cite{Dijkgraaf:1989pz} (see \cite{Cordova:2024jlk} for a review).

The fundamental connection between symmetry and topology was made precise in \cite{Gaiotto_2015} (see also \cite{Cordova:2022ruw,Shao:2023gho,Luo:2023ive,Bhardwaj:2023kri,Schafer-Nameki:2023jdn,Iqbal:2024pee,Brennan:2023mmt,McGreevy:2022oyu,Gomes:2023ahz,Costa:2024wks} for lecture notes and reviews). Every symmetry is defined intrinsically by an extended topological operator, and the selection rules can be deduced through the behavior of these operators in correlation functions. Subsequent developments have indicated an even more precise classification of finite symmetry: topological operators in a quantum field theory in spacetime dimension $D$ can be described by a topological quantum field theory (TQFT) in dimension $D+1$.
\cite{Witten:1998wy, Gaiotto_2015,kong2015boundarybulkrelationtopologicalorders,Gaiotto:2020iye,Freed:2022qnc,Kaidi:2022cpf,PhysRevResearch.2.043086}.  There are therefore at least two reasons to be interested in the topological operators in finite-group gauge theories.  First, these theories are an illuminating playground where we can make precise many of the abstract ideas that permeate more general models. And second, when $D=4$ and all particles are bosons,
it is expected that every unitary topological field theory is in fact a (possibly twisted) finite-group gauge theory \cite{PhysRevX.8.021074,PhysRevX.9.021005,Johnson-Freyd:2020usu}.\footnote{A closely related statement holds in $D=5$. Specifically, when all particles are bosons, any unitary TQFT is a finite-group gauge theory coupled to a theory of abelian two-form gauge fields \cite{Johnson-Freyd:2020ivj,Johnson-Freyd:2021tbq,Cordova:2023bja}.}

In the present work, we build on our previous study of defects in finite-group gauge theory \cite{Cordova:2024jlk}, focusing on the notion of condensation.  The idea of condensation refers to the process of appropriately summing over insertions of topological defects.  If the sum is carried out in all of spacetime, this condensation is a form of gauging the symmetry described by the summed defects.  In this case, the result of the condensation process is a new quantum field theory, and such constructions have been explored in detail in \cite{Kong:2013aya, Eliens:2013epa,Lan:2014uaa,Hung:2015hfa,Neupert:2016pjk,Burnell:2017otf, Tachikawa:2017gyf, Bhardwaj:2017xup,Cong:2017ffh,Hsin:2018vcg,Kaidi:2021gbs,Yu:2021zmu,Decoppet:2022dnz,Diatlyk:2023fwf, Cordova:2023jip, Zhang:2024bye,Cordova:2024goh, Kong:2024ykr,Perez-Lona:2024sds}. 

\begin{figure}
    \centering
    \includegraphics[width=0.75\textwidth]{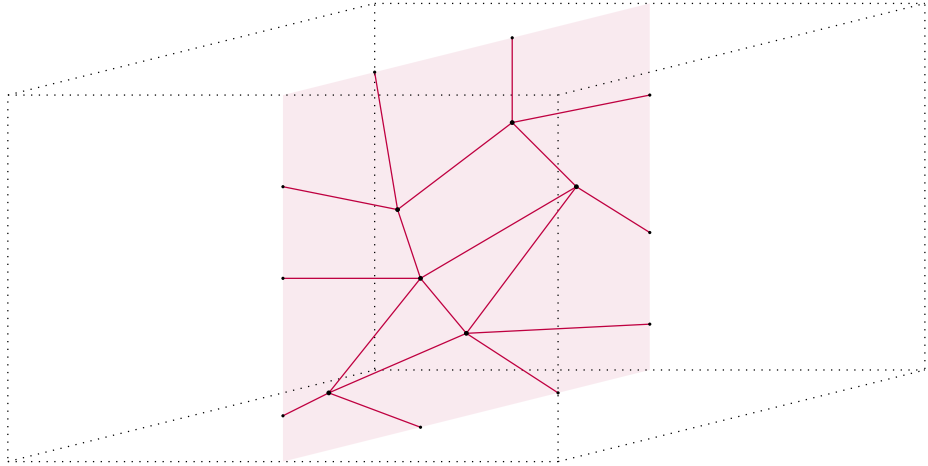}
        \caption{Illustration of a condensation defect slice, depicted as the red-shaded plane, formed by the insertion of lower-dimensional topological operators, represented by the lines in the illustration. The condensation defect is embedded within a region of spacetime, shown as the dotted cuboid.}
    \label{fig:mesh_of_lines}
\end{figure}

In contrast to these sums in all of spacetime, in this work, we discuss condensations where the sum is carried out over a submanifold of spacetime.  In this case, the result of the condensation procedure is an operator in the initial theory \cite{Carqueville:2017ono,Gaiotto:2019xmp,Kong:2020wmn,Roumpedakis:2022aik,Cui:2024cav,Cuiper:2024hvh,Ebisu:2024lie,Vandermeulen:2023smx,Lin:2022xod,Lyons:2024fsk,Carqueville:2018sld,Mulevicius:2020bat,Koppen:2021kry,Carqueville:2021dbv,Carqueville:2021edn,Choi:2023xjw, Choi:2023vgk,Buican:2023bzl}.  At an intuitive level, one may view the condensation defect as a fine mesh of lower-dimensional objects as illustrated in Fig.   \ref{fig:mesh_of_lines}.  This picture also accurately captures a key feature of such condensation operators, namely that they have trivial action via linking on other operators -- a hole may always be opened in the mesh, which allows the other operator to pass through and unlink.\footnote{More precisely, this is the case for the action by Hopf-linking, where a topological operator of dimension $(D-p)$ acts on operators of dimension $(p-1)$. Condensation defects can act non-trivially by multilinking. See \cite{Hsin:2019fhf,Barkeshli:2022edm} for examples.} In certain situations, a partial converse to this is also known; for instance, in unitary $D=3$ TQFTs with a unique identity operator, all surface operators are condensations \cite{Fuchs_2013}.   

The concept of condensation operators has been explored in a variety of contexts. In particular, they typically do not obey a group multiplication law and thus constitute an example of non-invertible symmetry \cite{Kapustin:2010if,Fuchs:2012dt, Bhardwaj:2017xup,Chang:2018iay,Komargodski:2020mxz, Choi:2021kmx, Kaidi:2021xfk,  Chang:2022hud}. In $D=3$ TQFTs there is a detailed dictionary between abstract algebra objects, more physically collections of anyons obeying certain algebraic properties, and surface operators \cite{Eliens:2013epa, Kong:2013aya, Lan:2014uaa, Hung:2015hfa, Neupert:2016pjk,Burnell:2017otf, Cong:2017ffh, Kaidi:2021gbs, Yu:2021zmu, Cordova:2023jip, Zhang:2024bye}.  This dictionary can also be extended to $D=2$ rational conformal field theories by viewing these as the boundaries of TQFTs (see e.g.\ \cite{Fuchs:2002cm,Fuchs:2003id, Fuchs:2004dz,Fuchs:2004xi,Frohlich:2004ef,Frohlich:2006ch,Diatlyk:2023fwf}).  Most closely connected to our approach below is the notion of higher gauging introduced in \cite{Roumpedakis:2022aik}.  In this context, the authors studied how sums over defects with abelian fusion rules give rise to a wide variety of defects.  This also allows one to exhibit the correspondence between the worldvolume action of the defect, itself often a (twisted) finite-group gauge theory, and the resulting defect operator presented as a bulk condensation.  Our work will directly build on the analysis of \cite{Roumpedakis:2022aik}, constructing more intricate examples of higher gauging.  

More specifically, in this paper, we will provide more examples of higher gauging and related condensation constructions by considering finite-group gauge theories in general spacetime dimension $D$, where the gauge groups can be abelian or non-abelian. For simplicity of illustration, we will focus on the theories without additional topological terms, i.e. untwisted topological gauge theories. In a recent analysis, \cite{Cordova:2024jlk}, we developed a method to construct invertible and non-invertible symmetries of finite-group gauge theories as domain walls on the lattice. In the present work, we demonstrate how to realize these symmetries as condensation defects, i.e., as suitable insertions of lower-dimensional topological operators. This realization allows us to derive properties of the non-invertible symmetries, including fusion rules and the action on other operators, in terms of the constituents that form the condensation defects. (For further examples of topological defects in finite-group gauge theories see e.g.\ \cite{Kitaev:1997wr,Kapustin:2010hk,Wen:2012hm,Barkeshli:2014cna,Hsin:2019fhf,Barkeshli:2022edm}.)

Our results can also be applied to describe invertible and non-invertible symmetry as defects on the lattice with continuous time with the Gauss law imposed energetically by Hamiltonian terms. For symmetry that corresponds to the condensation of particular electric and/or magnetic excitations that have energy cost from suitable Gauss law and flux terms in the Hamiltonian \cite{Kitaev:1997wr},
the symmetry defect can be constructed as a modification on the Hamiltonian with the Hamiltonian terms along the defect corresponding to the condensed excitations removed. Such a description has been used in the construction of Cheshire string defects where the electric charge condenses \cite{Else:2017yqj,Johnson-Freyd:2020twl}, and the fusion rules of Cheshire strings in $\mathbb{Z}_2$ gauge theory in $D=3+1$ spacetime dimension can be obtained in this way \cite{Kong:2020wmn}.\footnote{Cheshire strings have also been discussed in \cite{Chen:2023qst,Tantivasadakarn:2023zov}, but in terms of operators of linear depth called sequential circuits.} Our results can be extended to such description to general invertible or non-invertible symmetries in general untwisted gauge theories, which quantum double lattice models can describe \cite{Kitaev:1997wr}.

\subsection{Summary of results}

Let us now summarize our results in detail.  Throughout the following, $G$ denotes the finite gauge group, which in general may be abelian or non-abelian. In this work we aim to establish the correspondence between two descriptions of non-invertible symmetries in finite-group gauge theories: (1) symmetries as domain walls on the lattice \cite{Cordova:2024jlk}, and (2) symmetries as condensation defects \cite{Gaiotto:2019xmp}, i.e., as suitable insertions of lower dimensional topological operators. Let us briefly summarize the two perspectives:

\begin{itemize}
    \item[(1)] The first construction uses the folding trick to unfold the domain wall into a gapped boundary, which we review in Section \ref{sec:review}. The domain wall is then labeled by
    a subgroup $H\leq G\times G$ and a choice of topological action for the given subgroup $\alpha\in H^{D-1}(H,U(1))$. Given this data, we construct a codimension-one domain wall $\mathcal{D}_{H,\alpha}(\Sigma)$ supported on $\Sigma$ by having the gauge group elements on $\Sigma$ to be elements of $H\leq G\times G$ and by properly gluing them with the exterior of the domain wall. We further decorate $\Sigma$ with the topological action $\alpha$. Here, the left and right components of $H\leq G\times G$ correspond to the left and right of the domain wall respectively, which have a global meaning in terms of the orientation of the normal bundle $N\Sigma$. We remark that this construction applies to all finite gauge groups, including non-abelian ones, and in any spacetime dimension.
    
    \item[(2)] When the gauge group $G$ is abelian, condensation defects can be interpreted as the gauging of higher-form symmetries on the support of the 0-form symmetry (i.e., higher gauging defects), which we review in Section \ref{sec:highergauging}. In $D=2+1$ spacetime dimensions the higher-form symmetry generated by the electric Wilson lines and magnetic fluxes is a 1-form $G\times G$ symmetry. In this case, the condensation defects are classified by subgroups of the 1-form symmetry $H\lhd G\times G$ with a topological term (``discrete torsion") $f\in H^{2}(H,U(1))$ \cite{Roumpedakis:2022aik}. Given this data, one can construct a codimension-one topological operator $\mathcal{S}_{H,f}(\Sigma)$ by summing over all insertions of the global symmetry operators in the corresponding subgroup along $\Sigma$ with torsion equal to $f$\footnote{We set as our convention that the Wilson lines are always on the left of the magnetic defects in $\mathcal{S}_{H,f}(\Sigma)$ which corresponds to having them inserted on the left of the magnetic defect insertion.}. Here, the left and right components of $H\lhd G\times G$ correspond to the 1-form symmetry generated by the Wilson lines and magnetic defects respectively.
\end{itemize}

The data for the two classifications is the same in $D=2+1$ spacetime dimensions, but the meanings are remarkably different. This work aims to relate these two perspectives and extend the relation to the case where $G$ is non-abelian and the spacetime dimension is arbitrary.

\paragraph{Diagonal domain walls: electric condensations} Consider first the domain wall associated with the trivial subgroup and let us start the discussion with the abelian case. One can derive the following relation:
\begin{align}
    \mathcal{S}_{G\times 1}=\mathcal{D}_{1}.
    \label{eq:diagonalreview1}
\end{align}
In words, the higher gauging of the 1-form symmetry $G$ generated by the Wilson lines is equivalent to the domain wall that restricts the gauge group elements to be in the subgroup $1\lhd G\times G$. The equivalence between the two descriptions comes from two facts: 
\begin{itemize}
    \item The higher gauging of the 1-form symmetry $G$ generated by the Wilson lines can be reorganized as the product of insertions of the Wilson line in the regular representation,
    \begin{align}
        W_{\textrm{reg}}=\sum_{\rho\in \textrm{irreps}}d_{\rho}W_\rho,
    \end{align}
    along the generators of $\pi_1(\Sigma)$;
    \item The character of the regular representation has the following projection property:
\begin{align}
    \chi_{\textrm{reg}}(g)=\begin{cases} |G| & g=1,\\ 0 & \textrm{otherwise}. \end{cases}
\end{align}
\end{itemize}
Therefore, the operator insertion in the condensation defect $\mathcal{S}_{G\times 1}$ will project the gauge group elements to the trivial subgroup making the domain wall $\mathcal{D}_{1}$. To be concrete consider $G=\mathbb{Z}_N$, in this case we have:
\begin{align}
    \mathcal{S}_{\mathbb{Z}_{N}\times 1}(\Sigma) & =\frac{N}{|H_1(\Sigma,\mathbb{Z}_{N})|}\sum_{\gamma\in H_1({\Sigma},\mathbb{Z}_{N})} W(\gamma)=N\prod_{a=1}^{2g}
   \frac{1}{N} W_{\textrm{reg}}(\gamma_a)={\cal D}_{1}(\Sigma),
    \label{eq:diagonalreview2}
\end{align}
where $W_{\textrm{reg}}=1+W+W^2+\dots W^{N-1}$ is the Wilson line in the regular representation of $\mathbb{Z}_N$ and $a$ run over the generators of $\pi_1(\Sigma)$.

Note that the presentation of $\mathcal{D}_1$ in terms of insertions of $W_{\textrm{reg}}$ along the generators of $\pi_1(\Sigma)$ readily generalizes to non-abelian gauge groups and spacetime dimensions. Therefore, the extension of equation \eqref{eq:diagonalreview1} to the general setting is the leftmost equality in \eqref{eq:diagonalreview2}, which states that the domain wall $\mathcal{D}_1$ is equivalent to the condensation of the Wilson line in the regular representation.

These ideas also apply to the other diagonal domain walls, i.e., domain walls associated with diagonal subgroups $K^{(\textrm{id})}=\{(k,k):k\in K\}\lhd G\times G$ with $K\lhd G$. The character of the trivial representation of $K$ induced to $G$, that is $\operatorname{Ind}_K^G1$, generalizes the property of the regular representation in the sense that
\begin{align}
    \chi_{\operatorname{Ind}_K^G1}(g)=\begin{cases} \frac{|G|}{|K|} & g\in K, \\ 0 & \textrm{otherwise}, \end{cases}
\end{align} 
and $\rho_{\textrm{reg}}=\operatorname{Ind}_1^G1$. Therefore, the insertion of the Wilson line in the representation $\operatorname{Ind}_K^G1$ along the generators of $\pi_1(\Sigma)$ will project the gauge group elements to $K$, generating the domain wall $\mathcal{D}_{K^{(\textrm{id})}}(\Sigma)$. In the $G=\mathbb{Z}_N$ example, this discussion specializes to $\mathcal{S}_{\mathbb{Z}_{N/M}\times 1}=\mathcal{D}_{\mathbb{Z}_M^{(\textrm{id})}}$ for $M|N$.

Diagonal domain walls are orientation reversal invariant, which allows their generalization to higher-codimensional defects $\mathcal{D}_K(\Sigma_n)$ with $K\lhd G$ and $\Sigma_n$ a $n$-dimensional submanifold of spacetime (see Section \ref{sec:review} for details). From the perspective of their condensation expression, the fact that they can be generalized is particularly transparent. Wilson lines are one-dimensional objects, and correspondingly can be condensed along submanifolds of dimension greater than or equal to one. Specifically, the domain wall $\mathcal{D}_1(\Sigma_n)$, correspond to the insertion of $W_{\textrm{Ind}_K^G1}$ along the generators of $\pi_1(\Sigma_n)$.

At last, the condensation expression for the domain wall $\mathcal{D}_{K^{(\textrm{id})},\alpha}$ with $\alpha\in H^{D-1}(K,U(1))$ can be obtained by fusing $\mathcal{D}_{K^{(\textrm{id})}}$ with the general invertible electric defect $W_{\alpha}$, see \eqref{eq:invertibleelectricdect} for its definition.

\paragraph{Magnetic domain wall: magnetic condensation} To move to a more general case it is instructive to consider first the domain wall $\mathcal{D}_{G \times G}$. Again, let us start the discussion with the abelian $G$ case in $D=2+1$ spacetime dimensions. One can derive the following relation:
\begin{align}
    \mathcal{S}_{1\times G}=\mathcal{D}_{G\times G}.
    \label{eq:GGsummary1}
\end{align}
In words, the condensation of the 1-form symmetry $G$ generated by the magnetic defects is equivalent to the domain wall associated to the subgroup $G\times G\lhd G\times G$. Similarly to the diagonal case, the equivalence between the two descriptions comes from two facts:
\begin{itemize}
    \item That the condensation of the 1-form symmetry $G$ generated by the magnetic defects can be reorganized as the product of insertions of the magnetic operator:
    \begin{align}
        M_{\textrm{reg}}=\sum_{[g]\in \textrm{Cl}(G)}M_g,
    \end{align}
    along the generators of $\pi_1(\Sigma)$.
    \item That inserting $M_{\textrm{reg}}$ along a generator $\gamma\in\pi_1(\Sigma)$ will source flux for the dual generator $\tilde{\gamma}$ that has linking number one with $\gamma$ making an effective $G\times G$ boundary. See Fig. \ref{magneticcondensationglobal} for an illustration.
\end{itemize}
Therefore, the operator insertion in the condensation $\mathcal{S}_{1\times G}$ will source the extra holonomies to make the domain wall $\mathcal{D}_{G\times G}$. To be concrete consider again $G=\mathbb{Z}_N$, in this case we have:
\begin{align}
    \mathcal{S}_{1\times\mathbb{Z}_N}(\Sigma)=\frac{1}{N}\sum_{\gamma\in H_1(\Sigma,\mathbb{Z}_N)}M(\gamma)=\frac{1}{N}\prod_{a=1}^{2g}M_{\textrm{reg}}(\gamma_a)=\mathcal{D}_{\mathbb{Z}_N\times\mathbb{Z}_N}(\Sigma),
    \label{eq:GGsummary2}
\end{align}
where $M_{\textrm{reg}}=1+M+M^2+\dots+M^{p-1}$
is the magnetic operator analogous to the Wilson line in the regular representation. 

Just like the previous case, the presentation of $\mathcal{D}_{G\times G}$ in terms of insertions of $M_{\textrm{reg}}$ along the generators of $\pi_1(\Sigma)$ readily generalizes to non-abelian gauge groups and spacetime dimensions. Therefore, the extension of equation \eqref{eq:GGsummary1} to the general setting is the leftmost equality in \eqref{eq:GGsummary2}, which states that the domain wall $\mathcal{D}_{G\times G}$ is equivalent to the condensation of the magnetic object $M_{\textrm{reg}}$, which is the sum of all pure magnetic objects with unit coefficient. In higher spacetime dimensions, the insertions are over codimension-two closed submanifolds of $\Sigma$.

The domain wall $\mathcal{D}_{G\times G}$ cannot be generalized to a higher-codimensional defect. This is evident from its condensation construction. Magnetic defects are already codimension-two objects, and they cannot condense on submanifolds of higher codimension.

\paragraph{Untwisted normal domain walls: electric-magnetic sandwich} Here we invert the order of discussion. We start with general gauge group and spacetime dimensions and then consider particular cases. Specifically, the description as higher gauging symmetry defects when $G$ is abelian.

The condensation expression for a domain wall associated with a general subgroup $H\lhd G\times G$ is obtained by sandwiching the magnetic domain wall $\mathcal{D}_{G\times G}$ with the Wilson lines in the representation $\operatorname{Ind}_{H}^{G\times G}$. By doing this, one projects the $G\times G$ gauge group elements of $\mathcal{D}_{G\times G}$ to its $H$ subgroup. The expression one gets by following this procedure is:
\begin{align}
    \mathcal{D}_H(\Sigma)=\frac{|H|^{n-1}}{|G|^{2n-2}}\sum_{\substack{i_1,\dots,i_{n} \\ j_1,\dots ,j_{n}}} c_{i_1j_1}^HW_{i_1}(\gamma_1)\dots c_{i_{n}j_{n}}^HW_{i_{n}}(\gamma_{n})\mathcal{D}_{G\times G}(\Sigma)W_{{j_1}}(\gamma_1)\dots W_{j_{n}}(\gamma_{n}),
    \label{eq:generalsummary}
\end{align}
where $i_k,j_k$ run over irreducible representations of $G$ and $\gamma_1,\dots, \gamma_{n}$ are $n$ generators of $\pi_1(\Sigma)$. The coefficients are determined by the subgroup $H\lhd G\times G$ as:
\begin{align}
    c^H_{ij}=\langle\chi_{{\operatorname{Ind}_{H}^{G\times G}1}},\chi_{\rho_i}\chi_{\rho_j}\rangle=\frac{1}{|H|}\sum_{(g_L,g_R)\in H}\chi_{\rho_i}(g_L)\chi_{\rho_j}(g_R),
\end{align}
with $\rho_i$ the irreducible representations of $G$. We remark that in \eqref{eq:generalsummary}, writing the Wilson lines on the left and right of $\mathcal{D}_{G\times G}$ corresponds to having them inserted on the left and right of the magnetic insertion, respectively. Moving them from one side to the other costs a braiding phase.

\paragraph{Domain walls as higher gauging condensation defects in $\mathbb{Z}_p$ gauge theory} The previous results allow us to derive the dictionary between the domain walls and the condensation definition of codimension-one topological operators. In $D=2+1$ spacetime dimensions, given a particular group $H\lhd G\times G$ one can rewrite \eqref{eq:generalsummary} as the higher gauging of a subgroup of the 1-form global symmetry. By doing this for all subgroups of $\mathbb{Z}_p\times\mathbb{Z}_p$ when $p$ is prime we find all but the last two rows of the dictionary in Table \ref{tab:dictionaryZp}, which makes explicit the relation between domain walls and condensation defects in $\mathbb{Z}_p$ gauge theory. In Section \ref{sec:ZN} we derive these results in the general context of non-prime $p$ and arbitrary spacetime dimension (see Table \ref{tab:dictionaryZN} for a summary).
\begin{table}[ht]
    \centering
    \begin{tabular}{|c|c|}
    \hline Domain wall & Condensation\\ \hline
    $\mathcal{D}_{\mathbb{Z}_p^{(\textrm{id})}}$ & $\mathcal{S}_1$ \\ 
    $\mathcal{D}_{\mathbb{Z}_p^{(m)}}$  & 
    $\mathcal{S}_{\mathbb{Z}_{p}\times\mathbb{Z}_{p},\frac{m}{1-m}}$\\
    $\mathcal{D}_{1}$ & $\mathcal{S}_{\mathbb{Z}_{p}\times 1}$ \\
    $\mathcal{D}_{1\times \mathbb{Z}_p}$ & $\mathcal{S}_{\mathbb{Z}_p\times\mathbb{Z}_p}$\\
    $\mathcal{D}_{\mathbb{Z}_p\times1}$ & $\mathcal{S}_{\mathbb{Z}_p\times\mathbb{Z}_p,-1}$ \\
    $\mathcal{D}_{\mathbb{Z}_p\times\mathbb{Z}_p}$ & $\mathcal{S}_{1\times\mathbb{Z}_p}$ \\  
    $\mathcal{D}_{\mathbb{Z}_p\times\mathbb{Z}_p,1}$ & 
    $\mathcal{S}_{\mathbb{Z}_{p}^{(\textrm{id})}}$\\
    $\mathcal{D}_{\mathbb{Z}_p\times\mathbb{Z}_p,m}$ & 
    $\mathcal{S}_{\mathbb{Z}_{p}^{(m)}}$\\ \hline
    \end{tabular}
    \caption{Dictionary between higher gauging condensation defects and domain walls in $\mathbb{Z}_p$ gauge theory. Domain walls are classified by subgroups of the folded theory gauge group $H\lhd \mathbb{Z}_p\times \mathbb{Z}_p$ (with the left and right factors corresponding respectively to the left and right of the domain wall) with a choice of topological action $\alpha\in H^{D-1}(H,U(1))$. In $D=2+1$ spacetime dimensions, higher gauging condensation defects are classified by subgroups of the 1-form global symmetry $H\lhd \mathbb{Z}_p\times\mathbb{Z}_p$ (with the left and right factors generated respectively by Wilson lines and magnetic defects) with a choice of torsion term $f\in H^{2}(H,U(1))$. In the table, $\mathbb{Z}_p^{(m)}=\{(mn,n):n\in\mathbb{Z}_p\}\lhd \mathbb{Z}_p\times\mathbb{Z}_p$ with $1<m<p$, and $\mathbb{Z}_p^{(1)}=\mathbb{Z}_p^{(\textrm{id})}$. Note that, because $p$ is prime, $1-m$ is invertible and $f=\frac{m}{1-m}\in\mathbb{Z}_p$. See Table \ref{tab:dictionaryZN} for the non-prime $p$ and higher dimensional generalization.}
    \label{tab:dictionaryZp}
\end{table}

It remains, to understand the attachment of a topological action in the domain wall formalism and its relation to higher gauging. For $\mathbb{Z}_p$ there are $p-1$ such domain walls which are labeled by $H^{2}(\mathbb{Z}_p\times\mathbb{Z}_p,U(1))=\mathbb{Z}_p$, likewise there are $p-1$ different Dyons associated to the automorphism subgroups of the 1-form global symmetry. By viewing the attachment of the topological action as decorating the domain wall $\mathcal{D}_{\mathbb{Z}_p\times \mathbb{Z}_p}=\mathcal{S}_{1\times\mathbb{Z}_p}$ with suitable electric charges it becomes clear the correspondence between the two as displayed in Table \ref{tab:dictionaryZp}.

In spacetime dimension $D\neq2+1$, the higher-form symmetries generated by Wilson lines and magnetic defects have a different form degree, resulting in fewer subgroups compared to when they share the same form degree. For example, in this case, there are no diagonal subgroups corresponding to the condensation of Dyons. This feature leads to a mismatch in the number of domain walls and higher gauging condensation defects. This apparent problem is resolved by introducing the concept of ``sequential higher gauging". Sequential higher gauging involves the higher gauging of the $\mathbb{Z}_N\times\mathbb{Z}_N$ 1-form global symmetry generated by magnetic defects and the dual symmetry that emerges after gauging the $(D-2)$-form symmetry generated by Wilson lines. Since both factors have the same form degree, all subgroups can be considered and there is no number mismatch between the two constructions. In particular, one can generate the general invertible electric defects \eqref{eq:invertibleelectricdect} by this procedure. See Section \ref{sec:squentialhighergauging} for details.

\paragraph{Automorphism 0-form symmetry as higher gauging defects in abelian theories} As an application of our framework, we can derive an expression for the invertible automorphism domain walls in abelian gauge theories as higher gauging condensation defects. In Section \ref{sec:automorphism} we derive these expressions for the domain walls associated with a set of automorphism generators. Here, we present the dictionary for all automorphisms of $G=\mathbb{Z}_2\times\mathbb{Z}_2$ that can be derived using these expressions. 

The Klein four group $\mathbb{Z}_2\times\mathbb{Z}_2$ has 3 generators of order $2$, which we denote by $A=(1,0)$, $B=(0,1)$ and $C=(1,1)$. The automorphism group of $\mathbb{Z}_2\times\mathbb{Z}_2$ is isomorphic to ordered lists of the three generators $\textrm{Aut}(\mathbb{Z}_2\times\mathbb{Z}_2)\cong \{\sigma\cdot [A,B,C]:\sigma\in S_3\}\cong S_3$ with group product given by composing the $S_3$ elements. The bijection is obtained by identifying the first and second elements in the list with the image of $A=(1,0)$ and $B=(0,1)$ under the corresponding automorphism. For instance, the automorphism associated with the list $(23)\cdot[A,B,C]=[A,C,B]$ maps $(1,0)=A\mapsto A=(1,0)$ and $(0,1)=B\mapsto C=(1,1)$. We use this isomorphism to denote the elements of $\operatorname{Aut}(\mathbb{Z}_2\times\mathbb{Z}_2)$ by elements of $S_3=(\textrm{id},(13),(23),(12),(123),(132)\}$. Likewise, denote the automorphism subgroup associated with $\sigma\in S_3$ by $(\mathbb{Z}_2\times\mathbb{Z}_2)^{(\sigma)}=\{(\sigma\cdot X,X):X\in\mathbb{Z}_2\times\mathbb{Z}_2\}\leq (\mathbb{Z}_2\times\mathbb{Z}_2)\times(\mathbb{Z}_2\times\mathbb{Z}_2)$. Then, we can derive the dictionary presented in Table \ref{tab:Z2Z2aut}. See Section \ref{Z2Z2} for the convention used for the torsion term.

\begin{table}[ht]
    \centering
    \begin{tabular}{|c|c|}
         \hline Domain wall & Condensation \\ \hline
         $\mathcal{D}_{(\mathbb{Z}_2\times\mathbb{Z}_2)^{(\textrm{id})}}$ & $\mathcal{S}_1$ \\ 
         $\mathcal{D}_{(\mathbb{Z}_2\times\mathbb{Z}_2)^{(13)}}$ & $\mathcal{S}_{(1\times \mathbb{Z}_{2})\times(\mathbb{Z}_{2}\times 1),1}$\\  
        $\mathcal{D}_{(\mathbb{Z}_2\times\mathbb{Z}_2)^{(23)}}$ & $ \mathcal{S}_{(\mathbb{Z}_{2}\times 1)\times(1\times\mathbb{Z}_{2}),1}$\\ 
        $\mathcal{D}_{(\mathbb{Z}_2\times\mathbb{Z}_2)^{(12)}}$ & $ \mathcal{S}_{\mathbb{Z}_{2}^{\textrm{(id)}}\times\mathbb{Z}_{2}^{\textrm{(id)}},1}$ \\ 
        $\mathcal{D}_{(\mathbb{Z}_2\times\mathbb{Z}_2)^{(123)}}$ & $ \mathcal{S}_{(\mathbb{Z}_{2}\times \mathbb{Z}_{2})\times (\mathbb{Z}_{2}\times \mathbb{Z}_{2}),(0,1,1,1)}$\\ 
        $\mathcal{D}_{(\mathbb{Z}_2\times\mathbb{Z}_2)^{(132)}}$ & $ \mathcal{S}_{(\mathbb{Z}_{2}\times \mathbb{Z}_{2})\times (\mathbb{Z}_{2}\times \mathbb{Z}_{2}),(1,1,1,0)}$\\  \hline 
    \end{tabular}
    \caption{Dictionary between higher gauging condensation defects and $\operatorname{Aut}(\mathbb{Z}_2\times\mathbb{Z}_2)\cong S_3$ automorphism domain walls.}
    \label{tab:Z2Z2aut}
\end{table}

Using the higher gauging condensation expression for the domain walls, we can compute the fusion rules and the transformations of other operators by leveraging the algebraic properties of Wilson lines and magnetic defects. Notably, applying this higher gauging expression reveals that the computation of general fusion rules depends in subtle ways on specific number-theoretic properties. However, in all cases where we can compute these rules easily, they consistently agree with the more general fusion rules and transformations derived from the domain wall definition in \cite{Cordova:2024jlk}.

\paragraph{Gauging 0-form symmetry: $\mathbb{D}_4$ gauge theory from gauging swap symmetry} The group $\mathbb{D}_4\cong(\mathbb{Z}_2\times\mathbb{Z}_2)\rtimes\mathbb{Z}_2$ can be constructed as the group extension of $\mathbb{Z}_2\times\mathbb{Z}_2$ by $\mathbb{Z}_2$ with the $\mathbb{Z}_2$ acting by swapping generators. It follows that $\mathbb{D}_4$ gauge theory can be constructed by gauging symmetry generated by the automorphism domain wall $\mathcal{D}_{(\mathbb{Z}_2\times\mathbb{Z}_2)^{(23)}}$. By using its explicit condensation expression reviewed in Table \ref{tab:Z2Z2aut}, i.e., $ \mathcal{S}_{(\mathbb{Z}_{2}\times 1)\times(1\times\mathbb{Z}_{2}),1}$, one can explicitly derive the action:
\begin{align}
    S_{\mathbb{D}_4}=i\pi\int_\mathcal{M}( \tilde{a}^{(D-2)}\cup da^{(1)}+\tilde{b}^{(D-2)}\cup db^{(1)}+\tilde{c}^{(D-2)}\cup dc^{(1)}+a^{(1)}\cup \tilde{b}^{(D-2)}\cup c^{(1)}),
    \label{eq:D4action}
\end{align}
which is the action for $\mathbb{D}_4$. In Section \ref{sec:gaugingD8} we show in detail this derivation and explain how to understand the operators of $\mathbb{D}_4$ gauge theory as combinations of the $\mathbb{Z}_2$ factors associated with the corresponding group extension.

\section{Review of non-invertible symmetries as domain walls}
\label{sec:review}

Finite-group gauge theories can be defined by a path integral on the lattice \cite{Dijkgraaf:1989pz}. The partition function is given by a finite summation over gauge equivalence classes of flat gauge field configuration and is weighted by a topological action. Specifically, given a triangulation of a manifold $\mathcal{M}$ one has:
\begin{itemize}
    \item  A \textit{flat gauge field configuration} is an assignment of group elements $g_{ij}\in G$ to every oriented 1-simplex $[i,j]$ ($i<j$) such that $g_{ij}\cdot g_{jk}\cdot g_{ki}=1$ for every 2-simplex $[i,j,k]$.
    \item Two gauge field configurations are said to be \textit{gauge equivalent} if exists an assignment of group elements $h_i\in G$ to every vertex $v_i$ of the triangulation of $\mathcal{M}$ such that $g_{ij}^\prime=h_i\cdot g_{ij}\cdot h_j^{-1}$ for every 1-simplex $[i,j]$. 
    \item The total action is a product of local terms, i.e., a product of algebraic $D$-cocycles $[\alpha_D]\in H^D(G,U(1))$ that are evaluated on the gauge fields of each $D$-simplex.
\end{itemize}
See \cite{Cordova:2024jlk} for more details. 

Furthermore, finite-group gauge theories have three basic operators in generic dimensions:
\begin{itemize}
    \item Wilson lines: line operators $W_\rho$ labeled by representations $\rho$ of $G$.
    \item General invertible electric defects: dimension-$n$ operators $W_{\alpha_n}$ labeled by elements of $[\alpha_n]\in H^n(G,U(1))$ and defined by
    \begin{equation}
        W_{\alpha_n}(\Sigma_n) = \prod_{[v_{i_1},\dots,v_{i_{n+1}}]\in\Sigma_n}\alpha_n(g_{i_1i_2}\dots,g_{i_ni_{n+1}})^{\epsilon_i},
   \label{eq:invertibleelectricdect}
    \end{equation}
    with the product over the $n$-simplices of the $n$-dimensional submanifold $\Sigma_n$ and
    $\epsilon_i=\pm1$ depending on whether the orientation of the $n$-simplex agrees with that of $\mathcal{M}$.
    \item Magnetic defects: codimension-two operators $M_g$ labeled by conjugacy classes $[g]$ of $G$.
\end{itemize}
See \cite{Cordova:2024jlk} for a review. These operators will be used as the condensation building blocks for the domain walls.

\paragraph{Non-invertible symmetries as domain walls on the lattice}

In a recent work, we showed how to construct codimension-one topological operators on the lattice for untwisted finite-group $G$ gauge theory in generic dimensions $D$. The construction uses the data that classifies gapped boundaries in the folded theory:
\begin{itemize}
    \item A subgroup $H\leq G\times G$;
    \item A topological action $\alpha\in H^{D-1}(H,U(1))$ for the subgroup.
\end{itemize}
Given this data, the domain wall $\mathcal{D}_{H}(\Sigma)$ on a closed, orientable and connected codimension-one submanifold $\Sigma$ is defined by setting the gauge group elements and gauge transformations on $\Sigma$ to be in the subgroup $H\leq G\times G$ and by properly gluing the $H$ gauge group elements of $\Sigma$ with the $G$ gauge group elements of the rest of spacetime. See Fig. \ref{fig:definitionofholonomy} for an example of a valid flat gauge field configuration and Fig. \ref{definitiongaugetransformation} for an example of an equivalent gauge field configuration and \cite{Cordova:2024jlk} for more details. The domain wall  $\mathcal{D}_{H,\alpha}(\Sigma)$ is then obtained by attaching the topological action $\alpha\in H^{D-1}(H,U(1))$ on $\Sigma$. The left and right of the subgroup $H\leq G\times G$ are defined globally with respect to the orientation of the normal bundle $N\Sigma$. Orientation-reversal of a domain wall, denoted with a bar, flips the normal vector and exchanges the left and right.

\begin{figure}[ht]
    \centering
    \begin{minipage}{0.485\textwidth}
        \centering
        \includegraphics[width=\textwidth]{figures/latticedefinition.tex}
    \caption{An example of flat gauge field configuration in a patch of the domain wall associated with the subgroup $H\leq G\times G$. Note that the holonomies along $(v_2,v_3,v_4,v_2)$ and $(v_1,v_3,v_4,v_1)$ are trivial, but the holonomy along $(v_1,v_3,v_2,v_4,v_1)$ is not.}
    \label{fig:definitionofholonomy}
    \end{minipage}\hfill
    \begin{minipage}{0.485\textwidth}
        \centering
        \includegraphics[width=\textwidth]{figures/gaugetransformations.tex}
        \caption{An example of equivalent gauge field configuration for the same patch of the domain wall associated with the subgroup $H\leq G\times G$. They are related by a gauge transformation that has image $(k_L,k_R)\in H$ on $v_3$ and image $1\in H$ on $v_1,v_2,v_4$.}
        \label{definitiongaugetransformation}
    \end{minipage}
\end{figure}

\paragraph{Fusion rules of domain walls} 

Despite being simple, this definition is generic because it applies to any group $G$ and dimension $D$. Furthermore, it allows simple derivations of some of the fusion rules and the transformation of other operators. Particular interest was devoted to the following class of domain walls:
\begin{itemize}
    \item Automorphism domain walls: $\mathcal{D}_{G^{(\phi)}}$, with $\phi\in\operatorname{Aut}(G)$;
    \item Diagonal domain walls: $\mathcal{D}_{K^{(\textrm{id})},\alpha}$, with $K\lhd G$ and $\alpha\in H^{D-1}(K,U(1))$;\footnote{The topological action is evaluated on the right entry of $K^{(\phi)}$. More formally, as a topological action of $H^{D-1}(K^{(\phi)},U(1))$ it is $R^*\alpha$, i.e., the pullback of $\alpha\in H^{D-1}(K,U(1))$ by $R:K^{(\phi)}\rightarrow K$ defined by $R(\phi\cdot k,k)=k$.}
    \item Magnetic domain wall: $\mathcal{D}_{G\times G}$;
\end{itemize}
which generates a closed fusion algebra and obeys the following basic fusion rules:
\begin{align}
    \vphantom{\frac{1}{1}} \mathcal{D}_{G^{(\phi)}}\times\mathcal{D}_{G^{(\phi^\prime)}} & =
    \mathcal{D}_{G^{(\phi\circ \phi^\prime)}},\label{eq:fusionautomorphism}\\
    \vphantom{\frac{1}{1}} \mathcal{D}_{G^{(\phi)}}\times\mathcal{D}_{G\times G} =\mathcal{D}_{G\times G} \times\mathcal{D}_{G^{(\phi)}}&=
    \mathcal{D}_{G\times G} , \label{eq:GGanihilation}\\
    \vphantom{\frac{1}{1}}\mathcal{D}_{G^{(\phi)}}\times\mathcal{D}_{K^{(\textrm{id})},\alpha} & =
    \mathcal{D}_{K^{(\phi)},\alpha} ,\label{eq:autdiacommute1}\\
    \vphantom{\frac{1}{1}}\mathcal{D}_{K^{(\textrm{id})},\alpha}\times \mathcal{D}_{G^{(\phi)}}=\mathcal{D}_{G^{(\phi)}}\times \mathcal{D}_{\phi^{-1}(K),\phi^*\alpha} &=
    \mathcal{D}_{(\phi^{-1}(K))^{(\phi)},\phi^*\alpha} ,\label{eq:autdiacommute2}\\
    \vphantom{\frac{1}{1}} \mathcal{D}_{K^{(\textrm{id})},\alpha}\times\mathcal{D}_{K^{\prime(\textrm{id})},\alpha^\prime} &=
    \frac{|G|}{|K\cdot K^\prime|}\mathcal{D}_{(K\cap K^\prime)^{(\textrm{id})},\alpha\cdot\alpha^\prime}\label{eq:fusiondiag},\\
    \vphantom{\frac{1}{1}} \mathcal{D}_{K_L^{(\textrm{id})},\alpha_L}\times\mathcal{D}_{G\times G}\times \mathcal{D}_{K_R^{(\textrm{id})},\alpha_R} &=
    \mathcal{D}_{K_L\times K_R,\alpha_L\times\alpha_R}, \label{eq:facrtorizeddefinition}\\
    \vphantom{\frac{1}{1}} \mathcal{D}_{G\times G}\times\mathcal{D}_{K^{(\textrm{id})},\alpha}\times \mathcal{D}_{G\times G} & =    
    \mathcal{Z}(K,\alpha)\mathcal{D}_{G\times G} \label{eq:fusionGGdiagGG},
\end{align}
with $\phi\circ\phi^\prime$ the automorphism composition of $\phi,\phi^\prime\in\operatorname{Aut}(G)$; $\phi^{-1}(K)$ the image of $K$ under $\phi^{-1}$; $\phi^*\alpha$ the pullback of $\alpha:K^{D-1}\rightarrow U$ by $\phi:\phi^{-1}(K)\rightarrow K$; $\alpha\cdot\alpha^\prime\vert_{K\cap K^\prime}\in H^{D-1}(K\cap K^\prime,U(1))$; $|G|/|K\cdot K^\prime|$ the 0-form partition function of $G/K\cdot K^\prime$ gauge theory on $\Sigma$; and $\mathcal{Z}(K,\alpha)$ the partition function of $K$ gauge theory twisted by $\alpha$ on $\Sigma$. The fusion of factorized domain walls can be computed using associativity and the basic fusion rules above and is given by:
\begin{align}
        \mathcal{D}_{K_L\times K_R,\alpha_L\times\alpha_R}\times \mathcal{D}_{K_L^\prime\times K_R^\prime,\alpha_L^\prime\times\alpha_R^\prime}=\frac{|G|}{|K_R\cdot K_L^\prime|}\mathcal{Z}(K_R\cap K_L^\prime,\alpha_R\cdot\alpha_L^\prime)\mathcal{D}_{K_L\times K_R^\prime,\alpha_L\times\alpha_R^\prime}.
    \label{eq:fusionfactorized}
\end{align}
Under orientation-reversal we have
\begin{align}
    \overline{\mathcal{D}}_{G^{(\phi)}}=\mathcal{D}_{G^{(\phi^{-1})}},\qquad \overline{\mathcal{D}}_{K^{(\textrm{id})},\alpha}=\mathcal{D}_{K^{(\textrm{id})},\alpha},\qquad \overline{\mathcal{D}}_{G\times G}=\mathcal{D}_{G\times G}.
    \label{eq:orientationreversal}
\end{align}
Furthermore, orientation-reversal inverts the order in a fusion product. For instance:
\begin{align}
    \overline{\mathcal{D}}_{K^{(\phi)}}=\overline{\mathcal{D}}_{K^{(\textrm{id})}}\times \overline{\mathcal{D}}_{G^{(\phi)}}=\mathcal{D}_{G^{(\phi^{-1})}}\times\mathcal{D}_{\phi(K),\phi^{-1*}\alpha}=\mathcal{D}_{(\phi(K))^{(\phi^{-1})},\phi^{-1*}\alpha}.
\end{align}

\paragraph{Higher codimensional topological operators: Cheshire strings} In defining the domain wall $\mathcal{D}_{H,\alpha}(\Sigma)$, a crucial aspect is the orientation of the normal bundle $N\Sigma$. This orientation allows for the consistent global definition of `left' and `right', corresponding to the left and right components of the subgroup $H < G \times G$. Diagonal domain walls are invariant under orientation reversal and can be generalized as higher codimensional operators. The $n$-dimensional generalizations of diagonal domain walls are classified by subgroups $K < G$ and a topological action $\alpha \in H^n(K, U(1))$, and they obey the following fusion rule:
\begin{align}
    \vphantom{\frac{1}{1}}\mathcal{D}_{K,\alpha}(\Sigma_n)\times\mathcal{D}_{K,\alpha^\prime}(\Sigma_n) =
    \frac{|G|}{|K\cdot K^\prime|}\mathcal{D}_{(K\cap K^\prime),\alpha\cdot\alpha^\prime}(\Sigma_n)\label{eq:highercodimension}
\end{align}
with $\Sigma_n$ a $n$-dimensional submanifold of $\mathcal{M}$. This fusion rule generalizes the fusion rule of Cheshire strings \cite{Else:2017yqj,Johnson-Freyd:2020twl}.

\paragraph{Transformation of other operators}
From the modifications of gauge transformations along $\Sigma$ in the presence of $\mathcal{D}_{H}(\Sigma)$ one modifies which operators are gauge invariant. From this feature, we derive the transformations of other operators on the domain wall. Some examples are:
\begin{align}
    & \mathcal{D}_{1}\cdot W_{\rho_i}=\sum_{\rho_k\in\textrm{irreps}}d_id_kW_{\rho_k}, && \mathcal{D}_{1}\cdot M_g=0,\\
    & \mathcal{D}_{G\times G}\cdot W_{\rho_i}=0, && \mathcal{D}_{G\times G}\cdot M_g=\sum_{[k]\in \operatorname{Cl}(G)}M_{k},\label{eq:action_GG}\\
    & \mathcal{D}_{G^{(\phi)}}\cdot W_{\rho_i}=W_{\rho_i\cdot\phi^{-1}}, && \mathcal{D}_{G^{(\phi)}}\cdot M_g=M_{\phi(g)}\label{eq:action_automorphism}.
\end{align}
for all simple Wilson lines $W_{\rho_i}$ and magnetic defects $M_g$ where $d_i$ is the dimension of the irreducible representation $\rho_i$. We remark that the transformation of the invertible symmetry operator $\mathcal{D}_{G^{(\phi)}}$ preserves the Aharonov-Bohm braiding between the Wilson lines and the magnetic operator

\paragraph{Non-invertible electric-magnetic duality domain wall} Note that in the data specifying a domain wall, the dimension dependence arises from the topological action $\alpha \in H^{D-1}(H, U(1))$. By examining the specific case of $G = \mathbb{Z}_2$ gauge theory in $D = 3$, we compute the fusion and transformation of other operators for the domain wall associated with the subgroup $H = \mathbb{Z}_2 \times \mathbb{Z}_2 \lhd \mathbb{Z}_2 \times \mathbb{Z}_2$ with the non-trivial topological action $\alpha \in H^2(H, U(1)) = \mathbb{Z}_2$. We find:
\begin{align}
    & \mathcal{D}_{\mathbb{Z}_2\times\mathbb{Z}_2,\alpha}\times \mathcal{D}_{\mathbb{Z}_2\times\mathbb{Z}_2,\alpha}=1, \label{eq:nontrivialfusion}\\
    & \mathcal{D}_{\mathbb{Z}_2\times\mathbb{Z}_2,\alpha}\cdot W=M, && \mathcal{D}_{\mathbb{Z}_2\times\mathbb{Z}_2,\alpha}\cdot M=W \label{eq:nontrivialaction},
\end{align} 
showing that $\mathcal{D}_{\mathbb{Z}_2\times\mathbb{Z}_2,\alpha}$ is the electric-magnetic duality symmetry defect. The procedure we follow for the computation is more general and demonstrates that $\mathcal{D}_{G \times G, \alpha}$ extends electric-magnetic duality to higher dimensions, generic gauge groups $G$, and topological actions $\alpha \in H^{D-1}(G, U(1))$. This class of domain walls combines invertible electric and magnetic defects and generally obeys a non-invertible fusion rule, see \cite{Cordova:2024jlk} for examples.

\section{Domain walls as condensation defects}
\label{sec:operator}

In Section \ref{sec:review} we reviewed the construction of non-invertible symmetries of finite-group gauge theories as domain walls on the lattice \cite{Cordova:2024jlk}. Here, we show how to realize these symmetries as condensation defects, i.e., as suitable insertions of lower dimensional operators along a codimension-one submanifold \cite{Gaiotto:2019xmp}. Specifically, we show how to construct the domain walls $\mathcal{D}_{H}$ for $H\lhd G\times G$ as a condensation of Wilson lines and magnetic defects. In summary, we will show that the defect $\mathcal{D}_{G\times G}$ corresponds to the condensation of magnetic objects, and that $\mathcal{D}_H$ can be obtained by further sandwiching the magnetic insertion with suitable Wilson lines, thereby projecting the $G\times G$ gauge fields to its $H\lhd G\times G$ subgroup. Provided these condensation expressions, in Section \ref{sec:domainwallascondensationsfusion}, we compute the fusion of the symmetry defects using the algebraic properties of the lower dimensional objects that make them. This allows us to verify the fusion rules derived using the domain wall lattice definition in \cite{Cordova:2024jlk}.

\subsection{Diagonal domain walls: electric condensations}
\label{sec:diagonalcondensation}

To understand the general idea of using Wilson lines to project the gauge group elements to a particular subgroup it is useful to start with the diagonal domain walls. They are made out of Wilson lines that project the $G$ gauge group elements along $\Sigma$ to a particular subgroup $K\lhd G$.

\paragraph{Identity diagonal domain wall}

Let us consider what is arguably the simplest case, the domain wall associated with the trivial subgroup of $G\times G$, i.e., $\mathcal{D}_1(\Sigma)$. The domain wall $\mathcal{D}_1(\Sigma)$ is defined by restricting the gauge field configurations and its gauge transformations along $\Sigma$ to be elements of $1\leq G\times G$. This can be achieved in the following way:
\begin{itemize}
    \item The gauge fields on the wall are enforced to be trivial by inserting the Wilson line in the regular representation divided by $|G|$ in all closed loops of $\Sigma$ because:
    \begin{align}
        \chi_{\rho_{\textrm{reg}}}(g)=\textrm{Tr}[\rho_{\textrm{reg}}(g)]=\begin{cases}
        |G| & g =1, \\ 0 & \textrm{otherwise},
    \end{cases}~
    \label{eq:regularcharacter}
\end{align}
for the regular representation $\rho_{\textrm{reg}}$ of $G$ and above $g\in G$.

\item In addition to the projection of the gauge field, we need to correct for gauge transformations. In the definition of $\mathcal{D}_1$, gauge transformations are restricted to the subgroup $1\leq G$. Because we are not changing the gauge transformations when we insert the projector, we need to change the correction due to the volume of gauge transformations which gives $1/|G|$ for each disconnected component. To change $1/|G|$ to $1$ we need, therefore, to multiply by $|G|$.
\end{itemize}
 
Thus the domain wall can be described in the following equivalent ways:
\begin{equation}
{\cal D}_1(\Sigma)=    |G|\prod_{\text{closed loops }\gamma}\frac{1}{|G|}W_{\textrm{reg}}(\gamma)=|G|\prod_{\gamma\in\pi_1(\Sigma)}\frac{1}{|G|}W_{\textrm{reg}}(\gamma)=|G|\prod_{\gamma\in S}\frac{1}{|G|}W_{\textrm{reg}}(\gamma)~,
    \label{electriccondensationidentity}
\end{equation}
where the first product is over all closed loops on $\Sigma$ and $S$ is a generating set of $\pi_1(\Sigma)$. The equivalence between the different expressions can be obtained in the following way. First, one deforms all insertions over homotopy equivalent loops to some representative cycle. Then, one uses the fusion rule $W_{\textrm{reg}}(\gamma)W_{\textrm{reg}}(\gamma)=|G|W_{\textrm{reg}}(\gamma)$ to replace the multiple insertions to a single insertion. At last, consider $\gamma=\gamma_1\dots \gamma_k$ with $\gamma_i\in S$, which gives $g_\gamma=g_{\gamma_1}\dots g_{\gamma_k}$. If $g_{\gamma_i}=e$ for all $\gamma_i\in S$ then $g_\gamma=e$. Hence, $\frac{1}{|G|}W_{\textrm{reg}}(\gamma)$ equals 1 if $W_{\textrm{reg}}(\gamma_i)$ are inserted for all $\gamma_i\in S$.

\paragraph{General diagonal domain wall}

The case of general diagonal domain wall, $\mathcal{D}_{K^{(\textrm{id})}}$ with $H=K^{(\textrm{id})}=\{(k,k):k\in K\}\leq G\times G$ and $K\lhd G$, is similar. Consider the induced representation to $G$ from the trivial representation of $K\lhd G$, i.e., $\operatorname{Ind}_K^G1$. In the previous case $K=1$ and $\rho_\textrm{reg}=\operatorname{Ind}_1^G1$. Then, the character of this induced representation is the induced character to $G$ from the trivial character of $K\lhd G$ and is the permutation character of $G$ acting on the right cosets of $K$ by multiplication from the right. The induced character has the property
\begin{equation}
    \chi_{\operatorname{Ind}_K^G1}(g)=\begin{cases} \frac{|G|}{|K|} & g\in K, \\ 0 & \textrm{otherwise}. \end{cases}~
    \label{eq:projector_diagonal}
\end{equation}
Therefore, the diagonal domain wall can be described by the operator
\begin{equation}
{\cal D}_{K^{(\textrm{id})}}(\Sigma)= 
\frac{|G|}{|K|}
\prod_{\textrm{closed loops }\gamma}\frac{|K|}{|G|}W_{\operatorname{Ind}_K^G1}(\gamma),
    \label{eq:oprdiagwall} 
\end{equation}
where the pre-factor corrects for the volume of gauge transformations as explained in the previous case. Similarly to the $K=1$ case, the insertion defining $\mathcal{D}_{K^{(\textrm{id})}}$ can be simplified to an insertion over $\pi_1(\Sigma)$ or over a generating set of $\pi_1(\Sigma)$

In addition, we can also modify the operator by stacking with additional invertible electric operators \ref{eq:invertibleelectricdect} associated to $\alpha\in H^{D-1}(K,U(1))$ \cite{Barkeshli:2022edm}. This gives $\mathcal{D}_{K^{(\textrm{id})},\alpha}$, the twisted diagonal operators.

\paragraph{Higher codimensional operators}

We remark that the above definition is suitable for the generalization to higher dimensional topological operators discussed in \eqref{eq:highercodimension}. Instead of inserting along all closed loops of the codimension-one surface $\Sigma$, one inserts along all closed loops of the $n$-dimensional surface $\Sigma_n$. Also, to attach a topological action we fuse our defect with the corresponding $n$-dimensional invertible electric defect \eqref{eq:invertibleelectricdect}.

In \cite{Tantivasadakarn_2024} it was asked  ``Why did we not include Cheshire strings in the
description of $D=2+1$ topological order?". From \eqref{electriccondensationidentity} we see that in $D=2+1$ the codimension-one operator corresponding to the ``Cheshire strings" is the same as the Wilson line in the regular representation, i.e., $\mathcal{D}_{1}(\gamma)=W_{\textrm{reg}}(\gamma)$. In $D=3+1$, on the other hand, the equivalent codimension-one operator $\mathcal{D}_1(\Sigma_2)$ is a condensation of $W_{\textrm{reg}}$.

\paragraph{Frobenius algebra interpretation}

The composite line $\operatorname{Ind}_K^G1$ is associated with a symmetric separable Frobenius algebra of $\operatorname{Rep}(G)$ and \eqref{eq:oprdiagwall} should correspond to its generalized gauging along the codimension-one surface $\Sigma$ \cite{ostrik2001module,Diatlyk:2023fwf}. The character property \eqref{eq:projector_diagonal} connects the idea of generalized gauging and the domain wall lattice construction reviewed in Section \ref{sec:review}. Further decorating $\Sigma$ with the invertible electric defect defined in \eqref{eq:invertibleelectricdect} which is classified by $\alpha\in H^{D-1}(K,U(1))$ should be equivalent to different choices of $(D-1)$-topological junctions for $\operatorname{Ind}_K^G1$ on $\Sigma$.

\subsection{Factorized domain walls: magnetic condensation}

Consider factorized domain walls corresponding to the subgroups $H=K_L\times K_R\leq  G\times G$ for $K_L,K_R\lhd G$. The difference between having two diagonal domain walls $\mathcal{D}_{K_L^{(\textrm{id})}}$ and $\mathcal{D}_{K_R^{(\textrm{id})}}$ close together (which would fuse according to \eqref{eq:fusiondiag}, the diagonal fusion) and having the domain wall $\mathcal{D}_{K_L\times K_R}$ is the presence of magnetic defects on the middle. The insertion of magnetic defects comes from $\mathcal{D}_{G\times G}$ as in \eqref{eq:fusionGGdiagGG}.

\paragraph{Factorized domain wall with $H=G\times G$}

\begin{figure}[ht]
    \centering
    \begin{tikzpicture}
    \begin{knot}[
    clip width=3,
    flip crossing={2},
    ]
    \strand [ultra thick, purple ] (0,0) circle (2.0cm);
    \strand [ultra thick, cyan] (0,0) circle (0.6cm);
    \end{knot}
    \fill [yellow!20!white, opacity=.60,even odd rule] (0,0) circle[radius=2cm] circle[radius=0.6cm];
    \node at (-2.5,0) {$\sum\limits_{g_L}$};
    \node at (0,0) {\textbullet};
    \node at (0.3,0) {${}_{g_L}$};
    \node at (-1.25,0) {$M_{\textrm{reg}}$};
    \node at (-1.25,-0.5) {$\times$};
    \end{tikzpicture}
    \caption{The holonomy in the blue path is $g_L$. Now consider the magnetic operator $M_{g_Rg_L^{-1}}$. For this insertion, the holonomy around the red path is $g_R$. By summing over $g_L$ in the path integral and over conjugacy classes, in the insertion of $M_{\textrm{reg}}$, we will be summing over all $(g_L,g_R)\in G\times G$ for this non-trivial cycle which produces the $G\times G$ domain wall.}
    \label{magneticcondensationglobal}
\end{figure}
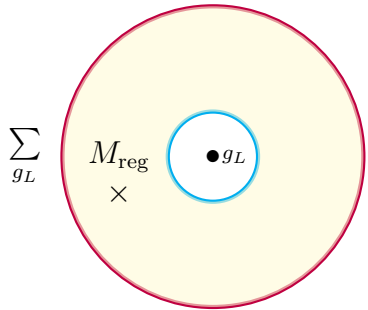

The holonomy for a contractible path that crosses the domain wall $\mathcal{D}_{G\times G}$ with group elements $(g_L,g_R)\in G\times G$ can be any element $g_Rg_L^{-1}\in G$ as one sees in the example of Fig. \ref{fig:definitionofholonomy}. This suggests that we should have insertions of all magnetic defects along $\Sigma$ to produce the $\mathcal{D}_{G\times G}$ domain wall. In addition, we showed that all pure magnetic defects can end on \eqref{eq:action_GG}, which again shows that they condense on the wall. The operator that condenses is:
\begin{equation}
M_{\textrm{reg}}=\sum_{[g]\in \textrm{Cl}(G)} M_{g},
\end{equation}
which is analogous to the Wilson line in the regular representation. By inserting $M_{\textrm{reg}}$ on codimension-two submanifolds of $\Sigma$ we can effectively double the holonomies for the paths in $\Sigma$. See Fig. \ref{magneticcondensationglobal} for an illustration.

\paragraph{General factorized domain wall}
The general factorized domain wall with $H=K_L\times K_R\leq G\times G$ can be obtained from ${\cal D}_{G\times G}$ by projecting the group elements on the left and right sides of the wall to the $K_L$ and $K_R$ subgroups respectively using $\mathcal{D}_{K_L^{(\textrm{id})}}$ and $\mathcal{D}_{K_R^{(\textrm{id})}}$, the explicit expression is given in \eqref{eq:facrtorizeddefinition}.

\subsection{Domain walls of a normal subgroup: electric-magnetic sandwich}

The discussion above can be extended to the general normal domain wall $\mathcal{D}_{H}$ associated with a general normal subgroup $H\lhd G\times G$. In this case, one can consider the projector $1_H^{G\times G}$ of elements of $(g_L,g_R)\in G\times G$ to the subgroup $H\lhd G\times G$. This projector decomposes in irreducible characters of $G\times G$ which are themselves products of irreducible characters of $G$, and has the property that:

\begin{align}
    \chi_{{\operatorname{Ind}_{H}^{G\times G}1}}(g_L,g_R)=\sum_{i,j\in\textrm{irreps of }G}c_{ij}^H\chi_{\rho_i}(g_L)\chi_{\rho_j}(g_R)=\begin{cases} \frac{|G|^2}{|H|} & (g_L,g_R)\in H,\\ 0 & \textrm{otherwise},\end{cases}
\end{align}
with coefficients
\begin{align}
    c^H_{ij}=\langle\chi_{{\operatorname{Ind}_{H}^{G\times G}1}},\chi_{\rho_i}\chi_{\rho_j}\rangle=\frac{1}{|H|}\sum_{(g_L,g_R)\in H}\chi_{\rho_i}(g_L)\chi_{\rho_j}(g_R),
    \label{coefficient}
\end{align}
that one can find by using the orthogonality relations of characters and the defining property of $\chi_{{\operatorname{Ind}_{H}^{G\times G}1}}$. Similar to the diagonal case, the above is the character of the trivial representation in $H$ induced to $G\times G$, \textit{i.e,} $\operatorname{Ind}_H^{G\times G}1$.

Sandwiching the domain wall $\mathcal{D}_{G\times G}$ with these characters for all loops of $\Sigma$ will perform the desired projection. Simplifying the insertion to be on the generators of $\pi_1(\Sigma)$ as done in \eqref{electriccondensationidentity}, the general expression can be written as
\begin{align}
    \mathcal{D}_H(\Sigma)=\frac{|H|^{n-1}}{|G|^{2n-2}}\sum_{\substack{i_1,\dots,i_{n} \\ j_1,\dots ,j_{n}}} c_{i_1j_1}^HW_{i_1}(\gamma_1)\dots c_{i_{n}j_{n}}^HW_{i_{n}}(\gamma_{n})\mathcal{D}_{G\times G}(\Sigma)W_{{j_1}}(\gamma_1)\dots W_{j_{n}}(\gamma_{n}),
\label{generaldomain}
\end{align}
where $i_k,j_k$ run over irreducible characters of $G$ and $\gamma_1,\dots, \gamma_{n}$ are $n$ generators of $\pi_1(\Sigma)$. Here, we also abbreviated $W_{\rho_{i_k}}$ to $W_{i_k}$ as a short hand notation. 

It is important to clarify the meaning of this sandwich construction \eqref{generaldomain}. On the lattice, to define magnetic defects along a codimension-one surface $\Sigma$, one needs to consider a thickened surface $\Sigma\times[0,1]$. If not, there are no 2-simplices perpendicular to $\Sigma$ in which one can fix the holonomies. Along this thickened surface with the inserted magnetic defect, there are two inequivalent ways to insert a Wilson line. It can be positioned either to the right, $\Sigma \times 0$, or to the left, $\Sigma \times 1$, of the magnetic operator. We adopt the convention that $W_\rho(\gamma) M_g(\Gamma)$ represents the Wilson line on the left of the magnetic defect, while $M_g(\Gamma) W_\rho(\gamma)$ places it on the right. Thus, $W_{i_n}(\gamma_n) \mathcal{D}_{G \times G}(\Sigma) W{j_n}(\gamma_n)$ indicates that the Wilson line $W_{i_n}$ is inserted along a loop $\gamma_n$ to the left of the magnetic insertions on $\Sigma \times 0$, while $W_{j_n}$ is inserted along the same loop $\gamma_n$, but to the right of the magnetic insertions on $\Sigma \times 1$.  Moving them from one side to the other costs a braiding phase.

As a particular example, for factorized domain wall the coefficient \eqref{coefficient} factorizes, i.e., $c_{ij}^{K_L\times K_R}=c_i^{K_L}c_j^{K_R}$ where $c_i^{K}$ is the coefficient of the irreducible representation $\rho_i$ in the expansion of $\operatorname{Ind}_{K_L}^G1$, and we get \eqref{eq:facrtorizeddefinition}.

\subsection{Domain walls in abelian theories as condensations} 
\label{sec:operatorabelian}

Every subgroup of $G\times G$ is normal when $G$ is abelian. Therefore, \eqref{generaldomain} can be applied to generate all untwisted domain walls of abelian theories. In this section, we will simplify \eqref{generaldomain} for the abelian case. The simplified expression will then be used to perform consistency checks and compute some fusion rules.

For gauge theories with abelian finite group $G$, Wilson lines in irreducible representations and simple magnetic defects obey the following commutation relation in a codimension-one submanifold $\Sigma$: 
\begin{align}
    W_\rho(\gamma)M_g(\Gamma)=\chi_\rho(g)^{\langle\gamma,\Gamma\rangle}M_g(\Gamma)W_\rho(\gamma),
     \label{eq:intersection}
 \end{align}
where $\chi_\rho(g)$ is the character of the irreducible representation $\rho$ evaluated in the conjugacy class associated with the element $g\in G$ and $\langle\gamma,\Gamma\rangle$ is the \textit{intersection number}\footnote{The intersection number $\langle\gamma,\Gamma\rangle$ is defined explicitly as the integral over $\Sigma$ of the cup product of the Poincaré dual of $\gamma\in H_1(\Sigma,\mathbb{Z})$ and $\Gamma\in H_{D-2}(\Sigma,\mathbb{Z})$.} 
of $\gamma\in H_1(\Sigma,\mathbb{Z})$ and $\Gamma\in H_{D-2}(\Sigma,\mathbb{Z})$ on $\Sigma$. The commutation relation \eqref{eq:intersection} can be understood from the discussion after \eqref{generaldomain} and it follows from the linking action of the two operators. Alternatively, one can think of $\Sigma$ as a time slice and \eqref{eq:intersection} follows from the canonical quantization of the BF-type action with the $(D-2)$ and $1$-form fields being canonically conjugate variables \cite{Witten:1998wy}.

To further simplify \eqref{generaldomain} we assume $\pi_1(\Sigma)$ is torsion-free. With this assumption one can derive the following relations:
\begin{align}
     \mathcal{Z}(G,\Sigma)=\frac{|H_{D-2}(\Sigma,G)|}{|G|}=\frac{|H_1(\Sigma,G)|}{|G|}=|G|^{b_1-1},
     \label{partitionfunction}
\end{align}
where $b_1$ is the first betti number of $\Sigma$. For the first term we used that $|G|\mathcal{Z}(G,\Sigma)=|\operatorname{Hom}(H_1(\Sigma,\mathbb{Z}),G)|=|G|^{b_1}$ where in the last equality we used that $H_1(\Sigma,\mathbb{Z})=\mathbb{Z}^{b_1}$ is torsion-free. For the second term we used Poincaré duality $H_{D-2}(\Sigma,G)\cong H^1(\Sigma,G)$ and the universal coefficient theorem for cohomology $H^1(\Sigma,G)\cong\operatorname{Hom}(H_1(\Sigma,\mathbb{Z}),G)=G^{b_1}$ where the assumption that $\Sigma$ is connected leads to $\operatorname{Ext}^1(H_0(\Sigma,\mathbb{Z}),G)=\operatorname{Ext}^1(\mathbb{Z},G)=0$. Finally, using the universal coefficient theorem for homology we have $H_1(\Sigma,\mathbb{Z})\otimes G=H_1(\Sigma,G)=G^{b_1}$, where we used that $\operatorname{Tor}_1(H_{0}(\Sigma,\mathbb{Z}),G)=\operatorname{Tor}_1(\mathbb{Z},G)=0$. In this context it is possible to choose a basis $\{\gamma_a:1\leq a\leq b_1\}$ and $\{\Gamma_a:1\leq a\leq b_1\}$ for $H_1(\Sigma,\mathbb{Z})=\mathbb{Z}^{b_1}$ and $H_{D-2}(\Sigma,\mathbb{Z})=\mathbb{Z}^{b_1}$ such that $\langle\gamma_a,\Gamma_b\rangle=\delta_{ab}$ where $\langle\cdot,\cdot\rangle:H_1(\Sigma,\mathbb{Z})\times H_{D-2}(\Sigma,\mathbb{Z})\rightarrow \mathbb{Z}$ is the intersection number defined by integrating over $\Sigma$ the cup product of the Poincaré dual of the homology classes. We are going to use relation \eqref{partitionfunction} and this fact to simplify some of the expressions in abelian theories in the constructions below.

When $G$ is abelian, the operator $M_{\textrm{reg}}$ obeys the same fusion rule as $W_{\textrm{reg}}$, i.e., $M_\textrm{reg}\times M_{\textrm{reg}}=\sum_gM_g\times \sum_{g^\prime}M_{g^\prime}=\sum_{g,g^\prime}M_{gg^\prime}=|G|\sum_g M_g=|G|M_{\textrm{reg}}$, where we used the fact that each element of $G$ appears once, and only once in each row and column of the Cayley table of $G$. Similarly to the derivation of the diagonal domain wall condensation \eqref{electriccondensationidentity}, we can write the operator $\mathcal{D}_{G\times G}$ explicitly in the following equivalent ways:
\begin{equation}
    {\cal D}_{G\times G}(\Sigma)=\mathcal{Z}(G)\prod_{\Gamma\in H_{D-2}(\Sigma,\mathbb{Z})}\frac{M_{\textrm{reg}}(\Gamma)}{|G|}=\mathcal{Z}(G)\prod_{\Gamma\in S}\frac{M_{\textrm{reg}}(\Gamma)}{|G|}=\frac{1}{|G|}\prod_{\Gamma\in S}M_\text{reg}(\Gamma)~,
    \label{eq:magneticdomainwall}
\end{equation}
where $S$ is a set of generators of $H_{D-2}(\Sigma,\mathbb{Z})$ and we used that $|G|\mathcal{Z}(G)=|H^1(\Sigma,G)|=|H_{D-2}(\Sigma,G)|$ by Poincaré duality. Choosing a othonormal basis for the insertion of magnetic and electric operators, \eqref{generaldomain} can be written more compactly as
\begin{equation}
    {\cal D}_H(\Sigma)= \frac{|G|}{|H|}\prod_{a=1}^{b_1}\frac{|H|}{|G\times G|}\sum_{i,j\in\textrm{irreps}}c_{ij}^HW_{\rho_i}(\gamma_a)M_\textrm{reg}(\Gamma_a)W_{\rho_j}(\gamma_a).
    \label{generaldomainbasis}
\end{equation}
where we used \eqref{eq:magneticdomainwall} and we commuted the operators with zero intersection. Let us now illustrate how this definition reproduces expression \eqref{electriccondensationidentity} for $\mathcal{D}_1$ and the fact that $\mathcal{D}_{G^{(\textrm{id})}}=1$.

\paragraph{Example: trivial group domain wall} Note that the trivial subgroup domain wall is a nice checking example because it is at the same time a factorized and diagonal domain wall. To show that \eqref{electriccondensationidentity} follows from \eqref{generaldomainbasis} we need to use the commutation relation \eqref{eq:intersection}. Provided the fact that $c_{ij}^{1}=d_id_j=1$, we have:
\begin{align}
    \begin{split}
    W_{\textrm{reg}}(\gamma)M_{\textrm{reg}}(\Gamma)W_{\textrm{reg}}(\gamma) & =\sum_{i,j,g}W_{\rho_i}(\gamma)M_g(\Gamma)W_{\rho_j}(\gamma),\\
    &= \sum_{g,i,j}\chi_{\rho_i}(g)M_g(\Gamma)W_{\rho_i\times \rho_j}(\gamma),\\
    & =\sum_g \chi_{\textrm{reg}}(g) M_g(\Gamma)W_{\textrm{reg}}(\gamma),\\
    & =|G|W_{\textrm{reg}}(\gamma),
    \end{split}
    \label{eq:general_to_diagonal}
\end{align}
where in the first equality we used \eqref{eq:intersection} assuming for simplicity that $\langle\gamma,\Gamma\rangle=1$, in the second equality we redefined the dummy variable over irreps and used the definition of the regular character and its property \eqref{eq:regularcharacter}. From this equation, it is easy to derive \eqref{electriccondensationidentity} from \eqref{generaldomainbasis}.

\paragraph{Example: automorphism domain walls}

Let us consider the domain wall that generates group automorphism symmetry for abelian gauge group $G$. The domain wall for the automorphism $\phi\in \text{Aut}(G)$ corresponds to the subgroup $G^{(\phi)}=\{(\phi\cdot g, g):g\in G\}\leq G\times G$. The condensation expression for the automorphism domain wall is given by (\ref{generaldomain}) with
\begin{equation}
       c_{ij}^{G^{(\phi)}}=\frac{1}{|G|}\sum_{g\in G}\chi_{\rho_i}(\phi\cdot g)\chi_{\rho_j}(g)=\langle \chi_{\rho_i\circ\phi},\chi_{\overline{\rho}_j}\rangle=\begin{cases} 1 & \overline{\rho}_j=\rho_i\circ\phi, \\ 0 & \textrm{otherwise}, \end{cases}
       \label{eq:cij_automorphism}
\end{equation}
where $\overline{\rho}_i$ is the complex conjugate representation of $\rho_i$\footnote{The complex conjugate representation $\overline{\rho}$ of a representation $\rho$ is the representation such that $\overline{\rho}_i(g)=\overline{\rho_i(g)}$ for all $g\in G$} and $\rho_j\cdot \phi$ is the representation obtained by the composition of $\phi$ and $\rho_j$.

In particular, the automorphism associated with the identity, i.e., $\phi=\textrm{id}$, should give 1, the trivial domain wall. Indeed, using \eqref{generaldomainbasis} for the identity automorphism we have:

\begin{align}
    \begin{split}
        \mathcal{D}_{G^{(\textrm{id})}}(\Sigma) & =\prod_{a=1}^{b_1}\frac{1}{|G|}\sum_{i\in\textrm{irreps}}W_{\overline{\rho}_i}(\gamma_a)M_{\textrm{reg}}(\Gamma_a)W_{\rho_i}(\gamma_a),\\
        & =\prod_{a=1}^{b_1}\frac{1}{|G|}\sum_{i\in\textrm{irreps}}\sum_{g\in G}\chi_{\overline{\rho}_i}(g)M_g(\Gamma_a)W_{\overline{\rho}_i}(\gamma_a)W_{\rho_i}(\gamma_a),\\
        & =\prod_{a=1}^{b_1}\frac{1}{|G|}\sum_{g\in G}\chi_{\rho_{\textrm{reg}}}(g)M_g(\Gamma_a),\\
        & = \prod_{a=1}^{b_1}\sum_{g\in G}\delta_{ge}M_g(\Gamma_a),\\
        & = 1,
    \end{split}
\end{align}
where above we used \eqref{eq:intersection} to permute the Wilson line and magnetic defects, we used that $W_\rho(\gamma)W_{\overline{\rho}}(\gamma)=1$%\footnote{Recall that here we are assuming that $G$ is abelian so that all irreducible representations are one dimensional and $\rho(g)\overline{\rho}(g)=1$.}
, and we used the regular representation character identity shown in \eqref{eq:regularcharacter}.

\subsection{Fusion rules of domain walls from condensation definition}
\label{sec:domainwallascondensationsfusion}

Here, we derive the fusion rules of domain walls reviewed in Section \ref{sec:review} by employing the condensation expressions obtained in Section \ref{sec:operator} and the algebraic properties of Wilson lines and magnetic defects. This approach provides a self-consistency check on the correctness of the condensation expressions and on the fusion rules of domain walls derived by other methods in \cite{Cordova:2024jlk}.

\paragraph{Diagonal domain walls}

Let us compute the fusion \eqref{eq:fusiondiag} for the diagonal domain walls with subgroups $K^{(\textrm{id})},K^{\prime(\textrm{id})}$ using \eqref{eq:oprdiagwall} and the tensor product of representations. We have:
\begin{align}
    \begin{split}  
        \mathcal{D}_{K^{(\textrm{id})}}\times \mathcal{D}_{K^{\prime(\textrm{id})}} & =\frac{|G|}{|K|}\frac{|G|}{|K^\prime|}\prod_{\textrm{closed loops }\gamma}\frac{|K||K^\prime|}{|G||G|}W_{\textrm{Ind}_K^G1}(\gamma)W_{\textrm{Ind}_{K^\prime}^G1}(\gamma),\\
        & =\frac{|G|}{|K\cdot K^\prime|}\frac{|G|}{|K\cap K^\prime|}\prod_{\textrm{closed loops }\gamma}\frac{|K\cap K^\prime|}{|G|}W_{\textrm{Ind}_{K\cap K^\prime}^G1}(\gamma),\\
        & =\frac{|G|}{|K\cdot K^\prime|}\mathcal{D}_{(K\cap K^\prime)^{(\textrm{id})}},
    \end{split}
\end{align}
where we used that
\begin{equation}
    \operatorname{Ind}_K^G1\otimes \operatorname{Ind}_{K^\prime}^G1=\frac{|G|}{|K\cdot K^\prime|}\operatorname{Ind}_{K\cap K^\prime}^G1,
\end{equation}
and $|K||K^\prime|=|K\cdot K^\prime||K\cap K^\prime|$. The same derivation applies to the fusion \eqref{eq:highercodimension} associated with the higher codimensional generalization of the diagonal domain wall.

\paragraph{Automorphism domain walls in abelian theories}

In this section, we derive the fusion rules \eqref{eq:fusionautomorphism} and \eqref{eq:GGanihilation} involving the domain walls with automorphism subgroups $G^{(\phi)}= \{(\phi\cdot g,g): g\in G\}\leq G\times G$ with $\phi\in\operatorname{Aut}(G)$ using the condensation definition \eqref{generaldomainbasis} with $c_{ij}^H$ given by \eqref{eq:cij_automorphism}. 

For the automorphism fusion rule \eqref{eq:fusionautomorphism} we have:
\begin{align}
    \begin{split}
        & \mathcal{D}_{G^{(\phi)}}\times \mathcal{D}_{G^{(\phi^{\prime})}}\\ & =\prod_{a=1}^{b_1}\frac{1}{|G|^{2}}\sum_{i,j\in\textrm{irreps}}W_{\overline{\rho}_i}(\gamma_a)M_{\textrm{reg}}(\Gamma_a)W_{\rho_i\circ\phi}(\gamma_a)W_{\overline{\rho}_j}(\gamma_a) M_{\textrm{reg}}(\Gamma_a)W_{\rho_j\circ\phi^\prime}(\gamma_a),\\
        & =\prod_{a=1}^{b_1}\frac{1}{|G|^{2}}\sum_{i,j\in\textrm{irreps}}\sum_{g,g^\prime\in G}\chi_{\rho_i\circ\phi}(g)\chi_{\overline{\rho}_j}(g)W_{\overline{\rho}_i}(\gamma_a)M_{gg^\prime}(\Gamma_a)W_{\rho_i\circ\phi}(\gamma_a)W_{\overline{\rho}_j}(\gamma_a)W_{\rho_j\circ\phi^\prime}(\gamma_a),\\
        & = \prod_{a=1}^{b_1}\frac{1}{|G|}\sum_{i\in\textrm{irreps}}W_{\overline{\rho}_i}(\gamma_a)M_{\textrm{reg}}(\Gamma_a)W_{\rho_i\circ\phi\circ\phi^\prime}(\gamma_a),\\
        & = \mathcal{D}_{G^{(\phi\circ \phi^{\prime})}},
    \end{split}
\end{align}
where from the second to the third line we redefined $g^\prime=g^{-1}\tilde{g}$ and performed the independent summation on $g$ using the character ortogonality condition, namely:
\begin{align}
    \frac{1}{|G|}\sum_{g\in G}\chi_{\rho_i\circ\phi}(g)\chi_{\overline{\rho}_j}(g)=\langle \chi_{\rho_i\circ\phi},\chi_{\rho_j}\rangle=   \begin{cases} 1 & \rho_j=\rho_i\circ\phi,\\
    0 & \textrm{otherwise},
    \end{cases}
\end{align}
and the fact that $W_\rho(\gamma)W_{\overline{\rho}}(\gamma)=1$. 

Similarly, for \eqref{eq:GGanihilation} we have:
\begin{align}
    \begin{split}
        & \mathcal{D}_{G\times G}\times \mathcal{D}_{G^{(\phi)}},\\
        & =\prod_{a=1}^{b_1}\frac{1}{|G|^2}\sum_{i\in \textrm{irreps}}M_{\textrm{reg}}(\Gamma_a)W_{\overline{\rho}_i}(\gamma_a)M_{\textrm{reg}}(\Gamma_a)W_{\rho_i\circ\phi}(\gamma_a),\\
        & =\prod_{a=1}^{b_1}\frac{1}{|G|^2}\sum_{i\in \textrm{irreps}}\sum_{g,g^\prime\in G}\chi_{\overline{\rho}_i}(g)W_{\overline{\rho}_i}(\gamma_a)M_g(\Gamma_a)M_{g^\prime}(\Gamma_a)W_{\rho_i\circ\phi}(\gamma_a),\\
        & =\prod_{a=1}^{b_1}\frac{1}{|G|^2}\sum_{i\in \textrm{irreps}}\sum_{g,\tilde{g}\in G}\chi_{\overline{\rho}_i}(g)W_{\overline{\rho}_i}(\gamma_a)M_{\tilde{g}}(\Gamma_a)W_{\rho_i\circ\phi}(\gamma_a),\\
        & =\prod_{a=1}^{b_1}\frac{1}{|G|}\sum_{i\in \textrm{irreps}}\delta_{i0}W_{\overline{\rho}_i}(\gamma_a)M_{\textrm{reg}}(\Gamma_a)W_{\rho_i\circ\phi}(\gamma_a),\\
        & =\prod_{a=1}^{b_1}\frac{1}{|G|}M_{\textrm{reg}}(\Gamma_a),\\
        & = \mathcal{D}_{G\times G},
    \end{split}
\end{align}
where we defined $\tilde{g}=g\cdot g^\prime$ and we used the character ortogonality condition $\langle \chi_{\overline{\rho}_i},\chi_{\rho_0}\rangle=\delta_{i0}$ with $\rho_0=1$ the trivial representation. The fusion $\mathcal{D}_{G^{(\phi)}}\times \mathcal{D}_{G\times G}=\mathcal{D}_{G\times G}$ can be derived in a similar way.

\paragraph{Factorized domain walls in abelian theories}

Let us derive the fusion rule \eqref{eq:fusionGGdiagGG} using the condensation expressions \eqref{eq:oprdiagwall} and \eqref{eq:magneticdomainwall}. Using the basis for the electric and magnetic operator insertions such that $\langle\gamma_a,\Gamma_b\rangle=\delta_{ab}$, we have:
\begin{align}
     \begin{split}
         \mathcal{D}_{G\times G}\times \mathcal{D}_{K^{(\textrm{id})}}\times \mathcal{D}_{G\times G} & =\frac{1}{|K||G|}\prod_{a=1}^{b_1}\frac{|K|}{|G|} M_{\textrm{reg}}(\Gamma_a)W_{\operatorname{Ind}_K^G1}(\gamma_a)M_{\textrm{reg}}(\Gamma_a),\\
        & =\frac{1}{|K||G|}\prod_{a=1}^{b_1} \frac{|K|}{|G|}\sum_{i,g,g^\prime} \chi_{\rho_i}(g)M_{gg^\prime}(\Gamma_a) c_iW_{\rho_i}(\gamma_a),\\
        & =\frac{|K|^{b_1}}{|K||G|}\prod_{a=1}^{b_1} M_{\textrm{reg}}(\Gamma_a),\\
        & = \mathcal{Z}(K)\mathcal{D}_{G\times G}.
     \end{split}
\end{align}
In the first line, we reorganized the product by commuting all operators and leaving just the ones with non-zero intersection number uncommuted. In the second, we used \eqref{eq:intersection} to permute the simple Wilson lines and magnetic defects and we fused the two simple magnetic defects. In the third line we used the character orthogonality condition:

\begin{align}
    \frac{1}{|G|}\sum_g \chi_{\rho_i}(g)=\langle\chi_{\rho_i},\chi_{1}\rangle=\begin{cases} 1 & \rho_i=1, \\ 0 & \textrm{otherwise}, \end{cases}
\end{align}
and our assumption about $\Sigma$ summarized before \eqref{generaldomainbasis} which implies that $\mathcal{Z}(K)=|K|^{b_1-1}$.

\section{Domain walls as higher gauging condensation defects in \texorpdfstring{$\mathbb{Z}_N$}{} gauge theory}
\label{sec:ZN}

In this section, we investigate $\mathbb{Z}_N$ gauge theory in general spacetime dimensions. First, we consider general untwisted domain walls and show that their condensation expression derived in Section \ref{sec:operator} can be recast using the concept of higher gauging \cite{Roumpedakis:2022aik}. Consequently, our condensation expression provides a dictionary between the domain wall \cite{Cordova:2024jlk} and the higher gauging \cite{Roumpedakis:2022aik} constructions of non-invertible symmetries in $\mathbb{Z}_N$ gauge theory. Next, we analyze twisted domain walls and demonstrate that they can be recast using a new concept that generalizes higher gauging, which we term \textit{sequential higher gauging}.

\subsection{Review of higher gauging condensation defects}
\label{sec:highergauging}

Higher gauging is a specific procedure used to construct condensation defects, broadly defined as defects obtained through suitable insertions of lower-dimensional topological operators. Gauging a non-anomalous abelian global symmetry involves summing over insertions of the symmetry generators across spacetime, usually resulting in a different theory. Higher gauging, however, involves gauging a non-anomalous global symmetry on a submanifold rather than the entire spacetime. This procedure creates a defect on the submanifold, known as a condensation defect \cite{Roumpedakis:2022aik}.

In this section, we will focus on $\mathbb{Z}_N$ gauge theory in general spacetime dimensions $D$ without a Dijkgraaf-Witten topological action. The theory has $\mathbb{Z}_N$ 1-form symmetry generated by the unit charge magnetic operator $M$ and $\mathbb{Z}_N$ $(D-2)$-form symmetry generated by the unit charge Wilson lines $W$. Higher gauging condensation defects, which we denote by $\mathcal{S}_{\mathbb{Z}_{q}\times \mathbb{Z}_{q^\prime},f}$, are classified by a choice of subgroup of the higher-form symmetry $\mathbb{Z}_{q}\times \mathbb{Z}_{q^\prime}\lhd \mathbb{Z}_N\times\mathbb{Z}_N$ generated respectively by Wilson lines and magnetic defects with $q|N$, $q^\prime|N$ and a choice of torsion term, $f\in H^{D-1}(K(\mathbb{Z}_q,D-2)\times K(\mathbb{Z}_{q^\prime},1),U(1))\cong\mathbb{Z}_{\gcd(N/q,N/q^\prime)}$ \footnote{Given a group $G$ and an integer $n$, the Eilenberg-Maclane space $K(G,n)$ is defined as a topological space that has $n$-th homotopy group $\pi_n(G)$ isomorphic to $G$ and all other homotopy groups trivial. In $D=2+1$, the cohomology group that classifies discrete torsion is the same as the group cohomology of $\mathbb{Z}_q\times \mathbb{Z}_{q^\prime}$, because an Eilenberg–MacLane space of type $K(G,1)$ is isomorphic to $BG$, the classifying space of $G$. We emphasize that the form-degree of the symmetries generated by Wilson lines and magnetic defects is such that $H^{D-1}(K(\mathbb{Z}_q,D-2)\times K(\mathbb{Z}_{q^\prime},1),U(1))$ is isomorphic to $\mathbb{Z}_{\gcd(N/q,N/q^\prime)}$ and does not depend on the spacetime dimension $D$.}. Explicitly we have:
\begin{align}
    \mathcal{S}_{\mathbb{Z}_{q}\times\mathbb{Z}_{q^\prime},f}(\Sigma)=C(\Sigma)\sum_{\substack{\gamma\in H_1(\Sigma,\mathbb{Z}_q) \\ \Gamma\in H_{D-2}(\Sigma,\mathbb{Z}_q^\prime)}}e^{\frac{2\pi i}{\gcd(N/q,N/q^\prime)}f\langle\gamma,\Gamma\rangle}W^{N/q}(\gamma)M^{N/q^\prime}(\Gamma),
    \label{eq:highergaugingdefinition}
\end{align}
with $C(\Sigma)$ a coefficient that depends on the topology of $\Sigma$. The conventions we are using is that the first subgroup appearing in the label of $\mathcal{S}_{\mathbb{Z}_q\times\mathbb{Z}_{q^\prime}}$ corresponds to the subgroup of the $(D-2)$-form symmetry generated by Wilson lines and the second to the subgroup of the $1$-form symmetry generated by magnetic defect. We also set as a convention that the algebraic expression for the higher gauging condensation defects should have the Wilson lines on the left, which corresponds to having the Wilson lines in the global left of the magnetic defects as reviewed in Section \ref{sec:review}. Note that in $D=2+1$, one can fuse the Wilson line and magnetic operator into a Dyon with electric and magnetic charge. Therefore, in $D=2+1$ one can gauge ``diagonal" 1-form subgroups generated by the Dyons. As we are going to show in Section \ref{sec:electricmagnetichighergauging} this generates the electric-magnetic symmetry reviewed in \eqref{eq:nontrivialfusion} and \eqref{eq:nontrivialaction}

One can use \eqref{eq:highergaugingdefinition} and the algebraic properties of Wilson lines and magnetic defects of \eqref{eq:intersection} to compute the fusion rules of higher gauging condensation defects. In this section, we will check again the fusion rules we derived in the previous sections by following this procedure. We deter to the next section a derivation of some general fusion rules, but we refer the reader to Appendix A of \cite{Roumpedakis:2022aik} for more details.

Similarly, one can use \eqref{eq:highergaugingdefinition} to compute the transformation of other operators. Here, we reinterpret Eqs. (5.16) and (5.43) of \cite{Roumpedakis:2022aik} to higher dimensions. We can think of the moves performed in their paper as done in a 3d slice, with the $1$-form magnetic defect filling all other dimensions. One can then derive:
\begin{align}
    \mathcal{S}_{\mathbb{Z}_{q}\times\mathbb{Z}_{q^\prime},f}\cdot L=\sum_{\substack{a,a^\prime=0\\ Q_1+\frac{(\gcd(N/q,N/q^\prime)+f)}{\gcd(q,q^\prime)} a^\prime q=0\mod q\\ Q_2-faq^\prime=0\mod q^\prime}}^{q-1,q^\prime-1}W^{Na/q}M^{Na^\prime/q^\prime}L.
    \label{eq:action_highergauging}
\end{align}
with $(Q_1, Q_2)$ the $\mathbb{Z}_q\times\mathbb{Z}_{q^\prime}$ higher-form symmetry charges of the operator L. This expression computes the transformation of operators ``from right to left". This comes from the convention used in \cite{Roumpedakis:2022aik} of having the surface defect oriented inwards. The transformation ``from left to right" is the same as the action of the orientation reversal \eqref{eq:orientationreversal}.

\subsection{Untwisted domain walls as higher gauging defects}

In this section, we derive the non-trivial correspondence presented in Table \ref{tab:dictionaryZN} between the untwisted domain walls and higher gauging condensation defects. We further use these higher gauging expressions to confirm the fusion rules and the transformation of other operators derived in \cite{Cordova:2024jlk} and reviewed in Section \ref{sec:review}.

\begin{table}[ht]
    \centering
    \begin{tabular}{|c|c|}
    \hline Domain wall & Condensation  \\ \hline
    $\mathcal{D}_{\mathbb{Z}_M^{(\textrm{id})}}$ & $\mathcal{S}_{\mathbb{Z}_{N/M}\times 1}$ \\
    $\mathcal{D}_{\mathbb{Z}_N\times\mathbb{Z}_N}$ & $\mathcal{S}_{1\times \mathbb{Z}_N}$ \\
    $\mathcal{D}_{1\times \mathbb{Z}_N}$ & $\mathcal{S}_{\mathbb{Z}_N\times\mathbb{Z}_N}$\\
    $\mathcal{D}_{\mathbb{Z}_N\times1}$ & $\mathcal{S}_{\mathbb{Z}_N\times\mathbb{Z}_N,-1}$ \\
    $\mathcal{D}_{\mathbb{Z}_N^{(\textrm{id})}}$ & $\mathcal{S}_1$ \\
    $\mathcal{D}_{\mathbb{Z}_N^{(m)}}$ & $\mathcal{S}_{\mathbb{Z}_{N/\ell}\times\mathbb{Z}_{N/\ell},\ell\frac{m}{1-m}}$ \\ \hline
    \end{tabular}
    \caption{Dictionary between higher gauging condensation defects and domain walls in $\mathbb{Z}_N$ gauge theory. Domain walls are classified by subgroups of the folded theory $K\lhd \mathbb{Z}_N\times \mathbb{Z}_N$ (with the left and right factors corresponding respectively to the left and right of the domain wall) with a choice of topological action $\alpha\in H^{D-1}(K,U(1))$. Higher gauging condensation defects are classified by subgroups of the higher-form global symmetry $\mathbb{Z}_q\times\mathbb{Z}_{q^\prime}\lhd \mathbb{Z}_N\times\mathbb{Z}_N$ (with left and right factors generated respectively by Wilson line and magnetic defects) with a choice of torsion term $f\in H^{D-1}(K(\mathbb{Z}_q,D-2)\times K(\mathbb{Z}_{q^\prime},1),U(1))=\mathbb{Z}_{\gcd(N/q,N/q^\prime)}$. In the last row, $\mathbb{Z}_N^{(m)}=\{(mn,n):n\in\mathbb{Z}_N\}$ is the automrphism subgroup associated with the automorphism $1\mapsto m\in\mathbb{Z}_N^\times$ and $\ell=\gcd(N,m-1)$. }
    \label{tab:dictionaryZN}
\end{table}

\subsubsection{Diagonal domain walls}
\label{diagonalmesh}

Let us start with the diagonal domain wall, $\mathcal{D}_{\mathbb{Z}_M^{(\textrm{id})}}$ associated with the subgroup $\mathbb{Z}_N^{(\textrm{id})}=\{(m,m):m\in\mathbb{Z}_M\}\lhd\mathbb{Z}_N\times\mathbb{Z}_N$ with $M|N$. The representation associated with the projector of \eqref{eq:projector_diagonal} can be written in terms of the unit charge Wilson line as
\begin{equation}
    \operatorname{Ind}_{\mathbb{Z}_M}^{\mathbb{Z}_N}1=\sum_{q=0}^{N/M-1}W^{qM}.
\end{equation}
Therefore, using \eqref{eq:oprdiagwall} we have
\begin{align}
    \begin{split}
        {\cal D}_{\mathbb{Z}_M^{(\text{id})}}(\Sigma)&=\frac{N}{M}\prod_{a=1}^{b_1}
   \frac{1}{|N/M|} 
    \sum_{q_a=0}^{N/M-1}W^{q_aM}(\gamma_a),\\
    & =\frac{N/M}{|H_1(\Sigma,\mathbb{Z}_{N/M})|}\sum_{q_1}\sum_{q_2}\cdots \sum_{q_{b_1}} 
    W^{q_1M}(\gamma_1)\cdots W^{q_{b}M}(\gamma_{b_1}),\\ 
    &=\frac{N/M}{|H_1(\Sigma,\mathbb{Z}_{N/M})|}
    \sum_{q_1}\sum_{q_2}\cdots \sum_{q_{b_1}} 
    W^{M}(q_1\gamma_1+\cdots q_{b_1}\gamma_{b_1}),\\
    &=\frac{N/M}{|H_1(\Sigma,\mathbb{Z}_{N/M})|}\sum_{\gamma\in H_1({\Sigma},\mathbb{Z}_{N/M})} W^M(\gamma)~,\\
    & = \mathcal{S}_{\mathbb{Z}_{N/M}\times1}(\Sigma),
    \end{split}
    \label{eq:diagonal_highergauging}
\end{align}
%Thus we find that the domain wall is a mesh given by summing over different Wilson loop operators of charges equal to a multiple of $M$. 
The Wilson line $W^M$ generate a $\mathbb{Z}_{N/M}$ $(D-2)$-form symmetry in $D$ spacetime dimension, and the domain wall is equivalent to gauging this symmetry on the support of the domain wall instead of the entire spacetime \cite{Roumpedakis:2022aik,Choi:2022zal}. The higher gauging expression in \eqref{eq:diagonal_highergauging} differs from the result of \cite{Roumpedakis:2022aik} by an Euler counterterm on the wall. Indeed, in $D=2+1$, $\Sigma$ is a surface and for a genus-g surface the coefficient in front equals $|N/M|^{1-2g}$. Rescaling by an Euler counter term $\lambda^{\chi(\Sigma)}$ with $\lambda=|N/M|^{-1/2}$ and $\chi(\mathcal{M})=2-2g$ the Euler number of $\Sigma$, the prefacto becomes $|N/M|^{-g}=1/\sqrt{|H_1(\Sigma,\mathbb{Z}_{N/M})|}$ which is the same normalization used in \cite{Roumpedakis:2022aik}. If instead we used $\lambda=|N/M|^{-1}$, the normalization would become $|N/M|^{-1}=1/|H^0(\mathcal{M},\mathbb{Z}_{N/M})|$ which is the same normalization used in \cite{Kaidi:2022cpf}. In $D=2+1$ because the Euler counterterm is $2-2g$ it is particularly hard to reason if there exists a more natural normalization. We believe our formalism gives an affirmative answer to this question and here we provide what we believe is the answer.

Consistently, we could also use the general expression \eqref{generaldomain} to derive the higher gauging expression \eqref{eq:diagonal_highergauging}. We postpone this to the next section after deriving the higher gauging expression for $\mathcal{D}_{\mathbb{Z}_N\times\mathbb{Z}_N}$. 

\paragraph{Fusion rule} One can use the right side of \eqref{eq:diagonal_highergauging} and the algebraic properties of Wilson lines and magnetic defects to compute its fusion rule. The detailed calculation takes about two pages and was presented in Appendix A of \cite{Roumpedakis:2022aik}. Adapting their derivation to our normalization leads one to
\begin{align}
    \mathcal{S}_{\mathbb{Z}_{N/M}\times 1}\times \mathcal{S}_{\mathbb{Z}_{N/M^\prime}\times 1}=N\frac{\gcd(M,M^\prime)}{MM^\prime}\mathcal{S}_{\mathbb{Z}_{N/\gcd(M,M^\prime)}\times 1}
\end{align}
where we used that $\gcd(N/M,N/M^\prime)=N/(MM^\prime/\gcd(M,M^\prime))$. This agrees with \eqref{eq:fusiondiag} because $\mathbb{Z}_{N/n}\cap \mathbb{Z}_{N/n^\prime}=\mathbb{Z}_{\gcd(N/n,N/n^\prime)}=\mathbb{Z}_{N/(nn^\prime/\gcd(n,n^\prime))}$ and  $\mathbb{Z}_{M}\cdot \mathbb{Z}_{M^\prime}=\mathbb{Z}_{MM^\prime/\gcd(M,M^\prime)}$. We see that \eqref{eq:fusiondiag} specializes to finite-group gauge theory but generalizes in dimensions and gauge group the expression Eq. (5.24) of \cite{Roumpedakis:2022aik}.

\paragraph{Transformation of other operators} We can use the condensation expression of \eqref{eq:diagonal_highergauging} to compute the transformation of other operators using the algebraic properties of Wilson lines and magnetic defects. Using \eqref{eq:action_highergauging} one can derive
\begin{align}
    \mathcal{S}_{\mathbb{Z}_{N/M}\times1}\cdot W^{k N/M}=\sum_{n=0}^{M-1}W^{nN/M},\qquad \mathcal{S}_{\mathbb{Z}_{N/M}\times1}\cdot M^{\ell M}=\sum_{n=0}^{N/M-1}M^{nM},
\end{align}
for all $0\leq k\leq M-1$ and $0\leq \ell\leq N/M-1$ and $\mathcal{S}_{\mathbb{Z}_{N/M}\times1}\cdot W^k=\mathcal{S}_{\mathbb{Z}_{N/M}\times1}\cdot M^\ell=0$ for the other lines. This is consistent with the transformation properties reviewed in Section \ref{sec:review}. Indeed, the lines $W^{kN/M}$ for $0\leq k,\leq M-1$ are gauge invariant when restricted to $\mathbb{Z}_M$. In particular, all Wilson lines can end on $\mathcal{D}_{1}$.

\subsubsection{Factorized domain walls}
\label{sec:factorizedhighergauging}

Let us consider the factorized domain wall with $K=\mathbb{Z}_{N}\times \mathbb{Z}_N$.
The domain wall operator is \eqref{eq:magneticdomainwall}: denote a basis for $(D-2)$-cycles on the support ${\Sigma}$ of the domain wall operator by $\{\Gamma_a\}$, and the basic magnetic operator operator by $M$, then
\begin{align}
    {\cal D}_{\mathbb{Z}_N\times\mathbb{Z}_N}(\Sigma)
    &=\frac{1}{N}\prod_{a=1}^{b_1}\sum_{q_a=0}^{N-1}M^{q_a}(\Gamma_a)
    =\frac{1}{N}
        \sum_{\Gamma \in H_{D-2}(\Sigma,\mathbb{Z}_N)}M(\Gamma)=\mathcal{S}_{1\times\mathbb{Z}_N}(\Sigma),
    \label{eq:GGhighergauging}
\end{align}
As before, we see that $\mathcal{D}_{G\times G}$ is the higher gauging of the $\mathbb{Z}_N$ 1-form magnetic symmetry. Equation \eqref{eq:GGhighergauging} differs from the expression in \cite{Roumpedakis:2022aik} by an Euler conterterm on the wall. Indeed, for a genus-g surface, the coefficient in front equals $|N|^{-1}$. Rescaling by an Euler counterterm $\lambda^{\chi(\Sigma)}$ with $\lambda=|N|^{1/2}$ and $\chi(\Sigma)=2-2g$ the Euler number of $\Sigma$, the prefactor becomes $|N|^{-g}=1/\sqrt{H_1(\Sigma,\mathbb{Z}_N)}$ which is the same normalization used in \cite{Roumpedakis:2022aik}.

From the fusion rule \eqref{eq:facrtorizeddefinition} it is now easy to construct the $\mathcal{D}_{\mathbb{Z}_N\times 1},\mathcal{D}_{1\times \mathbb{Z}_N}$. The difference between the two comes from the convention used for the ordering of the Wilson line and magnetic operator in the definition of higher gauging, see \eqref{eq:highergaugingdefinition}. With the convention we are using it follows that:
\begin{align}
    & \mathcal{D}_{1\times \mathbb{Z}_N}(\Sigma)=\frac{1}{|H_1(\Sigma,\mathbb{Z}_N)|}\sum_{\substack{\gamma\in H_1(\Sigma,\mathbb{Z}_N)\\ \Gamma\in H_{D-2}(\Sigma,\mathbb{Z}_N)}}W(\gamma)M(\Gamma)=\mathcal{S}_{\mathbb{Z}_N\times \mathbb{Z}_N}(\Sigma),\\
    & \mathcal{D}_{ \mathbb{Z}_N\times 1}(\Sigma)=\frac{1}{|H_1(\Sigma,\mathbb{Z}_N)|}\sum_{\substack{\gamma\in H_1(\Sigma,\mathbb{Z}_N)\\ \Gamma\in H_{D-2}(\Sigma,\mathbb{Z}_N)}}M(\Gamma)W(\gamma)=\mathcal{S}_{\mathbb{Z}_N\times \mathbb{Z}_N,-1}(\Sigma).
\end{align}
where in the second line we used \eqref{eq:intersection} to permute $M(\Gamma)W(\gamma)$ giving the torsion term $f=-1\in H^{D-1}(K(\mathbb{Z}_N,D-2)\times K(\mathbb{Z}_N,1),U(1))=\mathbb{Z}_N$.

Finally, consistently with the previous section, we can use the general expression \eqref{generaldomain} to derive the higher gauging expression for $\mathcal{D}_1$ similarly to what we did in \eqref{eq:general_to_diagonal}. Using \eqref{eq:diagonal_highergauging} and \eqref{eq:GGhighergauging} we can write
\begin{align}
    \begin{split}
        \mathcal{D}_1(\Sigma) & =\frac{N}{|H_1(\Sigma,\mathbb{Z}_N)|^2}\sum_{\substack{\gamma,\gamma^\prime\in H_1(\Sigma,\mathbb{Z}_N)\\ \Gamma\in H_{D-2}(\Sigma,\mathbb{Z}_N)}}W(\gamma)M(\Gamma)W(\gamma^\prime),\\
        & =\frac{N}{|H_1(\Sigma,\mathbb{Z}_N)|^2}\sum_{\substack{\gamma,\gamma^\prime\in H_1(\Sigma,\mathbb{Z}_N),\\ \Gamma\in H_{D-2}(\Sigma,\mathbb{Z}_N)}}e^{\frac{2\pi i}{N}\langle\gamma,\Gamma\rangle}M(\Gamma)W(\gamma+\gamma^\prime),\\
        & = \frac{N}{|H_1(\Sigma,\mathbb{Z}_N)|}\sum_{\gamma\in H_1(\Sigma,\mathbb{Z}_N)}W(\gamma),\\
        & = \mathcal{S}_{\mathbb{Z}_N\times 1}(\Sigma),
    \end{split}
\end{align}
where we used \eqref{eq:intersection} to permute the Wilson line and magnetic operator, we joined the Wilson lines, redefined $\gamma=\gamma-\gamma^\prime$ and summed over $\gamma^\prime$ forcing $\Gamma=0$ and giving the $|H_1(\Sigma,\mathbb{Z}_N)|$ factor.

\paragraph{Fusion rules} We can use the higher gauging expression for $\mathcal{D}_{\mathbb{Z}_N\times\mathbb{Z}_N}$ derived in \eqref{eq:GGhighergauging} and the algebraic properties of Wilson lines and magnetic defects to compute its fusion rule. For example:
\begin{align}
    \begin{split}
        & \mathcal{S}_{1\times\mathbb{Z}_N}\times \mathcal{S}_{1\times\mathbb{Z}_N} =\frac{1}{N^2}\sum_{\Gamma,\Gamma^\prime}M(\Gamma)M(\Gamma^\prime)= \mathcal{Z}(\mathbb{Z}_N,\Sigma) \mathcal{S}_{1\times\mathbb{Z}_N},\\
        & \mathcal{S}_{\mathbb{Z}_N\times1}\times \mathcal{S}_{\mathbb{Z}_N\times\mathbb{Z}_N}=\frac{N}{|H_1(\Sigma,\mathbb{Z}_N)|^2}\sum_{\gamma,\gamma^\prime,\Gamma}W(\gamma^\prime+\gamma)M(\Gamma)=N \mathcal{S}_{\mathbb{Z}_N\times\mathbb{Z}_N},\\
        & \mathcal{S}_{\mathbb{Z}_N\times1}\times \mathcal{S}_{\mathbb{Z}_N\times\mathbb{Z}_N,-1}=\frac{N}{|H_1(\Sigma,\mathbb{Z}_N)|^2}\sum_{\gamma,\gamma^\prime,\Gamma}e^{-\frac{2\pi i}{N}\langle \gamma,\Gamma\rangle}W(\gamma^\prime+\gamma)M(\Gamma)= \mathcal{S}_{\mathbb{Z}_N\times1}.
    \end{split}
\end{align}
For the first derivation we used $M(\Gamma)M(\Gamma^\prime)=M(\Gamma+\Gamma^\prime)$ and then we redefined the $(D-2)$-cycles, summed over the remaining one giving a factor of $|H_{D-2}(\Sigma,\mathbb{Z}_N)|=|H^1(\Sigma,\mathbb{Z}_N)|=N\mathcal{Z}(\mathbb{Z}_N,\Sigma)$. For the second, we redefined the 1-cyles and summed over the remaining one giving a factor of $|H_1(\Sigma,\mathbb{Z}_N)|$. For the third, we did the same, but the summation over the remaining cycle forced $\Gamma=0$ in addition to giving the factor of $|H_1(\Sigma,\mathbb{Z}_N)|$. This, and the other 13 fusion rules, agree with the elegant expression \eqref{eq:fusionfactorized}. For the derivation, one needs to use \eqref{partitionfunction}.

\paragraph{Transformation of other operators} Similarly, we can use the higher gauging expression for the domain walls to compute the transformation of other operators using the algebraic properties of Wilson lines and magnetic defects. Using \eqref{eq:action_highergauging} one can derive

\begin{align}
    \mathcal{S}_{1\times\mathbb{Z}_{N/M}}\cdot M^{k N/M}=\sum_{n=0}^{M-1}M^{nN/M},\qquad \mathcal{S}_{1\times\mathbb{Z}_{N/M}}\cdot W^{\ell M}=\sum_{n=0}^{N/M-1}W^{nM},
\end{align}
for all $0\leq k\leq M-1$ and $0\leq \ell\leq N/M-1$ and $\mathcal{S}_{1\times\mathbb{Z}_{N/M}}\cdot M^k=\mathcal{S}_{1\times\mathbb{Z}_{N/M}}\cdot W^\ell=0$ for the other lines. This is consistent with \eqref{eq:action_GG} reviewed in Section \ref{sec:review}. In particular, for $M=N$ all magnetic defects can end on $\mathcal{D}_{\mathbb{Z}_N\times\mathbb{Z}_N}$ and no electric lines can.

\subsubsection{Automorphism domain walls}
\label{sec:highergaugingautomorphism}

Consider the automorphism $m\in\mathbb{Z}_N^\times\cong\operatorname{Aut}(\mathbb{Z}_N)$ that sends $1\mapsto m$ and its associated subgroup $\mathbb{Z}_N^{(m)}=\{(mn,n):n\in\mathbb{Z}_N\}$. Let's first assume that $\gcd(1-m,N)=1$. Then

\begin{align}
    \begin{split}
        {\cal D}_{\mathbb{Z}_N^{(m)}}(\Sigma) & =\frac{1}{N^{b_1}}\sum_{\substack{i_1,i_2,\dots,i_{b_1}}} W_{\overline{\rho}_{i_1}}(A_1)\dots W_{\overline{\rho}_{j_{b_1}}}(A_{b_1})\mathcal{D}_{\mathbb{Z}_N\times \mathbb{Z}_N}(\Sigma)W_{\rho_{i_1}\cdot\phi}(A_1)\dots W_{\rho_{i_{b_1}\cdot\phi}}(A_{b_1})\\
        & = \frac{1}{N^{b_1}}\sum_{i_1=0}^{N-1}\dots \sum_{i_{b_1}=0}^{N-1} W^{-1}(i_1A_1+\dots +i_{b_1}A_{b_1})\mathcal{D}_{\mathbb{Z}_N\times \mathbb{Z}_N}(\Sigma)W^{m}(i_1A_1+\dots +i_{b_1}A_{b_1})\\
        & = \frac{1}{N^{b_1}}\sum_{\substack{\gamma\in H_1(\Sigma,\mathbb{Z}_N)\\ \Gamma\in H_{D-2}(\Sigma,\mathbb{Z}_N)}}W^{-1}(\gamma)M(\Gamma)W^{m
        }(\gamma)\\
        & = \frac{1}{N^{b_1}}\sum_{\substack{\gamma\in H_1(\Sigma,\mathbb{Z}_N)\\ \Gamma\in H_{D-2}(\Sigma,\mathbb{Z}_N)}}e^{-\frac{2\pi i}{N}m\langle\gamma,\Gamma\rangle}W^{m-1}(\gamma)M(\Gamma)\\
        & =\frac{1}{N^{b_1}}\sum_{\substack{\gamma\in H_1(\Sigma,\mathbb{Z}_N)\\ \Gamma\in H_{D-2}(\Sigma,\mathbb{Z}_N)}}e^{\frac{2\pi i}{N}\frac{m}{1-m}\langle\gamma,\Gamma\rangle}W(\gamma)M(\Gamma),\\
        & = \mathcal{S}_{\mathbb{Z}_N\times\mathbb{Z}_N,\frac{m}{1-m}}(\Sigma)
    \end{split}
\end{align}
In the first equality, we combined the paths. Then, we passed the Wilson lines through the magnetic operator getting a phase proportional to their intersection number, then we redefined $\gamma\rightarrow(1-m)\gamma$ (which is possible under the assumption that $\gcd(1-m,N)=1$). Notice that if $m=1$ the Wilson lines would be annihilated and the summation over $\gamma$ would enforce $\Gamma=0$ with an extra factor of $N^{b_1}$ resulting in $D_1(\Sigma)=1$, confirming the previous result. We derived that $\mathcal{D}_{\mathbb{Z}_N^{(m)}}$ is the higher gauging of the global higher-form symmetry $\mathbb{Z}_N\times\mathbb{Z}_N$ generated by the Wilson line and magnetic operator with torsion term given by $\frac{m}{1-m}$ \footnote{If we repeated the above derivation for the defect associated with the inverse automorphism $1\mapsto m^{-1}$, we would get the higher gauging of the electric and magnetic symmetries with a torsion term equal to $\frac{1}{m-1}$. This gives the deeper underlying reasons behind the ``crucial" change of variables $f=\frac{1}{m-1}$ mentioned in \cite{Roumpedakis:2022aik}.}.

Now, consider the more general case $\ell=\gcd(1-m ,N)\neq1$. We would instead have:

\begin{align}
    \begin{split}
        \mathcal{D}_{\mathbb{Z}^{(m)}_N}(\Sigma) & =\sum_{j_a,k_a=0}^{N-1}e^{-\frac{2\pi i}{N}mj_ak_a}W^{m-1}(k_a \gamma_a)M(j_a \Gamma_a)\\
        & = \sum_{l_a,s_a,j_a=0}^{\ell,N/\ell-1,N-1}e^{-\frac{2\pi i}{N}mj_a(l_a N/\ell+s_a)}W^{m-1}(s_a\gamma_a)M(j_a \Gamma_a),\qquad k_a=l_aN/\ell+s_a\\
        & = \ell^{b_1}\sum_{s_a,s_a^\prime=0}^{N/\ell-1}e^{-\frac{2\pi i}{N/\ell}ms^\prime_as_a}W^{m-1}(s_a\gamma_a)M(\ell s^\prime_a \Gamma_a),\\
        & =\frac{\ell^{b_1}}{N^{b_1}}\sum_{\substack{\gamma\in H_1(\Sigma,\mathbb{Z}_{N/\ell})\\ \Gamma\in H_{D-2}(\Sigma,\mathbb{Z}_{N/\ell})}}^{N/\ell-1}e^{\frac{2\pi i}{N/\ell}\ell\frac{m}{1-m}\langle\gamma,\Gamma\rangle}W^{\ell}(\gamma)M^\ell( \Gamma),\\
        & = S_{\mathbb{Z}_{N/\ell}\times\mathbb{Z}_{N/\ell},\ell \frac{m}{1-m}}(\Sigma),
    \end{split}
    \label{eq:automorphismhighergauging}
\end{align}
where from the second to the third line we performed the summation over $l_a$ giving the factor of $\ell^{b_1}$ and enforcing $j_aN/\ell\equiv 0\mod N$, that is $j_a=\ell s^\prime_a$ with $0\leq s^\prime\leq N/\ell-1$. Finally we redefined $s_a \rightarrow \frac{\ell}{1-m} s_a$ which is possible because $(1-m)/\ell$ and $N/\ell$ are coprime.

\paragraph{Fusion rule} One can use the condensation expression of \eqref{eq:automorphismhighergauging} and the algebraic properties of Wilson lines and magnetic defects to compute its fusion rule. As we are going to see in detail in a more general context in Section \ref{sec:highergauginggeneratorfusion}, the derivation depends in subtle ways on the number theoretic properties of the particular numbers involved, but in all cases we investigated we checked that the defect fuses according to the automorphism composition law as in \eqref{eq:fusionautomorphism}.

\paragraph{Transformation of other operators} Similarly, one can use the condensation expression of \eqref{eq:automorphismhighergauging} to compute the transformation of other operators of the automorphism domain wall using the algebraic properties of Wilson lines and magnetic defects. Using \eqref{eq:action_highergauging} one can derive
\begin{align}
    \mathcal{S}_{\mathbb{Z}_{N/\ell}\times\mathbb{Z}_{N/\ell},\ell \frac{m}{1-m}}\cdot M=M^{m},\qquad \mathcal{S}_{\mathbb{Z}_{N/\ell}\times\mathbb{Z}_{N/\ell},\ell \frac{m}{1-m}}\cdot W=W^{1/m}.
\end{align}
This is consistent with \eqref{eq:action_automorphism} reviewed in Section \ref{sec:review}. Indeed, let $\rho$ and $\rho^{1/m}$ be the representations associated to $W$ and $W^{1/m}$ respectively. Then $\rho^{1/m}\otimes \overline{\rho}$ is the trivial representation of $\mathbb{Z}_N\times\mathbb{Z}_N$ when restricted to the subgroup $\mathbb{Z}_N^{(m)}\lhd \mathbb{Z}_N\times\mathbb{Z}_N$. This means that the configuration with the Wilson line $W$ coming from the right and being permuted to $W^{1/m}$ after passing thru $\mathcal{D}_{\mathbb{Z}^{(m)}_N}$ is gauge invariant. Likewise, the configuration with magnetic operator $M$ coming from the right and being permute to $M^m$ is also gauge invariant. The fact that Eq. (5.43) of \cite{Roumpedakis:2022aik} also computes the transformation of other operators ``from right to left" corresponds to their convention of having the condensation defect oriented inwards. The transformation ``from left to right" would corresponds to have the condensation defect oriented outwards which is the same as the transformation of the orientation reversal \eqref{eq:orientationreversal}.

\paragraph{Consistency with electric-magnetic duality} The condensation expression \eqref{eq:automorphismhighergauging} is consistent with electromagnetic duality in $D=2+1$. Under electric-magnetic duality (i.e., $W\leftrightarrow M$) we should have $\mathcal{D}_{\mathbb{Z}_N^{(m)}}\mapsto \mathcal{D}_{\mathbb{Z}_N^{(1/m)}}$. Indeed, an electric-magnetic duality transformation will change the order of $M$ and $W$ in the above condensation. Then, switching the order back to the order conventionally used to define the higher gauging will change the phase to $\frac{m}{1-m}\mapsto \frac{m}{1-m}+1$. Finally, by swapping the paths $\langle\Gamma,\gamma\rangle\mapsto-\langle\gamma,\Gamma\rangle$ the overall transformation will exactly map $\mathcal{D}_{\mathbb{Z}^{(m)}}\mapsto \mathcal{D}_{\mathbb{Z}^{(1/m)}}$. Note that everything is consistent because $\ell\mapsto \gcd(N,1-\frac{1}{m})=\gcd(N,1-m)=\ell$ because $m$ is coprime with $N$.

\subsection{Twisted domain walls as sequential higher gauging defects}
\label{eq:highergauging_electricmagnetic}

In the previous section, we derived the correspondence between untwisted domain walls and higher gauging condensation defects. Here, we extend our investigation to twisted domain walls, i.e., domain walls decorated with a non-trivial topological action. To organize the investigation, we introduce the concept of sequential higher gauging. This notion involves the higher gauging of the $\mathbb{Z}_N\times\mathbb{Z}_N$ 1-form global symmetry generated by magnetic defects and the dual symmetry that emerges after gauging the $(D-2)$-form symmetry generated by Wilson lines.

\subsubsection{Electric-magnetic duality domain wall in \texorpdfstring{$D=2+1$}{}}
\label{sec:electricmagnetichighergauging}

To begin the discussion, let us show that in $D=2+1$, the $\mathbb{Z}_2$ electric-magnetic duality domain wall $\mathcal{D}_{\mathbb{Z}_2\times\mathbb{Z}_2,\alpha_2}$ with non-trivial topological action $\alpha_2\in H^2(\mathbb{Z}_2\times \mathbb{Z}_2,U(1))=\mathbb{Z}_2$ (see Section \ref{sec:review}), is equivalent to the higher gauging of the $\mathbb{Z}_2$ 1-form global symmetry generated by the Dyon, i.e.,
\begin{align}
    \mathcal{D}_{\mathbb{Z}_2\times\mathbb{Z}_2,\alpha_2}=\mathcal{S}_{\mathbb{Z}_2^{(\textrm{id})}}
    \label{eq:emdualityDyon}
\end{align}
with $\mathbb{Z}_2^{(\textrm{id})}\lhd \mathbb{Z}_2\times \mathbb{Z}_2$ the diagonal $\mathbb{Z}_2$ subgroup of the 1-form global symmetry.

One way to see this is by noticing that one can obtain $\mathcal{D}_{\mathbb{Z}_N\times\mathbb{Z}_N,\alpha_2}$ by attaching the non-trivial topological action $\alpha_2$ to $\mathcal{D}_{\mathbb{Z}_N\times\mathbb{Z}_N}$. We saw in Section \ref{sec:factorizedhighergauging} that the later is the higher gauging of the 1-form symmetry generated by magnetic defects. If we attach the topological action we should dress the magnetic defects with suitable electric charges \cite{Barkeshli:2022edm}. Therefore, $\mathcal{D}_{\mathbb{Z}_2\times\mathbb{Z}_2,\alpha_2}$ should be the condensation of the Dyon.

One can also derive this correspondence explicitly from the BF action (see Section 4.1 of \cite{Cordova:2024jlk} for notation and details). The action for the condensation of the Dyon on $\Sigma$ can be written as
\begin{align}
        \mathcal{S}_{\mathbb{Z}_2^{(\textrm{id})}}(\Sigma):\qquad\frac{i}{\pi}\int_{\mathcal{M}} \tilde{a}^{(1)}\wedge da^{(1)}+\frac{i}{\pi}\int_\Sigma (a^{(1)}+\tilde{a}^{(1)})\wedge A^{(1)}.
\end{align}
with $A^{(1)}$ the gauged background field. Having $\mathcal{M}=\mathcal{M}_L\cup \mathcal{M}_R$ such that $\partial \mathcal{M}_L=-\partial \mathcal{M}_R=\Sigma$, we define $a=a_L+a_R$ with $a_L$ defined on $\mathcal{M}_L$ and $a_R$ on $\mathcal{M}_R$ and the same for $\tilde{a}$. Then, one notes that the cross-terms that are integrated on $\mathcal{M}$ are zero except on $\Sigma$. However, integrating out $A$ forces $a\vert_\Sigma=\tilde{a}\vert_\Sigma$, therefore such terms cancel if we integrate one by part. Integration by parts generates the term $\tilde{a}_R\wedge a_L$ on $\Sigma$, and using the condition from integrating out $A^{(1)}$ we derive:
\begin{align}
    \mathcal{S}_{\mathbb{Z}_2^{(\textrm{id})}}(\Sigma):\qquad \frac{i}{\pi}\int_\mathcal{M} (\tilde{a}_L^{(1)}\wedge da_L^{(1)}+\tilde{a}_R^{(1)}\wedge da_R^{(1)})+\frac{i}{\pi}\int_\Sigma a_R^{(1)}\wedge a_L^{(1)},
\end{align}
which is precisely the defect associated to the $\mathbb{Z}_2\times\mathbb{Z}_2$ subgroup with the non-trivial topological action $\alpha_2\in H^2(\mathbb{Z}_2\times\mathbb{Z}_2,U(1))=\mathbb{Z}_2$ as showed in \cite{Cordova:2024jlk}. We conclude \eqref{eq:emdualityDyon}.

This result generalizes to $\mathbb{Z}_N$ where the $\mathbb{Z}_N^\times$ different Dyons corresponds to the different non-trivial topological actions $H^2(\mathbb{Z}_N,U(1))=\mathbb{Z}_N$.

\subsubsection{Diagonal twisted domain walls and sequential higher gauging}
\label{sec:squentialhighergauging}

There is a counting incompatibility between domain walls and higher gauging condensation defects that appears in spacetime dimensions different from $D=2+1$. The global symmetry $\mathbb{Z}_N\times\mathbb{Z}_N$ generated by Wilson lines and magnetic defects have different form degrees if $D\neq2+1$, which has fewer subgroups than the case when they are of the same form degree. For example, they do not have diagonal subgroups corresponding to Dyons when $D\neq2+1$. However, there are still domain walls $\mathcal{D}_H$ associated to every subgroup of $H\lhd \mathbb{Z}_N\times\mathbb{Z}_N$, including diagonal ones. This incompatibility appears because our domain wall classification encompasses general invertible electric operators \eqref{eq:invertibleelectricdect} obtained by stacking a topological action, and this objects are not generated by the higher gauging of Wilson lines and magnetic defects. However, there is a generalization of the notion of higher gauging that naturally produces them, and also solves the counting incompatibility mentioned before. 

\paragraph{Sequential higher gauging} If one has a theory with an 1-gaugable abelian $q$-form global symmetry $G^{(q)}$, gauging $G^{(q)}$ in a codimension-one submanifold $\Sigma$ generates a higher gauging condensation defect. Similarly to what happens when gauging on the whole spacetime \cite{Bhardwaj:2017xup}, there emerges a dual symmetry $\overline{G}^{(D-1-q)}$ after gauging $G^{(q)}$ on $\Sigma$. Gauging $\overline{G}^{(D-1-q)}$ cancels the original gauging procedure leading one back to the trivial defect. However, there might be different ways of gauging $\overline{G}^{(D-1-q)}$, labeled by a choice of discrete torsion. Gauging the dual symmetry with a non-trivial choice of discrete torsion produces a non-trivial defect.

\vspace{5mm}
Provided this sequential higher gauging notion, the counting problem we started with is resolved because we can consider any subgroup $H\lhd \mathbb{Z}_N\times \mathbb{Z}_N$ with a choice of discrete torsion $ H^{D-1}(H,U(1))$. Here, the left factor of $H$ is the 1-form symmetry generated by the magnetic defects, and the right factor is the 1-form dual symmetry that emerges when gauging the $(D-2)$-form symmetry generated by Wilson lines on $\Sigma$. In this setup, it is possible to generate the general electric operators of \eqref{eq:invertibleelectricdect}. Let us show that with a particular example.

Consider the generalization to higher codimension of the diagonal defects  (see \eqref{eq:highercodimension} and the discussion before that). For every $1\leq n\leq D$ and given $K=\mathbb{Z}_2$ they are classified by
\begin{align}
    H^{n}(\mathbb{Z}_2,U(1))\cong\begin{cases}
        \mathbb{Z}_2 & n\ \textrm{odd},\\
        1 & \textrm{otherwise},
    \end{cases}
\end{align}
so we just need to consider the case in which $n$ is odd. The domain wall $\mathcal{D}_{\mathbb{Z}_2^{(\textrm{id})},\alpha_n}(\Sigma_{n})$ is obtained by staking the non-trivial topological action $\alpha_n\in H^n(\mathbb{Z}_2,U(1))$ along some $n$-dimensional submanifold $\Sigma_{n}$. In terms of 1-cocyles, the topological action can be written as $\alpha_n=\frac{i}{\pi} \int_{\Sigma_n}a^{(1)}\wedge(da^{(1)})^{\frac{n-1}{2}}$. One can obtain this defect by gauging the electric $(D-2)$-form $\mathbb{Z}_N$ symmetry along $\Sigma_n$, and then one gauges the dual symmetry with the non-trivial discrete torsion. For example, for $n=3$ we have:
\begin{align}
    \mathcal{D}_{\mathbb{Z}_2^{(\textrm{id})},\alpha_3}(\Sigma_3)=e^{\frac{i}{\pi}\int_{\Sigma_3} a^{(1)}\wedge da^{(1)}}=\sum_{\substack{A^{(2)}\in H^2(\Sigma_3,\mathbb{Z}_N)\\ \tilde{A}^{(1)}\in H^1(\Sigma_3,\mathbb{Z}_2)}}e^{\frac{i}{\pi}\int_{\Sigma_3}\big(a^{(1)}\wedge A^{(2)}+A^{(2)}\wedge \tilde{A}^{(1)}+ \tilde{A}^{(1)}\wedge d\tilde{A}^{(1)}\big)}.
\end{align}
Above, $a^{(1)}$ is the 1-form gauge field of a $\mathbb{Z}_2$ gauge theory, $A^{(2)}$ is the gauged background field for the electric $(D-2)$-form $\mathbb{Z}_2$ symmetry supported on $\Sigma_3$, and $\tilde{A}^{(1)}$ is the gauged background field for the dual 1-form global symmetry associated to $A^{(2)}$ that emerged along $\Sigma_3$. We see that gauging the electric $(D-2)$-form $\mathbb{Z}_2$ on $\Sigma_3$ and then gauging the dual symmetry with the non-trivial torsion term is the same as adding a topological action for $a$ along $\Sigma_3$. The equality above generalizes in $n$ and to $\mathbb{Z}_N$.

To conclude, the notion of sequential higher gauging solves the apparent counting problem involving our lattice domain wall formalism and higher gauging. 

\subsubsection{Non-invertible electric-magnetic duality domain wall in \texorpdfstring{$D=3+1$}{}}

Consider the domain wall $\mathcal{D}_{\mathbb{Z}_2\times\mathbb{Z}_2,\alpha_3}$ with $\alpha_3$ the non-factorized topological action of $H^3(\mathbb{Z}_2\times\mathbb{Z}_2,U(1))=\mathbb{Z}_2\times\mathbb{Z}_2\times\mathbb{Z}_2$. This domain wall is equivalent to the higher gauging of the generalization of a Dyon, that is, the magnetic operator dressed with the invertible electric operator \eqref{eq:invertibleelectricdect}:
\begin{align}
    M_{\beta}(\Gamma)\equiv M(\Gamma)W_\beta(\Gamma)=e^{i\int_\Gamma (\tilde{a}^{(2)}+da^{(1)})},
    \label{eq:higherdimensionaldyon}
\end{align}
with $\beta$ the non-trivial element of $H^3(\mathbb{Z}_2,U(1))=\mathbb{Z}_2$. From the action for the condensation of $M_\beta$, namely:
\begin{align}
    \mathcal{D}_{\mathbb{Z}_2\times\mathbb{Z}_2,\alpha_3}:\qquad\frac{i}{\pi}\int_\mathcal{M}a^{(1)}\wedge d\tilde{a}^{(1)}+\frac{i}{\pi}\int_\Sigma(da^{(1)}+\tilde{a}^{(2)})\wedge A^{(1)},
\end{align}
we can repeat the same manipulations of Section \ref{sec:electricmagnetichighergauging} to derive
\begin{align}
    \mathcal{D}_{\mathbb{Z}_2\times\mathbb{Z}_2,\alpha_3}:\qquad \frac{i}{\pi}\int_\mathcal{M}(a_L^{(1)}\wedge d\tilde{a}_L^{(1)}+a_R^{(1)}\wedge d\tilde{a}_R^{(1)})+\frac{i}{\pi}\int_\Sigma a_R^{(1)}\wedge da_L^{(1)},
\end{align}
which corresponds to the domain wall associated to the $\mathbb{Z}_2\times\mathbb{Z}_2$ subgroup with the non-factorized topological action $\alpha_3\in H^3(\mathbb{Z}_2\times\mathbb{Z}_2,U(1))$ \cite{Cordova:2024jlk}.

Complementarily, we can also use the idea of sequential higher gauging to describe the above domain wall. In this language, the domain wall $\mathcal{D}_{\mathbb{Z}_2\times\mathbb{Z}_2,\alpha_3}$ corresponds to the higher gauging of the full 1-form symmetry $\mathbb{Z}_2\times\mathbb{Z}_2$ (where the right factor is the dual 1-form symmetry that emerges when gauging the electric symmetry) with the non-factorized element of $H^3(\mathbb{Z}_2\times\mathbb{Z}_2,U(1))\cong\mathbb{Z}_2\times\mathbb{Z}_2\times\mathbb{Z}_2$. More precisely,
\begin{align}
    \begin{split}
        & \mathcal{D}_{\mathbb{Z}_2\times\mathbb{Z}_2,\alpha_3}(\Sigma)=\sum_{\Gamma\in H_2(\Sigma,\mathbb{Z}_2)}M_\beta(\Gamma),\\
        & =\sum_{\substack{ A^{(2)}\in H^2(\Sigma,\mathbb{Z}_2)\\ \bar{A}^{(1)},\tilde{A}^{(1)}\in H^1(\Sigma,\mathbb{Z}_2)}}\exp\Big(\frac{i}{\pi}\int a^{(1)}\wedge A^{(2)}+A^{(2)}\wedge \bar{A}^{(1)}+\tilde{a}^{(2)}\wedge \tilde{A}^{(1)}+\tilde{A}^{(1)}\wedge d\bar{A}^{(1)}\Big),
    \end{split}
\end{align}
where $M_\beta$ is defined in \eqref{eq:higherdimensionaldyon}, $A^{(2)}$ is a background field for the electric 2-form symmetry, $\tilde{A}^{(1)}$ is a background field for the 1-form magnetic symmetry, and $\bar{A}^{(1)}$ is the background field for the 1-form dual symmetry that emerged after gauging the electric symmetry. To go from the second line to the first, we simply integrate out $A^{(2)}$, forcing $a^{(1)}=\bar{A}^{(1)}$ and then we rewrite the summation over $\tilde{A}^{(1)}$ as a summation over its Poncaré dual $\Gamma=\textrm{PD}(\tilde{A}^{(1)})$.

\section{Automorphism symmetry as higher gauging defects in abelian theories}
\label{sec:automorphism}

We generalize the discussion from the previous section to encompass general finite abelian gauge theories without topological terms. We focus on the invertible 0-form symmetries corresponding to the automorphisms of the gauge group. Our main result is the derivation of the higher gauging expression for a set of automorphism generators. As an illustration, we present the higher gauging condensation defects for the automorphism symmetries of $\mathbb{Z}_9$ and $\mathbb{Z}_2\times \mathbb{Z}_2$, i.e., $\operatorname{Aut}(\mathbb{Z}_9)\cong \mathbb{Z}_6$ and $\operatorname{Aut}(\mathbb{Z}_2\times\mathbb{Z}_2)\cong S_3$.

\subsection{Generators for automorphism symmetry of finite abelian groups}

By the fundamental theorem of finite abelian groups, a finite abelian group is isomorphic to a product of $H_p=\mathbb{Z}_{p^{e_1}}\times\dots\times \mathbb{Z}_{p^{e_L}}$ for different prime $p$ and $1\leq e_1\leq\dots\leq e_L$ are positive integers. From here on, we will denote $p^{e_i}=p_i$. Furthermore, $\operatorname{Aut}(H\times K)=\operatorname{Aut}(H)\times \operatorname{Aut}(K)$ for two finite groups $H,K$ with relatively prime orders, therefore, one just needs to find the automorphism of $H_p$ (see \cite{https://doi.org/10.48550/arxiv.math/0605185} for references). 

The automorphisms of $H_p$ are determined by where the $L$ generators are mapped to. Let $M_i$ be the generator of the $\mathbb{Z}_{p_i}$ factor. The automorphisms of $H_p$ are generated by the family $\phi_{(i,j,m)}:H_p\rightarrow H_p$ with $1\leq i,j\leq L$ and $m\in\mathbb{Z}_{p_i}$, or $\mathbb{Z}_{p_j}$, or $\mathbb{Z}_{p_i}^\times$ if $j>i,j<i$ or $j=i$ respectively. The automorphism $\phi_{(i,j,m)}$ is defined as $\phi_{(i,j,m)}\cdot M_k=M_k$ for $k\neq i$ and as 
\begin{align}
    \phi_{(i,j,m)}\cdot M_i=\begin{cases} 
        M_iM_j^{mp^{\epsilon_j-\epsilon_i}} & i<j,\\
        M_i^m & i=j,\\
        M_iM_j^{m} & i>j.\\
    \end{cases}
    \label{eq:automorphismgenerator}
\end{align}
For $i=j$ it is simply an automorphism of the $\mathbb{Z}_{p_i}$ factor. For $i\neq j$ it can be thought of as ``dressing" the $M_i$ generator with a subgroup generated by $M_j$. Furthermore, the power of $\frac{p_j}{p_i}=p^{e_j-e_i}$ for $j>i$ is necessary so that the image is a generator of order $p_i$.

If $p_j\neq p_i$ for all $j\neq i$, our claim is trivial to check seems compositions of \eqref{eq:automorphismgenerator} can be used to map $M_i$ to the most generic generator of order $p_i$. More explicitly, a generic generator of order $p_i$ has the form

\begin{align}
    M_1^{m_1}\dots M_{i-1}^{m_{i-1}}M_i^{m_i} M_{i+1}^{m_{i+1}p^{e_{i+1}-e_i}}\dots M_{L}^{m_L p^{e_L-e_i}},
\end{align}
with $m_i\in\mathbb{Z}_{p_i}^\times$, $m_j\in\mathbb{Z}_{p_i}$ for $j>i$ and $m_j\in\mathbb{Z}_{p_j}$ for $j<i$. And an automorphism that sends $M_i$ to the generic generator above is easily seem to be a composition of \eqref{eq:automorphismgenerator}. % For example, one can compose $\mathcal{S}_{(i,i,m_i)}$ with a product of $\mathcal{S}_{(i,j,m_j/m_i)}$ for all $j\neq i$.
Now, suppose exist $j$ such that $p_i=p_j$. In this case, it is necessary to check that the automorphisms above can permute generators with the same order. But the symmetric group is generated by transpositions \cite{Conrad_generatingsets}, so we just need to show that \eqref{eq:automorphismgenerator} can generate the transposition of two such generators. Let $M_i$ and $M_j$ be generators of order $q$, then their transposition is equal to the composition
\begin{align}
    \phi_{(j,j,q-1)}\circ \phi_{(i,j,1)}\circ \phi_{(j,i,1)}\circ \phi_{(i,i,q-1)}\circ \phi_{(i,j,1)}.
\end{align}
Above, we used the fact that $q-1$ and $q$ are always coprimes, i.e., $\gcd(q,q-1)=1$, so that, $q-1\in \mathbb{Z}_q^\times$. We conclude that \eqref{eq:automorphismgenerator} can generate all automorphisms of $H_p$.

\subsection{Domain walls of automorphism generators as higher gauging defects}
\label{sec:highergauginggenerator}

Now, consider a gauge theory with gauge group $H_p=\mathbb{Z}_{p_1}\times\dots\times \mathbb{Z}_{p_L}$. The theory has a $H_p$ $(D-2)$-form and $1$-form symmetries with $W_i$ and $M_i$ the Wilson line and magnetic operator associated to the $\mathbb{Z}_{p_i}$ factors, respectively. From the discussion in the previous section, we can construct all $\operatorname{Aut}(H_p)$ condensation defects by composition if we have the ones corresponding to the family of generators \eqref{eq:automorphismgenerator}. Given $p_j|q$ define the subgroup $\mathbb{Z}_{q,j}=\{(z_1,\dots,z_L):z_j\in\mathbb{Z}_q,z_i=1\ \forall\  i\neq j\}\lhd H_p$ so that, for example:
\begin{align}
    \mathcal{S}_{\mathbb{Z}_{q,i}\times\mathbb{Z}_{q,j},f}(\Sigma)\equiv\frac{1}{q^{b_1}}\sum_{\substack{\gamma\in H_1(\Sigma,\mathbb{Z}_{q})\\ \Gamma\in H_{D-2}(\Sigma,\mathbb{Z}_{q})}}e^{\frac{2\pi i}{q}f\langle\gamma,\Gamma\rangle}W_i^{p_i/q}(\gamma)M_j^{p_j/q}(\Gamma)
\end{align}
In words, the above is the higher gauging of the subgroup of $(D-2)$-form symmetry with all factors trivial except the $\mathbb{Z}_{p_i}$ factor that has subgroup $\mathbb{Z}_q$ gauged, and the subgroup of the 1-form symmetry with all factors trivial except the $\mathbb{Z}_{p_j}$ factor that has subgroup $\mathbb{Z}_{q}$ gauged. This gauging is performed with a choice of discrete torsion \cite{VAFA1986592,Vafa:1994rv} equal to $f \in \mathbb{Z}_q$.

Now denote the automormophism subgroup associated with \eqref{eq:automorphismgenerator} by $H_p^{(i,j,m)}=\{(\phi_{(i,j,k)}\cdot h, h):h\in H_p\}\lhd H_p\times H_p$. Similarly to the derivation of  \eqref{eq:automorphismhighergauging} we have
\begin{align}
    \mathcal{D}_{H_p^{(i,j,m)}}(\Sigma)=\begin{cases}
    \mathcal{S}_{\mathbb{Z}_{p_{i}/\gcd(m,p_{i}),i}\times\mathbb{Z}_{p_{i}/\gcd(m,p_{i}),j},-\frac{\gcd(m,p_{i})}{m}}(\Sigma), & i<j,\\
    \mathcal{S}_{\mathbb{Z}_{p_{i}/\gcd(1-m,p_{i}),i}\times\mathbb{Z}_{p_{i}/\gcd(1-m,p_{i}),j},\frac{\gcd(1-m,p_{i})m}{1-m}}(\Sigma), & i=j,\\
    \mathcal{S}_{\mathbb{Z}_{p_{j}/\gcd(m,p_{j}),i}\times\mathbb{Z}_{p_{j}/\gcd(m,p_{j}),j},-\frac{\gcd(m,p_{j})}{m}}(\Sigma), & i>j.
    \end{cases}
    \label{eq:generatorhighergauging}
\end{align}
with a well defined torsion term because $(1-m)/\gcd(1-m,p_{i})$ is coprime with $p_{i}/\gcd(1-m,p_{i})$ and has an inverse and likewise for $m/\gcd(m,p_j)$ and $m/\gcd(m,p_i)$. It is striking how the simplicity of the automorphism domain wall translates into a complex higher gauging expression that depends in subtle ways on particular number theoretic properties.

\subsubsection{Fusion rules}
\label{sec:highergauginggeneratorfusion}

We can use the condensation expression of \eqref{eq:generatorhighergauging} to compute the fusion rule of the automorphism domain wall using the algebraic properties of Wilson lines and magnetic defects. The derivation depends on particular number theoretic properties of the numbers in consideration. By making number theoretic assumptions we can find expressions that agree with the automorphism composition law reviewed in \eqref{eq:fusionautomorphism}.

Let us illustrate our claim by following similar steps to (A.21) from \cite{Roumpedakis:2022aik}. We have for example:
\begin{align}
    \begin{split}
        & \mathcal{S}_{\mathbb{Z}_{q,i}\times\mathbb{Z}_{q,j},f}\times \mathcal{S}_{\mathbb{Z}_{q,i}\times\mathbb{Z}_{q,j},f^\prime}\\
        & =\frac{1}{q^{2b_1}}\sum_{\gamma,\gamma^\prime,\Gamma,\Gamma^\prime}e^{\frac{2\pi i}{q}(f\langle\gamma,\Gamma\rangle+f^\prime\langle\gamma^\prime,\Gamma^\prime\rangle)}W_i^{\frac{p}{q}}(\gamma)M_j^{\frac{p}{q}}(\Gamma)W_i^{\frac{p}{q}}(\gamma^\prime)M_j^{\frac{p}{q}}(\Gamma^\prime),\\
        & =\frac{1}{q^{2b_1}}\sum_{\gamma,\gamma^\prime,\Gamma,\Gamma^\prime}e^{\frac{2\pi i}{q}(f\langle\gamma,\Gamma\rangle+f^\prime\langle\gamma^\prime,\Gamma^\prime\rangle-\frac{p}{q}\delta_{ij}\langle\gamma^\prime,\Gamma\rangle)}W_i^{\frac{p}{q}}(\gamma)W_i^{\frac{p}{q}}(\gamma^\prime)H_i^{\frac{p}{q}}(\Gamma)M_j^{\frac{p}{q}}(\Gamma),\\
        & =\frac{1}{q^{2b_1}}\sum_{\gamma,\gamma^\prime,\Gamma,\Gamma^\prime}e^{\frac{2\pi i}{q}(f\langle\gamma,\Gamma\rangle+f^\prime\langle\gamma^\prime,\Gamma^\prime\rangle-\frac{p}{q}\delta_{ij}\langle\gamma^\prime,\Gamma\rangle)}W_i^{\frac{p}{q}}(\gamma+\gamma^\prime)M_j^{\frac{p}{q}}(\Gamma+\Gamma^\prime),\\ 
        & =\frac{1}{q^{2b_1}}\sum_{\gamma,\gamma^\prime,\Gamma,\Gamma^\prime}e^{\frac{2\pi i}{q}(f\langle\gamma,\Gamma\rangle+(f+f^\prime+\frac{p}{q}\delta_{ij})\langle\gamma^\prime,\Gamma^\prime\rangle-(f+\frac{p}{q}\delta_{ij})\langle\gamma^\prime,\Gamma\rangle-f\langle\gamma,\Gamma^\prime\rangle)}W_i^{\frac{p}{q}}(\gamma)M_j^{\frac{p}{q}}(\Gamma),\\
        & =\frac{1}{q^{b_1}}\sum_{\gamma,\Gamma}e^{\frac{2\pi i}{q}\frac{ff^\prime}{f+f^\prime+\frac{p}{q}\delta_{ij}}\langle\gamma,\Gamma\rangle}W_i^{\frac{p}{q}}(\gamma)M_j^{\frac{p}{q}}(\Gamma),\\
        & =\mathcal{S}_{\mathbb{Z}_{q,j}\times\mathbb{Z}_{q,i},\frac{ff^\prime}{f+f^\prime+\frac{p}{q}\delta_{ij}}},
    \end{split}
\end{align}
where we assumed that $\gcd(f+f^\prime+\frac{p}{q}\delta_{ij},q)=1$ and summed over $\Gamma^\prime$ resulting in a Kronecker delta that forced $\gamma^\prime=\frac{f}{f+f^\prime+\frac{p}{q}\delta_{ij}}\gamma$, giving a factor of $q^{b_1}$.

If we assume that $p_i,p_j,m$ and $m^\prime$ are such that the conditions of the derivation above are satisfied, by plugging the values for $f$ and $f^\prime$ from the relation to the automorphism domain wall derived in \eqref{eq:generatorhighergauging} this fusion rule translates to:
\begin{align}
    & \mathcal{D}_{H_p^{(i,j,m)}}\times \mathcal{D}_{H_p^{(i,j,m^\prime)}}=\begin{cases} \mathcal{D}_{H_p^{(i,j,mm^\prime)}} & i=j,\\
    \mathcal{D}_{H_p^{(i,j,m+m^\prime)}} & i\neq j,\end{cases}
\end{align}
These, of course, agrees with the automorphisms composition law derived in \eqref{eq:fusionautomorphism}. But note that this particular derivation is correct just for numbers satisfying the assumptions.

Following the approach outlined in Appendix A of \cite{Roumpedakis:2022aik}, a more general expression for the fusion of condensation defects can be derived in roughly two pages of calculation, namely:

\begin{align}
    \mathcal{S}_{\mathbb{Z}_{q,i}\times\mathbb{Z}_{q,j},f}\times \mathcal{S}_{\mathbb{Z}_{q^\prime,i}\times\mathbb{Z}_{q^\prime,j},f^\prime}=\mathcal{S}_{\mathbb{Z}_{\lcm(q,q^\prime),i}\times\mathbb{Z}_{\lcm(q,q^\prime),j},\frac{ff^\prime}{\big(\frac{fq+f^\prime q^\prime+p\delta_{ij}}{\gcd(q,q^\prime)}\big)}}
    \label{generalfusion}
\end{align}
which holds as long as
\begin{align}
    \gcd\Big(\frac{fq+f^\prime q^\prime+p\delta_{ij}}{\gcd(q,q^\prime)},\lcm(q,q^\prime)\Big)=1.
    \label{condition}
\end{align}
In Section \ref{Z9} we will use this expression to show that the automorphism domain walls of $\mathbb{Z}_9$ fuse according to the automorphism composition as expected.

\subsubsection{Transformations of other operators}
\label{sec:highergaugingactionon}

We can use the condensation expression of \eqref{eq:generatorhighergauging} to compute the transformation of other operators on the automorphism domain wall using the algebraic properties of Wilson lines and magnetic defects. Using \eqref{eq:action_highergauging} and the value for $q,q^\prime$ and $f$ associated to the higher gauging expression of $\mathcal{D}_{H_p^{(i,j,m)}}$ derived in \eqref{eq:generatorhighergauging} and the fact that $(Q_1,Q_2)=(0,1)$ for the unit charge Wilson line and $(1,0)$ for the unit charge magnetic operator we derive:
\begin{align}
    \mathcal{D}_{H_p^{(i,j,m)}}(\Sigma)\cdot M_i=\begin{cases} M_iM_j^{m\frac{p_j}{p_i}},\\ 
    M_i^{m},\\
    M_iM_j^m,\end{cases}\quad \mathcal{D}_{H_p^{(i,j,m)}}(\Sigma)\cdot W_j=\begin{cases} W_jW_i^{-m}, & i<j,\\
    W_j^{\frac{1}{m}}, & i=j,\\
    W_jW_i^{-m\frac{p_i}{p_j}}, & i> j,\end{cases}
    \label{eq:generator_action_highergauging}
\end{align}
with the transformation of other operators being trivial. 

The derivation above shows that the defect \eqref{eq:generatorhighergauging} permutes the magnetic defects according to the automorphism \eqref{eq:automorphismgenerator} and that the representation of Wilson lines are permuted by composition with the inverse of \eqref{eq:automorphismgenerator}. This is consistent with \eqref{eq:action_automorphism} in particular with our remark that the transformation preserves the Aharonov-Bohm braiding between the Wilson lines and the magnetic operator.

Now consider the $i\neq j$ and for concreteness the automorphism $\phi^{(1,2,m)}:\mathbb{Z}_{p_1}\times\mathbb{Z}_{p_2}\rightarrow \mathbb{Z}_{p_1}\times\mathbb{Z}_{p_2}$ for some $m\in\mathbb{Z}_{p_1}$. This automorphism acts on generators according to \eqref{eq:automorphismgenerator} which is the same as the action from right to left on magnetic defects, i.e., $\mathcal{D}_{H_p^{(1,2,m)}}\cdot M_1=M_1M_2^{mp_2/p_1}$ and $M_2\mapsto M_2$. In terms of its action on group elements, the same automorphism is defined as $\phi^{(1,2,m)}(n_1,n_2)=(n_1,n_2+mn_1p_2/p_1)$. Now, consider the irreducible representations $\rho_i(n_1,n_2)= e^{\frac{2\pi i}{p_i}n_i}$ with $i=1,2$. It follows that $\rho_1\cdot\phi^{-1}=\rho_1$ and $\rho_2\cdot\phi^{-1}(n_1,n_2)=e^{\frac{2\pi i}{p_2}(n_2-mn_1p_2/p_1)}=\rho_2\otimes\rho_1^{-m}(n_1,n_2)$. Therefore, $\mathcal{D}_{H_p^{(1,2,m)}}\cdot W_2=W_2W_1^{-m}$ and $W_1\mapsto W_1$. This is but a particular case of the general expression \eqref{eq:generator_action_highergauging}.

\subsection{Examples}

Let us investigate in more detail some particular examples.

\subsubsection{\texorpdfstring{$\mathbb{Z}_9$: the $\mathbb{Z}_6$ higher gauging condensation defects}{}}
\label{Z9}

Here we will construct the symmetry defects associated to the automorphisms of $\mathbb{Z}_9$ gauge theory and we check their fusion rules. An automorphism of $\mathbb{Z}_9$ is a map that sends a generator $M$ to $M^m$ with $m\in\mathbb{Z}_9^\times=\{1,2,4,8,7,5\}\cong \mathbb{Z}_6$.  All this automorphisms are elements of the family defined in \eqref{eq:generatorhighergauging}. Denote the automorphism subgroup associated to $m$ by $\mathbb{Z}_9^{(m)}=\{(mn,n):n\in\mathbb{Z}_9\}$. Then we have the correspondence presented in Table \ref{tab:Z9automorphism}.

\begin{table}[ht]
    \centering
    \begin{tabular}{|c|c|}
         \hline Domain wall & Condensation \\ \hline
         $\mathcal{D}_{\mathbb{Z}_9^{(2)}}$ & $\mathcal{S}_{\mathbb{Z}_9\times\mathbb{Z}_9,1}$\\  
        $\mathcal{D}_{\mathbb{Z}_9^{(4)}}$ & $ \mathcal{S}_{\mathbb{Z}_3\times\mathbb{Z}_3,1}$\\ 
        $\mathcal{D}_{\mathbb{Z}_9^{(8)}}$ & $\mathcal{S}_{\mathbb{Z}_9\times\mathbb{Z}_9,4}$\\ 
        $\mathcal{D}_{\mathbb{Z}_9^{(7)}}$ & $ \mathcal{S}_{\mathbb{Z}_3\times\mathbb{Z}_3,2}$\\ 
        $\mathcal{D}_{\mathbb{Z}_9^{(5)}}$ & $ \mathcal{S}_{\mathbb{Z}_9\times\mathbb{Z}_9,7}$  \\ \hline 
    \end{tabular}
    \caption{Dictonary between higher gauging condensation defects and $\operatorname{Aut}(\mathbb{Z}_9)\cong \mathbb{Z}_6$ automorphism domain walls.}
    \label{tab:Z9automorphism}
\end{table}
\paragraph{Fusion rules}

Using the fusion rule \eqref{generalfusion}, we can check that the $5^2=25$ fusion rules agree with the composition law of automorphisms. For example, 

\begin{align}
    \mathcal{D}_{\mathbb{Z}_9^{(7)}}\times \mathcal{D}_{\mathbb{Z}_9^{(8)}}=\mathcal{S}_{\mathbb{Z}_3\times\mathbb{Z}_3,2}\times \mathcal{S}_{\mathbb{Z}_9\times\mathbb{Z}_9,4}=\mathcal{S}_{\mathbb{Z}_9\times\mathbb{Z}_9,1}=\mathcal{D}_{\mathbb{Z}_9^{(2)}}.
\end{align}
which agrees with the automorphism composition rule, seems $7\times 8=56\equiv 2\mod 9$.

\subsubsection{\texorpdfstring{$\mathbb{Z}_2\times\mathbb{Z}_2$: the $S^3$ higher gauging condensation defects}{}}
\label{Z2Z2}

Here we construct the automorphism condensation defects associated with the Klein four-group $\mathbb{Z}_2\times\mathbb{Z}_2$ and check that their fusion rules obey the $\operatorname{Aut}(\mathbb{Z}_2\times\mathbb{Z}_2)\cong S_3$ group law as in \eqref{eq:fusionautomorphism}. 

The Klein four group $\mathbb{Z}_2\times\mathbb{Z}_2$ has 3 generators of order $2$, which we denote by $A=(1,0)$, $B=(0,1)$ and $C=(1,1)$. The automorphism group of $\mathbb{Z}_2\times\mathbb{Z}_2$ is isomorphic to ordered lists of the three generators $\textrm{Aut}(\mathbb{Z}_2\times\mathbb{Z}_2)\cong \{\sigma\cdot [A,B,C]:\sigma\in S_3\}\cong S_3$ with group product given by the group product of the $S_3$ elements. The bijection is obtained by identifying the first and second elements in the corresponding list with the image of $A=(1,0)$ and $B=(0,1)$ under the corresponding automorphism. For instance, the automorphism associated with the list $(23)\cdot[A,B,C]=[A,C,B]$ maps $(1,0)=A\mapsto A=(1,0)$ and $(0,1)=B\mapsto C=(1,1)$. We use this isomorphism to denote the elements of $\operatorname{Aut}(\mathbb{Z}_2\times\mathbb{Z}_2)$ by elements of $S_3=\{\textrm{id},(13),(23),(12),(123),(132)\}$. Likewise, denote the automorphism subgroup associated to $\sigma\in S_3$ by $(\mathbb{Z}_2\times\mathbb{Z}_2)^{(\sigma)}=\{(\sigma\cdot n,n):n\in\mathbb{Z}_2\times\mathbb{Z}_2\}\leq \mathbb{Z}_2\times\mathbb{Z}_2\times\mathbb{Z}_2\times\mathbb{Z}_2$.

As shown in Section \ref{sec:automorphism}, the automorphisms of abelian groups can be generated by the family \eqref{eq:automorphismgenerator}. The two automorphisms in this family for the Klein four group are $(13),(23)$. Using \eqref{eq:generatorhighergauging} to write them and the fact that $(12)=(13)(23)(13)$, $(123)=(23)(13)$ and $(132)=(13)(23)$, we find the correspondence presented in Table \ref{tab:Z2Z2automorphism}.

\begin{table}[ht]
    \centering
    \begin{tabular}{|c|c|}
         \hline Domain wall & Condensation \\ \hline
         $\mathcal{D}_{(\mathbb{Z}_2\times\mathbb{Z}_2)^{(\textrm{id})}}$ & $\mathcal{S}_1$ \\ 
         $\mathcal{D}_{(\mathbb{Z}_2\times\mathbb{Z}_2)^{(13)}}$  & $\mathcal{S}_{(1\times \mathbb{Z}_{2})\times(\mathbb{Z}_{2}\times 1),1}$\\  
        $\mathcal{D}_{(\mathbb{Z}_2\times\mathbb{Z}_2)^{(23)}}$  & $ \mathcal{S}_{(\mathbb{Z}_{2}\times 1)\times(1\times\mathbb{Z}_{2}),1}$\\ 
        $\mathcal{D}_{(\mathbb{Z}_2\times\mathbb{Z}_2)^{(12)}}$ & $ \mathcal{S}_{\mathbb{Z}_{2}^{\textrm{(id)}}\times\mathbb{Z}_{2}^{\textrm{(id)}},1}$ \\ 
        $\mathcal{D}_{(\mathbb{Z}_2\times\mathbb{Z}_2)^{(123)}}$ & $ \mathcal{S}_{(\mathbb{Z}_{2}\times \mathbb{Z}_{2})\times (\mathbb{Z}_{2}\times \mathbb{Z}_{2}),(0,1,1,1)}$\\ 
        $\mathcal{D}_{(\mathbb{Z}_2\times\mathbb{Z}_2)^{(132)}}$  & $ \mathcal{S}_{(\mathbb{Z}_{2}\times \mathbb{Z}_{2})\times (\mathbb{Z}_{2}\times \mathbb{Z}_{2}),(1,1,1,0)}$\\  \hline 
    \end{tabular}
    \caption{Dictionary between higher gauging condensation defects and $\operatorname{Aut}(\mathbb{Z}_2\times\mathbb{Z}_2)\cong S_3$ automorphism domain walls.}
    \label{tab:Z2Z2automorphism}
\end{table}

The torsion in the last two rows have elements associated with the pairing of $\langle\gamma_1,\Gamma_1\rangle$, $\langle\gamma_1,\Gamma_2\rangle$, $\langle\gamma_2,\Gamma_1\rangle$ and $\langle\gamma_2,\Gamma_2\rangle$. For example:
\begin{align}
    \mathcal{D}_{(\mathbb{Z}_2\times\mathbb{Z}_2)^{(132)}}= \frac{1}{|\mathbb{Z}_2|^{2b_1}}\sum_{\gamma_1,\gamma_2,\Gamma_1,\Gamma_2}e^{\frac{2\pi i}{2}(\langle \gamma_1,\Gamma_1\rangle+\langle \gamma_1,\Gamma_2\rangle+\langle \gamma_2,\Gamma_1\rangle)}W_1(\gamma_1)W_2(\gamma_2)M_1(\Gamma_1)M_2(\Gamma_2).
\end{align}
which we denoted by $\mathcal{S}_{\mathbb{Z}_2\times\mathbb{Z}_2\times \mathbb{Z}_2\times \mathbb{Z}_2,(1,1,1,0)}$ with torsion equal to $(1,1,1,0)$.

\paragraph{Fusion rules} Because we constructed the symmetry defects from the generators $\mathcal{D}_{(\mathbb{Z}_2\times\mathbb{Z}_2)^{(13)}}$ and $\mathcal{D}_{(\mathbb{Z}_2\times\mathbb{Z}_2)^{(23)}}$, we just need to check their fusion. Transpositions are their own inverse, $(13)^2=(23)^2=1$, and we do have
\begin{align*}
    \begin{split}
        \mathcal{S}_{\mathbb{Z}_{2,2}\times\mathbb{Z}_{2,1},1}\times \mathcal{S}_{\mathbb{Z}_{2,2}\times\mathbb{Z}_{2,1},1} & =\frac{1}{|\mathbb{Z}_2|^{2b_1}}\sum_{\gamma_i,\Gamma_i^\prime}e^{\frac{2\pi i}{2}\langle \gamma,\Gamma\rangle}e^{\frac{2\pi i}{2}\langle \gamma^\prime,\Gamma^\prime\rangle}W_2(\gamma)M_1(\Gamma)W_2(\gamma^\prime)M_1(\Gamma^\prime)\\
        & =\frac{1}{|\mathbb{Z}_2|^{2b_1}}\sum_{\gamma,\gamma^\prime,\Gamma,\Gamma^\prime}e^{\frac{2\pi i}{2}\langle \gamma,\Gamma\rangle}e^{\frac{2\pi i}{2}\langle \gamma^\prime,\Gamma^\prime\rangle}W_2(\gamma+\gamma^\prime)M_1(\Gamma+\Gamma^\prime)\\
        & =\frac{1}{|\mathbb{Z}_2|^{2b_1}}\sum_{\gamma,\gamma^\prime,\Gamma,\Gamma^\prime}e^{\frac{2\pi i}{2}(\langle \gamma,\Gamma\rangle-\langle \gamma^\prime,\Gamma\rangle+\langle \gamma^\prime,\Gamma^\prime\rangle-\langle \gamma^\prime,\Gamma\rangle)}W_2(\gamma)M_1(\Gamma^\prime)\\
        & =\frac{1}{|\mathbb{Z}_2|^{2b_1}}\sum_{\gamma,\gamma^\prime,\Gamma,\Gamma^\prime}e^{\frac{2\pi i}{2}(\langle \gamma,\Gamma\rangle+\langle \gamma^\prime,\Gamma^\prime\rangle)}W_2(\gamma)M_1(\Gamma^\prime)\\
        & = 1,
    \end{split}
    \label{squareto1}
\end{align*}
where we integrate out $\Gamma$ and $\gamma^\prime$ forcing $\gamma=\Gamma^\prime=0$. The same derivation applies to $\mathcal{D}_{(\mathbb{Z}_2\times\mathbb{Z}_2)^{(23)}}$. Consistently, it also applies to $\mathcal{D}_{(\mathbb{Z}_2\times\mathbb{Z}_2)^{(13)}}$, because

\begin{align}
    W_\psi M_\psi=W_1W_2M_1M_2=(-1)M_1W_1W_2M_2=(-1)^2M_1M_2W_1W_2=M_\psi W_\psi.
\end{align}
At last, because $(23)(13)=(123)\neq(132)=(13)(23)$, we should have $\mathcal{D}_{(\mathbb{Z}_\times\mathbb{Z}_2)^{(23)}}\times\mathcal{D}_{(\mathbb{Z}_\times\mathbb{Z}_2)^{(13)}}=\mathcal{D}_{(\mathbb{Z}_\times\mathbb{Z}_2)^{(123)}}\neq \mathcal{D}_{(\mathbb{Z}_\times\mathbb{Z}_2)^{(132)}}=\mathcal{D}_{(\mathbb{Z}_\times\mathbb{Z}_2)^{(13)}}\times\mathcal{D}_{(\mathbb{Z}_\times\mathbb{Z}_2)^{(23)}}$, which is also the case for their higher gauging expression. This non-commutativity of the condensation defects comes from the fact that Wilson lines and magnetic defects associated to the same cyclic subgroup braids non-trivially.

\section{Gauging 0-form symmetry: \texorpdfstring{$\mathbb{D}_4$}{} gauge theory from gauging swap symmetry}
\label{sec:gaugingD8}

Given a homomorphism $\rho:H\rightarrow \text{Aut}(N)$, where $H$ is a group and $\text{Aut}(N)$ is the automorphism group of $N$, we can construct a group $G$ as a semidirect product of $N$ and $H$, denoted $G = N \rtimes_\rho H$. This semidirect product uses the homomorphism $\rho$ to describe how elements of $H$ act on elements of $N$. In the context of finite-group gauge theories, one can obtain $G$ gauge theory by gauging the $H$ 0-form symmetry of $N$ gauge theory, realized through automorphism domain walls $\mathcal{D}_{N^{(\phi)}}$ with $\phi\in \text{Aut}(H)$.

As a concrete example, the dihedral group of order $8$,
\begin{align}
    \mathbb{D}_4=\langle a,b,c | a^2 = b^2 = c^2 =(ac)^4 =1, ab=ba=cac,bc=cb=aca\rangle,
    \label{D4presentation}
\end{align}
is isomorphic to $(\mathbb{Z}_2\times\mathbb{Z}_2)\rtimes \mathbb{Z}_2$ with the $\mathbb{Z}_2=\{1,c\}$ acting under conjugation by swapping two generators, i.e., $cac=ab$ and $cbc=b$. Therefore, $\mathbb{D}_4$ gauge theory can be obtained by gauging the $\mathbb{Z}_2$ 0-form swap symmetry of $\mathbb{Z}_2\times\mathbb{Z}_2$ associated with the automorphism $\varphi\in\operatorname{Aut}(\mathbb{Z}_2\times\mathbb{Z}_2)\cong S_3$ that maps $\varphi:(0,1)\mapsto(1,1)\in\mathbb{Z}_2\times\mathbb{Z}_2$. This example has been discussed in many places, e.g. \cite{Barkeshli:2014cna,Tantivasadakarn:2022ceu}, and is also related with fracton topological orders \cite{Prem:2019etl}. The group $H^2_\varphi(\mathbb{Z}_2\times\mathbb{Z}_2,\mathbb{Z}_2)$ is trivial, which means that $\mathbb{D}_4$ is the only extension in this setting. 

The presentation \eqref{D4presentation} can be related to the more standard one $\mathbb{D}_4=\langle r,s|r^4=s^2=1, srs=r^{-1}\rangle$ with the identification $a=s$, $b=r^2$ and $c=sr$. This more standard presentation make it explicit that $\mathbb{D}_4$ is also isomorphic to $\mathbb{Z}_4\rtimes \mathbb{Z}_2$ with the $\mathbb{Z}_2$ acting under conjugation by the inversion map, i.e., $srs=r^{-1}$. As a consequence, $\mathbb{D}_4$ gauge theory can also be obtained by gauging the $\mathbb{Z}_2$ 0-form symmetry of $\mathbb{Z}_4$ gauge theory associated with the automorphism $\phi\in\operatorname{Aut}(\mathbb{Z}_4)\cong\mathbb{Z}_2$ that maps $\phi:1\mapsto 3\in\mathbb{Z}_4$. The group $H^2_\phi(\mathbb{Z}_4,\mathbb{Z}_2)$ is isomorphic to $\mathbb{Z}_2$, which means that in addition to $\mathbb{D}_4$, there is a non-split extension in this setting which is $\mathbb{Q}_8$, the quaternion group. Yet another presentation of $\mathbb{D}_4$ is the non-split extension of $\mathbb{Z}_2\cong Z(\mathbb{D}_4)$ by $\mathbb{Z}_2\times\mathbb{Z}_2$ acting trivially with a non-trivial extension class $H^2(\mathbb{Z}_2,\mathbb{Z}_2\times\mathbb{Z}_2)=\mathbb{Z}_2\times\mathbb{Z}_2\times\mathbb{Z}_2$. The other non-trivial extensions classes produce $\mathbb{Z}_2\times\mathbb{Z}_4$ and $\mathbb{Q}_8$. These other presentations are associated with other dual descriptions of $\mathbb{D}_4$ gauge theories. We leave the study of these dual settings to future works.

\subsection{The symmetry defect for \texorpdfstring{$\mathbb{Z}_2\times\mathbb{Z}_2$}{} swap symmetry}

From Table \ref{tab:Z2Z2automorphism}, the domain wall associated with the automorphism that maps $(0,1)\rightarrow(1,1)$ can be written as
\begin{align}
    \begin{split}
        \mathcal{D}_{(\mathbb{Z}_2\times \mathbb{Z}_2)^{(23)}} & =\mathcal{S}_{\mathbb{Z}_{2,1}\times\mathbb{Z}_{2,2}}(\Sigma),\\
        & =\sum_{\substack{\gamma\in H_1(\Sigma,\mathbb{Z}_2)\\ \Gamma\in H_{D-2}(\Sigma,\mathbb{Z}_2)}}(-1)^{\langle\gamma,\Gamma\rangle}W_{1}(\gamma)M_2(\Gamma),\\
        & = \sum_{\substack{A\in H^{(D-2)}(\Sigma,\mathbb{Z}_2),\\ \tilde{A}\in H^{1}(\Sigma,\mathbb{Z}_2)}}e^{\pi i\int_\Sigma a^{(1)}\cup A^{(D-2)}+\tilde{b}^{(D-2)}\cup \tilde{A}^{(1)}+A^{(D-2)}\cup \tilde{A}^{(1)}},\\
        & =e^{\pi i\int_\Sigma a^{(1)}\cup \tilde{b}^{(D-2)}},
        \label{eq:swapautomorphism}
    \end{split}
\end{align}
where we integrated out $\tilde{A}$ forcing $A=\tilde{b}$. In the second line, we are treating $W(\gamma)$ ad $M(\Gamma)$ as commuting field insertions in a path integral and used the Poincaré duals $A$ and $\tilde{A}$ of $\gamma$ and $\Gamma$ respectively.

\subsection{The action from gauging swap symmetry}

Gauging the $\mathbb{Z}_2$ 0-form symmetry associated with the swap automorphism \eqref{eq:swapautomorphism} corresponds to the following operator insertion:
\begin{align}
    \sum_{\Sigma\in H_{D-1}(\mathcal{M},\mathbb{Z}_2)}\mathcal{D}_{(\mathbb{Z}_2\times \mathbb{Z}_2)^{(12)}}(\Sigma)=\sum_{c^{(1)}\in H^1(\mathcal{M},\mathbb{Z}_2)}e^{\pi i\int_\mathcal{M} a^{(1)}\cup\tilde{b}^{(D-2)}\cup c^{(1)}}.
\end{align}
where $c$ is the Poincaré dual of $\Sigma$. Introducing an auxiliary gauge field $\tilde{c}\in C^{D-2}(\mathcal{M},\mathbb{Z}_2)$ to make $c$ flat dynamically leads to the action:
\begin{align}
        S_{\mathbb{D}_4}=i\pi\int_\mathcal{M}( \tilde{a}^{(D-2)}\cup da^{(1)}+\tilde{b}^{(D-2)}\cup db^{(1)}+\tilde{c}^{(D-2)}\cup dc^{(1)}+a^{(1)}\cup \tilde{b}^{(D-2)}\cup c^{(1)}).
        \label{eq:dihedralaction}
\end{align}
For $D=2+1$, this is the action for $\mathbb{D}_4$ gauge theory we studied in \cite{Cordova:2024jlk}. In the literature it was realized that the modular data of twisted $\mathbb{Z}_2\times\mathbb{Z}_2\times\mathbb{Z}_2$ gauge theory is the same as that of $\mathbb{D}_4$ \cite{deWildPropitius:1997wu,Coste_2000}. Here we derived a more general relation that holds in all dimensions by explicitly gauging the swap symmetry of $\mathbb{Z}_2\times\mathbb{Z}_2$ gauge theory. In the next section, we will explain the mapping between the operators of the two theories.

One can explicitly check that the action is invariant under the following gauge transformations:
\begin{align}
    & a\rightarrow a+d\alpha, && \tilde{a}\rightarrow \tilde{a}+d\tilde{\alpha}+\epsilon \tilde{b}+\tilde{\beta}c+\epsilon d\tilde{\beta}, \\
    & b\rightarrow b+d\beta+\epsilon a+\alpha c+\alpha d\epsilon, && \tilde{b}\rightarrow \tilde{b}+d\tilde{\beta},\\
    & c\rightarrow c+d\epsilon, && \tilde{c}\rightarrow \tilde{c}+d\tilde{\epsilon}+\alpha\tilde{b}+\tilde{\beta}a+\alpha d\tilde{\beta},
\end{align}
with $\alpha,\beta,\epsilon\in C^0(\mathcal{M},\mathbb{Z}_2)$ and $\tilde{\alpha},\tilde{\beta},\tilde{\epsilon}\in C^{D-3}(\mathcal{M},\mathbb{Z}_2)$ (and $\tilde{\alpha}=\tilde{\beta}=\tilde{\epsilon}=0$ if $D=2$). The fact that the gauge transformations of $c$ shift $b$ by $a$ descends from the $\mathbb{Z}_2$ action that maps $(0,1)\rightarrow(1,1)$ and leaves $(1,0)$ unchanged.

\subsection{Operators and fusion rules}

The set of Wilson lines and pure magnetic defects of $\mathbb{D}_4$ gauge theory, as well as their linking action, are summarized in Table \ref{tab:D4charactertable}. Here we are going to show how to construct these operators in terms of gauge invariant combinations of the Wilson lines and magnetic defects of the $\mathbb{Z}_2$ factors that enter in the group extensions $(\mathbb{Z}_2\times\mathbb{Z}_2)\rtimes\mathbb{Z}_2$ summarized in the presentation \eqref{D4presentation} (the identification $b=r^2$, $a=s$ and $c=sr$, relates this presentation with the more standard one). Provided these expressions, one can deduce the entries of the character table from the linking action of the $\mathbb{Z}_2$ constituents.
\begin{table}[ht]
    \centering
    \begin{tabular}{|c|ccccc|}
  \hline Class & $\{e\}$ & $\{b\}=\{r^2\}$ & $\{a,ab\}=\{s,sr^2\}$ & $\{c,cb\}=\{sr,sr^3\}$ & $\{ac,abc\}=\{r,r^3\}$ \\ \hline
1 & 1 & 1 & 1 & 1 & 1 \\
$\rho_a=\rho_{s}$ & 1 & 1 & -1 & 1 & -1 \\
$\rho_c=\rho_{sr}$ & 1 & 1 & 1 & -1 & -1 \\
$\rho_{a+c}=\rho_{r}$ & 1 & 1 & -1 & -1 & 1 \\
$\rho_b=\rho_{r^2}$ & 2 & -2 & 0 & 0 & 0 \\ \hline 
\end{tabular}
    \caption{Character table of $\mathbb{D}_4$.}
    \label{tab:D4charactertable}
\end{table}

Let us start with the Wilson lines. The gauge transformations yield the following gauge invariant, one-dimensional Wilson lines:
\begin{align}
    \mathbb{W}_a(\gamma)=W_a(\gamma)=e^{\pi i\oint_\gamma a},\quad \mathbb{W}_c(\gamma)=W_c(\gamma)=e^{\pi i\oint_\gamma c},\quad \mathbb{W}_{a+c}(\gamma)=\mathbb{W}_a(\gamma)\mathbb{W}_c(\gamma),
    \label{eq:1dWilson}
\end{align}
we use boldface letters to denote the operators of $\mathbb{D}_4$ and no-boldface for their $\mathbb{Z}_2$ constituents. These Wilson lines are associated with the three one-dimensional irreducible representations of $\mathbb{D}_4$ gauge theory, as shown in Table \ref{tab:D4charactertable}, with the obvious identification. The integral of $b$ on closed loops is not invariant under gauge transformations because $b$ shifts by $a$ and $c$. To make it gauge invariant, we impose Dirichlet boundary conditions for $a$ and $c$ on $\gamma$, which can be achieved by dressing the operator with the condensation of $a$ and $c$. Specifically, we have:
\begin{align}
    \mathbb{W}_b=\frac{1}{2}W_b(1+W_a)(1+W_c)=\frac{1}{2}(W_b+W_{a+b})(1+W_c),
    \label{eq:twodimensionalW}
\end{align}
where $W_b$ and $W_{a+b}$ are defined in the same way as \eqref{eq:1dWilson}. In Section \ref{sec:review}, we defined the diagonal domain walls and their higher codimensional generalizations through a normal subgroup $K\lhd G$. The condensation of Wilson lines projects onto this subgroup as described in Section \ref{sec:diagonalcondensation}. Here, the subgroup that is preserved by the condensation of $\mathbb{W}_a$ and $\mathbb{W}_c$ is the center $Z(\mathbb{D}_4)=\mathbb{Z}_2=\{1,b\}=\{1,r^2\}\lhd \mathbb{D}_4$, i.e.,
\begin{align}
    \mathcal{D}_{\mathbb{Z}_2}(\gamma)=1+W_a(\gamma)+W_c(\gamma)+W_{a+c}(\gamma).
    \label{eq:diagZ2}
\end{align}
We remark that the condensation of $W_a$ and $W_c$ in a loop equals the insertion of $1+W_a$ and $1+W_c$ on that loop \eqref{eq:oprdiagwall}. Note that \eqref{eq:twodimensionalW} can also be interpreted as the orbit of $W_b$ under the swap symmetry, dressed with the condensation of $W_c$. This two-dimensional Wilson line is associated with the two-dimensional irreducible representation of $\mathbb{D}_4$ denoted by $\rho_b$ in Table \ref{tab:D4charactertable}.

Provided the above expressions, one can directly compute the fusion rules of the Wilson lines of $\mathbb{D}_4$ gauge theory. The one-dimensional irreducible representations have a $\mathbb{Z}_2\times\mathbb{Z}_2$ fusion structure and the two-dimensional representation obeys:
\begin{align}
    & \mathbb{W}_b\times \mathbb{W}_a=\mathbb{W}_a\times \mathbb{W}_b=\mathbb{W}_b\times \mathbb{W}_c=\mathbb{W}_c\times \mathbb{W}_b= \mathbb{W}_b,\\
    & \mathbb{W}_b\times \mathbb{W}_b=(1+W_c)(1+W_a)=1+\mathbb{W}_a+\mathbb{W}_c+\mathbb{W}_{a+c},
\end{align}
which forms a $\mathbb{Z}_2\times\mathbb{Z}_2$ Tambara-Yamagami category \cite{Tambara:1998vmj}.

Now, let us consider the magnetic defects. In this case, the gauge transformations yield the following gauge invariant one-dimensional magnetic defect associated with the conjugacy class $\{b\}=\{r^2\}$:
\begin{align}
    \mathbb{M}_b(\Gamma)=M_b(\Gamma)=e^{i\pi\oint_{\Gamma}\tilde{b}}.
    \label{eq:1dMagnetic}
\end{align}
Similarly to the two-dimensional Wilson line, the magnetic defects associated with $a$, $c$, and $a+c$ need to be dressed with the condensation of $M_b$ and suitable condensations of Wilson lines to be made gauge invariant. These magnetic defects correspond to the three two-dimensional conjugacy classes of $\mathbb{D}_4$ displayed in Table \ref{tab:D4charactertable}. Specifically, we have\footnote{In more detail: $\tilde{a}$ shifts by $c$ so it is dressed with $\mathcal{S}_c$, $\tilde{c}$ shifts by $a$ so it is dressed with $\mathcal{S}_a$, and $\tilde{a}+\tilde{c}$ shifts by $a+c$ so it is dressed with $\mathcal{S}_{a+c}$.}
\begin{align}
    \mathbb{M}_a=\frac{1}{2}M_a(1+M_b)\mathcal{S}_c,\quad \mathbb{M}_c=\frac{1}{2}M_c(1+M_b)\mathcal{S}_a,\quad \mathbb{M}_{a+c}=\frac{1}{2}M_{a+c}(1+M_b)\mathcal{S}_{a+c},
    \label{eq:D4magnetic}
\end{align}
where $M_a$, $M_{c}$ and $M_{a+c}$ are defined in the same way as \eqref{eq:1dMagnetic} and
\begin{align}
    \mathcal{S}_{a+c}(\Gamma)=\frac{2}{|H_1(\Gamma,\mathbb{Z}_2)|}\sum_{\gamma\in H_1(\Gamma,\mathbb{Z}_2)}\mathbb{W}_{a+c}(\gamma)=\mathcal{D}_{\mathbb{Z}_4}(\Gamma),
    \label{eq:Sa}
\end{align}
is the condensation of $\mathbb{W}_{a+c}$ on the codimension-two submanifold $\Gamma$ and similarly for $\mathcal{S}_a$ and $\mathcal{S}_{c}$. In terms of the higher codimensional generalization of the diagonal domain walls of $\mathbb{D}_4$ gauge theory reviewed in section \ref{sec:review}, the condensation $\mathcal{S}_{a+c}$ corresponds to the normal subgroup $\mathbb{Z}_4=\{1,ac,(ac)^2,(ac)^3\}=\{1,r,r^2,r^3\}$ of $\mathbb{D}_4$. Likewise, the $\mathcal{S}_a$ and $\mathcal{S}_{c}$ condensations correspond to the two normal Klein four subgroups, i.e., $V_4=\{1,a,b,ba\}=\{1,s,r^2,r^2s\}$ and $V_4^\prime=\{1,c,b,cb\}=\{1,sr,r^2,sr^3\}$ respectively. Note that $\mathcal{S}_{a+c}\neq\mathcal{S}_a\mathcal{S}_c$, from \eqref{eq:highercodimension} we have
\begin{align}
    \mathcal{D}_{\mathbb{Z}_2}(\Gamma)=\mathcal{D}_{V_4\cap V_4^\prime}=\mathcal{D}_{\mathbb{Z}_4}\times\mathcal{D}_{V_4}=\mathcal{S}_a\times S_c=\frac{4}{|H_1(\Gamma,\mathbb{Z}_2)|^2}\sum_{\gamma,\gamma^\prime\in H_1(\Gamma,\mathbb{Z}_2)}W_a(\gamma)W_c(\gamma^\prime),
\end{align}
with $\mathbb{Z}_2=\{1,b\}=\{1,r^2\}$, the center of $\mathbb{D}_4$, which is equal to \eqref{eq:diagZ2} when we consider the condensation on a one-dimensional submanifold. Notably, the expressions in \eqref{eq:D4magnetic} are consistent even in $D=2$. In this degenerate case, the magnetic defects are local operators and $\mathcal{S}_a=\mathcal{S}_a=\mathcal{S}_{a+c}=2$. As a result, $\mathbb{M}_a$, $\mathbb{M}_c$ and $\mathbb{M}_{a+c}$ are dressed just with the condensation of $M_b$. Consistently, $\tilde{a}$ and $\tilde{c}$ are not shifted by $c$ and $a$ because $\tilde{\beta}=0$. 

Provided the above expressions, one can directly compute the fusion rules of the magnetic defects of $\mathbb{D}_4$ gauge theory. We have, for example:
\begin{align}
    \mathbb{M}_{a}(\Gamma)\times \mathbb{M}_{a}(\Gamma)=\mathcal{S}_c(\Gamma)+\mathbb{M}_{b}(\Gamma) \times\mathcal{S}_c(\Gamma),
    \label{eq:correctfusion}
\end{align}
i.e., we obtain the condensation of $\mathbb{W}_c$ in the codimension-two surface $\Gamma$, plus the magnetic defect $\mathbb{M}_{b}$ fused with this condensation. These description helps to clarify some of the results in \cite{Heidenreich:2021xpr,Arias-Tamargo:2022nlf}. We see that the appearance of the condensation of Wilson lines in the fusion channel of pure fluxes follows from gauge invariance. In summary, the ``would-be" pure fluxes are made gauge invariant by dressing them with suitable condensations. Whenever the ``would-be" magnetic fluxes fuse to the identity, what remains is the dressing by the electric condensation. Note that in $D=2+1$ spacetime dimensions we have $\mathcal{S}_a=1+W_a$, so \eqref{eq:correctfusion} is equal to:
\begin{align}
    \mathbb{M}_a\times \mathbb{M}_a=1+\mathbb{W}_c+\mathbb{M}_b+\mathbb{M}_b \times\mathbb{W}_c.
    \label{eq:D3}
\end{align}
However, for $D\neq2+1$ the fusion rule \eqref{eq:D3} is not correct because, in this case, Wilson lines and magnetic defects are operators of different dimension. On the other hand, the fusion rule \eqref{eq:correctfusion} is correct in arbitrary spacetime dimensions because all $\mathcal{S}_c$, $\mathbb{M}_a$ and $\mathbb{M}_b$ have codimension-two.

By expressing the Wilson lines and magnetic defects of $\mathbb{D}_4$ in terms of gauge-invariant combinations of Wilson lines and magnetic defects from the $\mathbb{Z}_2$ factors, one can directly deduce the elements of the $\mathbb{D}_4$ character table (see Table \ref{tab:D4charactertable}) through the linking actions of these components. For instance, $\mathbb{W}_b$ has a linking of zero with $\mathbb{M}_a$, $\mathbb{M}_c$ and $\mathbb{M}_{a+c}$ because $M_a$, $M_c$ and $M_{a+c}$ have linking zero with $(1+W_a)(1+W_b)$. However, $\mathbb{W}_b$ has a linking of $-2$ with $\mathbb{M}_b$ because $M_b$ has a linking number of $-1$ with $W_b$ and $4$ with $(1+W_a)(1+W_c)$. The other entries can be determined similarly.

In $D=2+1$, by relabeling $b\leftrightarrow \tilde{b}$, one can view the action \eqref{eq:dihedralaction} as describing $\mathbb{Z}_2\times\mathbb{Z}_2\times\mathbb{Z}_2$ gauge theory twisted by the cocycle associated with the diagonal $\mathbb{Z}_2$. Then, from the gauge transformations, one can check that the theory has 8 one-dimensional line operators:
\begin{align}
    \mathbb{W}_a(\gamma)=W_a(\gamma)=e^{\pi i\oint_\gamma a},\quad \mathbb{W}_b(\gamma)=W_b(\gamma)=e^{\pi i\oint_\gamma b},\quad \mathbb{W}_c(\gamma)=W_c(\gamma)=e^{\pi i\oint_\gamma c},  
\end{align}
and their products, forming a $\mathbb{Z}_2\times\mathbb{Z}_2\times\mathbb{Z}_2$ fusion ring. Additionally, there are 14 two-dimensional line operators such as:
\begin{align}
    \mathbb{M}_a(\gamma)=\frac{1}{2}M_a(\gamma)\mathcal{S}_b(\gamma)\mathcal{S}_c(\gamma),\qquad \mathbb{M}_a(\gamma)\mathbb{W}_a(\gamma).
\end{align}
which obey, e.g.:
\begin{align}
    & \mathbb{M}_a\times \mathbb{W}_b= \mathbb{M}_a\times \mathbb{W}_c= \mathbb{M}_a,\\
    & \mathbb{M}_a\times \mathbb{M}_a=\mathcal{S}_b\mathcal{S}_c=1+\mathbb{W}_b+\mathbb{W}_c+\mathbb{W}_{b+c},\\
    & \mathbb{M}_a\times \mathbb{M}_b=\frac{1}{2}M_aM_b\mathcal{S}_a \mathcal{S}_b\mathcal{S}_c=\mathbb{M}_{a+b}+\mathbb{M}_{a+b}\times \mathbb{W}_a.
\end{align}
The relation to $\mathbb{D}_4$ arises by switching back $b\leftrightarrow \tilde{b}$, which amounts to identifying $\mathbb{M}_b$ with the two-dimensional Wilson line and $\mathbb{W}_b$ with the magnetic defect associated with the one-dimensional conjugacy class of $\mathbb{D}_4$.

\section*{Acknowledgement}
We thank Maissam Barkeshli, Xie Chen, Arpit Dua, Ryohei Kobayashi, Michael Levin,  Zhu-Xi Luo, Wilbur Shirley, and Carolyn Zhang for discussions. 
D.B.C. and C.C. are supported by the US Department
of Energy Early Career program 5-29073, the Sloan Foundation, and the Simons Collaboration on
Global Categorical Symmetries.
P.-S.H. is supported by Department of Mathematics King's College London.

\bibliographystyle{utphys}
\bibliography{biblio}

\providecommand{\href}[2]{#2}\begingroup\raggedright\begin{thebibliography}{100}

\bibitem{Dijkgraaf:1989pz}
R.~Dijkgraaf and E.~Witten, ``{Topological Gauge Theories and Group
  Cohomology},'' \href{http://dx.doi.org/10.1007/BF02096988}{{\em Commun. Math.
  Phys.} {\bfseries 129} (1990) 393}.

\bibitem{Cordova:2024jlk}
C.~C\'{o}rdova, D.~B. Costa, and P.-S. Hsin, ``{Non-invertible symmetries in
  finite group gauge theory},''
  \href{http://arxiv.org/abs/2407.07964}{{\ttfamily arXiv:2407.07964
  [cond-mat.str-el]}}.

\bibitem{Gaiotto_2015}
D.~Gaiotto, A.~Kapustin, N.~Seiberg, and B.~Willett, ``Generalized global
  symmetries,'' \href{http://dx.doi.org/10.1007/jhep02(2015)172}{{\em Journal
  of High Energy Physics} {\bfseries 2015} no.~2, (Feb., 2015) }.
  \url{http://dx.doi.org/10.1007/JHEP02(2015)172}.

\bibitem{Cordova:2022ruw}
C.~C\'{o}rdova, T.~T. Dumitrescu, K.~Intriligator, and S.-H. Shao, ``{Snowmass
  White Paper: Generalized Symmetries in Quantum Field Theory and Beyond},'' in
  {\em {Snowmass 2021}}.
\newblock 5, 2022.
\newblock \href{http://arxiv.org/abs/2205.09545}{{\ttfamily arXiv:2205.09545
  [hep-th]}}.

\bibitem{Shao:2023gho}
S.-H. Shao, ``{What's Done Cannot Be Undone: TASI Lectures on Non-Invertible
  Symmetries},'' \href{http://arxiv.org/abs/2308.00747}{{\ttfamily
  arXiv:2308.00747 [hep-th]}}.

\bibitem{Luo:2023ive}
R.~Luo, Q.-R. Wang, and Y.-N. Wang, ``{Lecture notes on generalized symmetries
  and applications},''
  \href{http://dx.doi.org/10.1016/j.physrep.2024.02.002}{{\em Phys. Rept.}
  {\bfseries 1065} (2024) 1--43},
  \href{http://arxiv.org/abs/2307.09215}{{\ttfamily arXiv:2307.09215
  [hep-th]}}.

\bibitem{Bhardwaj:2023kri}
L.~Bhardwaj, L.~E. Bottini, L.~Fraser-Taliente, L.~Gladden, D.~S.~W. Gould,
  A.~Platschorre, and H.~Tillim, ``{Lectures on generalized symmetries},''
  \href{http://dx.doi.org/10.1016/j.physrep.2023.11.002}{{\em Phys. Rept.}
  {\bfseries 1051} (2024) 1--87},
  \href{http://arxiv.org/abs/2307.07547}{{\ttfamily arXiv:2307.07547
  [hep-th]}}.

\bibitem{Schafer-Nameki:2023jdn}
S.~Schafer-Nameki, ``{ICTP lectures on (non-)invertible generalized
  symmetries},'' \href{http://dx.doi.org/10.1016/j.physrep.2024.01.007}{{\em
  Phys. Rept.} {\bfseries 1063} (2024) 1--55},
  \href{http://arxiv.org/abs/2305.18296}{{\ttfamily arXiv:2305.18296
  [hep-th]}}.

\bibitem{Iqbal:2024pee}
N.~Iqbal, ``{Jena lectures on generalized global symmetries: principles and
  applications},''
\newblock 7, 2024.
\newblock \href{http://arxiv.org/abs/2407.20815}{{\ttfamily arXiv:2407.20815
  [hep-th]}}.

\bibitem{Brennan:2023mmt}
T.~D. Brennan and S.~Hong, ``{Introduction to Generalized Global Symmetries in
  QFT and Particle Physics},''
  \href{http://arxiv.org/abs/2306.00912}{{\ttfamily arXiv:2306.00912
  [hep-ph]}}.

\bibitem{McGreevy:2022oyu}
J.~McGreevy, ``{Generalized Symmetries in Condensed Matter},''
  \href{http://dx.doi.org/10.1146/annurev-conmatphys-040721-021029}{{\em Ann.
  Rev. Condensed Matter Phys.} {\bfseries 14} (2023) 57--82},
  \href{http://arxiv.org/abs/2204.03045}{{\ttfamily arXiv:2204.03045
  [cond-mat.str-el]}}.

\bibitem{Gomes:2023ahz}
P.~R.~S. Gomes, ``{An introduction to higher-form symmetries},''
  \href{http://dx.doi.org/10.21468/SciPostPhysLectNotes.74}{{\em SciPost Phys.
  Lect. Notes} {\bfseries 74} (2023) 1},
  \href{http://arxiv.org/abs/2303.01817}{{\ttfamily arXiv:2303.01817
  [hep-th]}}.

\bibitem{Costa:2024wks}
D.~Costa {\em et~al.}, ``{Simons Lectures on Categorical Symmetries},''
\newblock 11, 2024.
\newblock \href{http://arxiv.org/abs/2411.09082}{{\ttfamily arXiv:2411.09082
  [math-ph]}}.

\bibitem{Witten:1998wy}
E.~Witten, ``{AdS / CFT correspondence and topological field theory},''
  \href{http://dx.doi.org/10.1088/1126-6708/1998/12/012}{{\em JHEP} {\bfseries
  12} (1998) 012}, \href{http://arxiv.org/abs/hep-th/9812012}{{\ttfamily
  arXiv:hep-th/9812012}}.

\bibitem{kong2015boundarybulkrelationtopologicalorders}
L.~Kong, X.-G. Wen, and H.~Zheng, ``Boundary-bulk relation for topological
  orders as the functor mapping higher categories to their centers,'' 2015.
\newblock \url{https://arxiv.org/abs/1502.01690}.

\bibitem{Gaiotto:2020iye}
D.~Gaiotto and J.~Kulp, ``{Orbifold groupoids},''
  \href{http://dx.doi.org/10.1007/JHEP02(2021)132}{{\em JHEP} {\bfseries 02}
  (2021) 132}, \href{http://arxiv.org/abs/2008.05960}{{\ttfamily
  arXiv:2008.05960 [hep-th]}}.

\bibitem{Freed:2022qnc}
D.~S. Freed, G.~W. Moore, and C.~Teleman, ``{Topological symmetry in quantum
  field theory},'' \href{http://arxiv.org/abs/2209.07471}{{\ttfamily
  arXiv:2209.07471 [hep-th]}}.

\bibitem{Kaidi:2022cpf}
J.~Kaidi, K.~Ohmori, and Y.~Zheng, ``{Symmetry TFTs for Non-invertible
  Defects},'' \href{http://dx.doi.org/10.1007/s00220-023-04859-7}{{\em Commun.
  Math. Phys.} {\bfseries 404} no.~2, (2023) 1021--1124},
  \href{http://arxiv.org/abs/2209.11062}{{\ttfamily arXiv:2209.11062
  [hep-th]}}.

\bibitem{PhysRevResearch.2.043086}
L.~Kong, T.~Lan, X.-G. Wen, Z.-H. Zhang, and H.~Zheng, ``Algebraic higher
  symmetry and categorical symmetry: A holographic and entanglement view of
  symmetry,'' \href{http://dx.doi.org/10.1103/PhysRevResearch.2.043086}{{\em
  Phys. Rev. Res.} {\bfseries 2} (Oct, 2020) 043086}.
  \url{https://link.aps.org/doi/10.1103/PhysRevResearch.2.043086}.

\bibitem{PhysRevX.8.021074}
T.~Lan, L.~Kong, and X.-G. Wen, ``Classification of
  $\mathbf{(}3+1\mathbf{)}\mathrm{D}$ bosonic topological orders: The case when
  pointlike excitations are all bosons,''
  \href{http://dx.doi.org/10.1103/PhysRevX.8.021074}{{\em Phys. Rev. X}
  {\bfseries 8} (Jun, 2018) 021074}.
  \url{https://link.aps.org/doi/10.1103/PhysRevX.8.021074}.

\bibitem{PhysRevX.9.021005}
T.~Lan and X.-G. Wen, ``Classification of $3+1\mathrm{D}$ bosonic topological
  orders (ii): The case when some pointlike excitations are fermions,''
  \href{http://dx.doi.org/10.1103/PhysRevX.9.021005}{{\em Phys. Rev. X}
  {\bfseries 9} (Apr, 2019) 021005}.
  \url{https://link.aps.org/doi/10.1103/PhysRevX.9.021005}.

\bibitem{Johnson-Freyd:2020usu}
T.~Johnson-Freyd, ``{On the Classification of Topological Orders},''
  \href{http://dx.doi.org/10.1007/s00220-022-04380-3}{{\em Commun. Math. Phys.}
  {\bfseries 393} no.~2, (2022) 989--1033},
  \href{http://arxiv.org/abs/2003.06663}{{\ttfamily arXiv:2003.06663
  [math.CT]}}.

\bibitem{Johnson-Freyd:2020ivj}
T.~Johnson-Freyd and M.~Yu, ``{Fusion 2-categories With no Line Operators are
  Grouplike},'' \href{http://dx.doi.org/10.1017/S0004972721000095}{{\em Bull.
  Austral. Math. Soc.} {\bfseries 104} no.~3, (2021) 434--442},
  \href{http://arxiv.org/abs/2010.07950}{{\ttfamily arXiv:2010.07950
  [math.QA]}}.

\bibitem{Johnson-Freyd:2021tbq}
T.~Johnson-Freyd and M.~Yu, ``{Topological Orders in (4+1)-Dimensions},''
  \href{http://dx.doi.org/10.21468/SciPostPhys.13.3.068}{{\em SciPost Phys.}
  {\bfseries 13} no.~3, (2022) 068},
  \href{http://arxiv.org/abs/2104.04534}{{\ttfamily arXiv:2104.04534
  [hep-th]}}.

\bibitem{Cordova:2023bja}
C.~C\'{o}rdova, P.-S. Hsin, and C.~Zhang, ``{Anomalies of Non-Invertible
  Symmetries in (3+1)d},'' \href{http://arxiv.org/abs/2308.11706}{{\ttfamily
  arXiv:2308.11706 [hep-th]}}.

\bibitem{Kong:2013aya}
L.~Kong, ``{Anyon condensation and tensor categories},''
  \href{http://dx.doi.org/10.1016/j.nuclphysb.2014.07.003}{{\em Nucl. Phys. B}
  {\bfseries 886} (2014) 436--482},
  \href{http://arxiv.org/abs/1307.8244}{{\ttfamily arXiv:1307.8244
  [cond-mat.str-el]}}.

\bibitem{Eliens:2013epa}
I.~S. Eli\"ens, J.~C. Romers, and F.~A. Bais, ``{Diagrammatics for Bose
  condensation in anyon theories},''
  \href{http://dx.doi.org/10.1103/PhysRevB.90.195130}{{\em Phys. Rev. B}
  {\bfseries 90} no.~19, (2014) 195130},
  \href{http://arxiv.org/abs/1310.6001}{{\ttfamily arXiv:1310.6001
  [cond-mat.str-el]}}.

\bibitem{Lan:2014uaa}
T.~Lan, J.~C. Wang, and X.-G. Wen, ``{Gapped Domain Walls, Gapped Boundaries
  and Topological Degeneracy},''
  \href{http://dx.doi.org/10.1103/PhysRevLett.114.076402}{{\em Phys. Rev.
  Lett.} {\bfseries 114} no.~7, (2015) 076402},
  \href{http://arxiv.org/abs/1408.6514}{{\ttfamily arXiv:1408.6514
  [cond-mat.str-el]}}.

\bibitem{Hung:2015hfa}
L.-Y. Hung and Y.~Wan, ``{Generalized ADE classification of topological
  boundaries and anyon condensation},''
  \href{http://dx.doi.org/10.1007/JHEP07(2015)120}{{\em JHEP} {\bfseries 07}
  (2015) 120}, \href{http://arxiv.org/abs/1502.02026}{{\ttfamily
  arXiv:1502.02026 [cond-mat.str-el]}}.

\bibitem{Neupert:2016pjk}
T.~Neupert, H.~He, C.~von Keyserlingk, G.~Sierra, and B.~A. Bernevig, ``{Boson
  Condensation in Topologically Ordered Quantum Liquids},''
  \href{http://dx.doi.org/10.1103/PhysRevB.93.115103}{{\em Phys. Rev. B}
  {\bfseries 93} no.~11, (2016) 115103},
  \href{http://arxiv.org/abs/1601.01320}{{\ttfamily arXiv:1601.01320
  [cond-mat.str-el]}}.

\bibitem{Burnell:2017otf}
F.~J. Burnell, ``{Anyon condensation and its applications},''
  \href{http://dx.doi.org/10.1146/annurev-conmatphys-033117-054154}{{\em Ann.
  Rev. Condensed Matter Phys.} {\bfseries 9} (2018) 307--327},
  \href{http://arxiv.org/abs/1706.04940}{{\ttfamily arXiv:1706.04940
  [cond-mat.str-el]}}.

\bibitem{Tachikawa:2017gyf}
Y.~Tachikawa, ``{On gauging finite subgroups},''
  \href{http://dx.doi.org/10.21468/SciPostPhys.8.1.015}{{\em SciPost Phys.}
  {\bfseries 8} no.~1, (2020) 015},
  \href{http://arxiv.org/abs/1712.09542}{{\ttfamily arXiv:1712.09542
  [hep-th]}}.

\bibitem{Bhardwaj:2017xup}
L.~Bhardwaj and Y.~Tachikawa, ``{On finite symmetries and their gauging in two
  dimensions},'' \href{http://dx.doi.org/10.1007/JHEP03(2018)189}{{\em JHEP}
  {\bfseries 03} (2018) 189}, \href{http://arxiv.org/abs/1704.02330}{{\ttfamily
  arXiv:1704.02330 [hep-th]}}.

\bibitem{Cong:2017ffh}
I.~Cong, M.~Cheng, and Z.~Wang, ``{Hamiltonian and Algebraic Theories of Gapped
  Boundaries in Topological Phases of Matter},''
  \href{http://dx.doi.org/10.1007/s00220-017-2960-4}{{\em Commun. Math. Phys.}
  {\bfseries 355} (2017) 645--689},
  \href{http://arxiv.org/abs/1707.04564}{{\ttfamily arXiv:1707.04564
  [cond-mat.str-el]}}.

\bibitem{Hsin:2018vcg}
P.-S. Hsin, H.~T. Lam, and N.~Seiberg, ``{Comments on One-Form Global
  Symmetries and Their Gauging in 3d and 4d},''
  \href{http://dx.doi.org/10.21468/SciPostPhys.6.3.039}{{\em SciPost Phys.}
  {\bfseries 6} no.~3, (2019) 039},
  \href{http://arxiv.org/abs/1812.04716}{{\ttfamily arXiv:1812.04716
  [hep-th]}}.

\bibitem{Kaidi:2021gbs}
J.~Kaidi, Z.~Komargodski, K.~Ohmori, S.~Seifnashri, and S.-H. Shao, ``{Higher
  central charges and topological boundaries in 2+1-dimensional TQFTs},''
  \href{http://dx.doi.org/10.21468/SciPostPhys.13.3.067}{{\em SciPost Phys.}
  {\bfseries 13} no.~3, (2022) 067},
  \href{http://arxiv.org/abs/2107.13091}{{\ttfamily arXiv:2107.13091
  [hep-th]}}.

\bibitem{Yu:2021zmu}
M.~Yu, ``{Gauging Categorical Symmetries in 3d Topological Orders and Bulk
  Reconstruction},'' \href{http://arxiv.org/abs/2111.13697}{{\ttfamily
  arXiv:2111.13697 [hep-th]}}.

\bibitem{Decoppet:2022dnz}
T.~D. D\'ecoppet and M.~Yu, ``{Gauging noninvertible defects: a 2-categorical
  perspective},'' \href{http://dx.doi.org/10.1007/s11005-023-01655-1}{{\em
  Lett. Math. Phys.} {\bfseries 113} no.~2, (2023) 36},
  \href{http://arxiv.org/abs/2211.08436}{{\ttfamily arXiv:2211.08436
  [math.CT]}}.

\bibitem{Diatlyk:2023fwf}
O.~Diatlyk, C.~Luo, Y.~Wang, and Q.~Weller, ``{Gauging non-invertible
  symmetries: topological interfaces and generalized orbifold groupoid in 2d
  QFT},'' \href{http://dx.doi.org/10.1007/JHEP03(2024)127}{{\em JHEP}
  {\bfseries 03} (2024) 127}, \href{http://arxiv.org/abs/2311.17044}{{\ttfamily
  arXiv:2311.17044 [hep-th]}}.

\bibitem{Cordova:2023jip}
C.~C\'{o}rdova and D.~Garc\'\i{}a-Sep\'ulveda, ``{Non-Invertible Anyon
  Condensation and Level-Rank Dualities},''
  \href{http://arxiv.org/abs/2312.16317}{{\ttfamily arXiv:2312.16317
  [hep-th]}}.

\bibitem{Zhang:2024bye}
C.~Zhang, A.~Vishwanath, and X.-G. Wen, ``{Hierarchy construction for
  non-abelian fractional quantum Hall states via anyon condensation},''
  \href{http://arxiv.org/abs/2406.12068}{{\ttfamily arXiv:2406.12068
  [cond-mat.str-el]}}.

\bibitem{Cordova:2024goh}
C.~C\'{o}rdova and D.~Garc\'\i{}a-Sep\'ulveda, ``{Topological Cosets via Anyon
  Condensation and Applications to Gapped $\mathrm{\bf{QCD_{2}}}$},''
  \href{http://arxiv.org/abs/2412.01877}{{\ttfamily arXiv:2412.01877
  [hep-th]}}.

\bibitem{Kong:2024ykr}
L.~Kong, Z.-H. Zhang, J.~Zhao, and H.~Zheng, ``{Higher condensation theory},''
  \href{http://arxiv.org/abs/2403.07813}{{\ttfamily arXiv:2403.07813
  [cond-mat.str-el]}}.

\bibitem{Perez-Lona:2024sds}
A.~Perez-Lona, D.~Robbins, E.~Sharpe, T.~Vandermeulen, and X.~Yu, ``{Notes on
  gauging noninvertible symmetries, part 2: higher multiplicity cases},''
  \href{http://arxiv.org/abs/2408.16811}{{\ttfamily arXiv:2408.16811
  [hep-th]}}.

\bibitem{Carqueville:2017ono}
N.~Carqueville, I.~Runkel, and G.~Schaumann, ``{Line and surface defects in
  Reshetikhin\textendash{}Turaev TQFT},''
  \href{http://dx.doi.org/10.4171/qt/121}{{\em Quantum Topol.} {\bfseries 10}
  no.~3, (2018) 399--439}, \href{http://arxiv.org/abs/1710.10214}{{\ttfamily
  arXiv:1710.10214 [math.QA]}}.

\bibitem{Gaiotto:2019xmp}
D.~Gaiotto and T.~Johnson-Freyd, ``{Condensations in higher categories},''
  \href{http://arxiv.org/abs/1905.09566}{{\ttfamily arXiv:1905.09566
  [math.CT]}}.

\bibitem{Kong:2020wmn}
L.~Kong, Y.~Tian, and Z.-H. Zhang, ``{Defects in the 3-dimensional toric code
  model form a braided fusion 2-category},''
  \href{http://dx.doi.org/10.1007/JHEP12(2020)078}{{\em JHEP} {\bfseries 12}
  (2020) 078}, \href{http://arxiv.org/abs/2009.06564}{{\ttfamily
  arXiv:2009.06564 [cond-mat.str-el]}}.

\bibitem{Roumpedakis:2022aik}
K.~Roumpedakis, S.~Seifnashri, and S.-H. Shao, ``{Higher Gauging and
  Non-invertible Condensation Defects},''
  \href{http://arxiv.org/abs/2204.02407}{{\ttfamily arXiv:2204.02407
  [hep-th]}}.

\bibitem{Cui:2024cav}
W.~Cui, B.~Haghighat, and L.~Ruggeri, ``{Non-Invertible Surface Defects in 2+1d
  QFTs from Half Spacetime Gauging},''
  \href{http://arxiv.org/abs/2406.09261}{{\ttfamily arXiv:2406.09261
  [hep-th]}}.

\bibitem{Cuiper:2024hvh}
B.~V.-D. Cuiper and J.~Garre-Rubio, ``{Systematic construction of stabilizer
  codes via gauging abelian boundary symmetries},''
  \href{http://arxiv.org/abs/2410.09044}{{\ttfamily arXiv:2410.09044
  [quant-ph]}}.

\bibitem{Ebisu:2024lie}
H.~Ebisu and B.~Han, ``{Non-invertible duality defects in one, two, and three
  dimensions via gauging spatially modulated symmetry},''
  \href{http://arxiv.org/abs/2409.16744}{{\ttfamily arXiv:2409.16744
  [cond-mat.str-el]}}.

\bibitem{Vandermeulen:2023smx}
T.~Vandermeulen, ``{Gauge Defects},''
  \href{http://arxiv.org/abs/2310.08626}{{\ttfamily arXiv:2310.08626
  [hep-th]}}.

\bibitem{Lin:2022xod}
L.~Lin, D.~G. Robbins, and E.~Sharpe, ``{Decomposition, Condensation Defects,
  and Fusion},'' \href{http://dx.doi.org/10.1002/prop.202200130}{{\em Fortsch.
  Phys.} {\bfseries 70} no.~11, (2022) 2200130},
  \href{http://arxiv.org/abs/2208.05982}{{\ttfamily arXiv:2208.05982
  [hep-th]}}.

\bibitem{Lyons:2024fsk}
A.~Lyons, C.~F.~B. Lo, N.~Tantivasadakarn, A.~Vishwanath, and R.~Verresen,
  ``{Protocols for Creating Anyons and Defects via Gauging},''
  \href{http://arxiv.org/abs/2411.04181}{{\ttfamily arXiv:2411.04181
  [quant-ph]}}.

\bibitem{Carqueville:2018sld}
N.~Carqueville, I.~Runkel, and G.~Schaumann, ``{Orbifolds of Reshetikhin-Turaev
  TQFTs},'' {\em Theor. Appl. Categor.} {\bfseries 35} (2020) 513--561,
  \href{http://arxiv.org/abs/1809.01483}{{\ttfamily arXiv:1809.01483
  [math.QA]}}.

\bibitem{Mulevicius:2020bat}
V.~Mulevi\v{c}ius and I.~Runkel, ``{Constructing modular categories from
  orbifold data},'' \href{http://dx.doi.org/10.4171/qt/170}{{\em Quantum
  Topol.} {\bfseries 13} no.~3, (2023) 459--523},
  \href{http://arxiv.org/abs/2002.00663}{{\ttfamily arXiv:2002.00663
  [math.QA]}}.

\bibitem{Koppen:2021kry}
V.~Koppen, V.~Mulevicius, I.~Runkel, and C.~Schweigert, ``{Domain Walls Between
  3d Phases of Reshetikhin\textendash{}Turaev TQFTs},''
  \href{http://dx.doi.org/10.1007/s00220-022-04489-5}{{\em Commun. Math. Phys.}
  {\bfseries 396} no.~3, (2022) 1187--1220},
  \href{http://arxiv.org/abs/2105.04613}{{\ttfamily arXiv:2105.04613
  [hep-th]}}.

\bibitem{Carqueville:2021dbv}
N.~Carqueville, V.~Mulevicius, I.~Runkel, G.~Schaumann, and D.~Scherl,
  ``{Orbifold graph TQFTs},'' \href{http://arxiv.org/abs/2101.02482}{{\ttfamily
  arXiv:2101.02482 [math.QA]}}.

\bibitem{Carqueville:2021edn}
N.~Carqueville, V.~Mulevicius, I.~Runkel, G.~Schaumann, and D.~Scherl,
  ``{Reshetikhin\textendash{}Turaev TQFTs Close Under Generalised Orbifolds},''
  \href{http://dx.doi.org/10.1007/s00220-024-05068-6}{{\em Commun. Math. Phys.}
  {\bfseries 405} no.~10, (2024) 242},
  \href{http://arxiv.org/abs/2109.04754}{{\ttfamily arXiv:2109.04754
  [math.QA]}}.

\bibitem{Choi:2023xjw}
Y.~Choi, B.~C. Rayhaun, Y.~Sanghavi, and S.-H. Shao, ``{Remarks on boundaries,
  anomalies, and noninvertible symmetries},''
  \href{http://dx.doi.org/10.1103/PhysRevD.108.125005}{{\em Phys. Rev. D}
  {\bfseries 108} no.~12, (2023) 125005},
  \href{http://arxiv.org/abs/2305.09713}{{\ttfamily arXiv:2305.09713
  [hep-th]}}.

\bibitem{Choi:2023vgk}
Y.~Choi, D.-C. Lu, and Z.~Sun, ``{Self-duality under gauging a non-invertible
  symmetry},'' \href{http://dx.doi.org/10.1007/JHEP01(2024)142}{{\em JHEP}
  {\bfseries 01} (2024) 142}, \href{http://arxiv.org/abs/2310.19867}{{\ttfamily
  arXiv:2310.19867 [hep-th]}}.

\bibitem{Buican:2023bzl}
M.~Buican and R.~Radhakrishnan, ``{Invertibility of Condensation Defects and
  Symmetries of 2 + 1d QFTs},''
  \href{http://dx.doi.org/10.1007/s00220-024-05096-2}{{\em Commun. Math. Phys.}
  {\bfseries 405} no.~9, (2024) 217},
  \href{http://arxiv.org/abs/2309.15181}{{\ttfamily arXiv:2309.15181
  [hep-th]}}.

\bibitem{Hsin:2019fhf}
P.-S. Hsin and A.~Turzillo, ``{Symmetry-enriched quantum spin liquids in (3 +
  1)$d$},'' \href{http://dx.doi.org/10.1007/JHEP09(2020)022}{{\em JHEP}
  {\bfseries 09} (2020) 022}, \href{http://arxiv.org/abs/1904.11550}{{\ttfamily
  arXiv:1904.11550 [cond-mat.str-el]}}.

\bibitem{Barkeshli:2022edm}
M.~Barkeshli, Y.-A. Chen, P.-S. Hsin, and R.~Kobayashi, ``{Higher-group
  symmetry in finite gauge theory and stabilizer codes},''
  \href{http://arxiv.org/abs/2211.11764}{{\ttfamily arXiv:2211.11764
  [cond-mat.str-el]}}.

\bibitem{Fuchs_2013}
J.~Fuchs, C.~Schweigert, and A.~Valentino, ``Bicategories for boundary
  conditions and for surface defects in 3-d tft,''
  \href{http://dx.doi.org/10.1007/s00220-013-1723-0}{{\em Communications in
  Mathematical Physics} {\bfseries 321} no.~2, (May, 2013) 543–575}.
  \url{http://dx.doi.org/10.1007/s00220-013-1723-0}.

\bibitem{Kapustin:2010if}
A.~Kapustin and N.~Saulina, ``{Surface operators in 3d Topological Field Theory
  and 2d Rational Conformal Field Theory},''
  \href{http://arxiv.org/abs/1012.0911}{{\ttfamily arXiv:1012.0911 [hep-th]}}.

\bibitem{Fuchs:2012dt}
J.~Fuchs, C.~Schweigert, and A.~Valentino, ``{Bicategories for boundary
  conditions and for surface defects in 3-d TFT},''
  \href{http://dx.doi.org/10.1007/s00220-013-1723-0}{{\em Commun. Math. Phys.}
  {\bfseries 321} (2013) 543--575},
  \href{http://arxiv.org/abs/1203.4568}{{\ttfamily arXiv:1203.4568 [hep-th]}}.

\bibitem{Chang:2018iay}
C.-M. Chang, Y.-H. Lin, S.-H. Shao, Y.~Wang, and X.~Yin, ``{Topological Defect
  Lines and Renormalization Group Flows in Two Dimensions},''
  \href{http://dx.doi.org/10.1007/JHEP01(2019)026}{{\em JHEP} {\bfseries 01}
  (2019) 026}, \href{http://arxiv.org/abs/1802.04445}{{\ttfamily
  arXiv:1802.04445 [hep-th]}}.

\bibitem{Komargodski:2020mxz}
Z.~Komargodski, K.~Ohmori, K.~Roumpedakis, and S.~Seifnashri, ``{Symmetries and
  strings of adjoint QCD$_{2}$},''
  \href{http://dx.doi.org/10.1007/JHEP03(2021)103}{{\em JHEP} {\bfseries 03}
  (2021) 103}, \href{http://arxiv.org/abs/2008.07567}{{\ttfamily
  arXiv:2008.07567 [hep-th]}}.

\bibitem{Choi:2021kmx}
Y.~Choi, C.~C\'{o}rdova, P.-S. Hsin, H.~T. Lam, and S.-H. Shao,
  ``{Noninvertible duality defects in 3+1 dimensions},''
  \href{http://dx.doi.org/10.1103/PhysRevD.105.125016}{{\em Phys. Rev. D}
  {\bfseries 105} no.~12, (2022) 125016},
  \href{http://arxiv.org/abs/2111.01139}{{\ttfamily arXiv:2111.01139
  [hep-th]}}.

\bibitem{Kaidi:2021xfk}
J.~Kaidi, K.~Ohmori, and Y.~Zheng, ``{Kramers-Wannier-like Duality Defects in
  (3+1)D Gauge Theories},''
  \href{http://dx.doi.org/10.1103/PhysRevLett.128.111601}{{\em Phys. Rev.
  Lett.} {\bfseries 128} no.~11, (2022) 111601},
  \href{http://arxiv.org/abs/2111.01141}{{\ttfamily arXiv:2111.01141
  [hep-th]}}.

\bibitem{Chang:2022hud}
C.-M. Chang, J.~Chen, and F.~Xu, ``{Topological defect lines in two dimensional
  fermionic CFTs},''
  \href{http://dx.doi.org/10.21468/SciPostPhys.15.5.216}{{\em SciPost Phys.}
  {\bfseries 15} no.~5, (2023) 216},
  \href{http://arxiv.org/abs/2208.02757}{{\ttfamily arXiv:2208.02757
  [hep-th]}}.

\bibitem{Fuchs:2002cm}
J.~Fuchs, I.~Runkel, and C.~Schweigert, ``{TFT construction of RCFT correlators
  1. Partition functions},''
  \href{http://dx.doi.org/10.1016/S0550-3213(02)00744-7}{{\em Nucl. Phys. B}
  {\bfseries 646} (2002) 353--497},
  \href{http://arxiv.org/abs/hep-th/0204148}{{\ttfamily arXiv:hep-th/0204148}}.

\bibitem{Fuchs:2003id}
J.~Fuchs, I.~Runkel, and C.~Schweigert, ``{TFT construction of RCFT
  correlators. 2. Unoriented world sheets},''
  \href{http://dx.doi.org/10.1016/j.nuclphysb.2003.11.026}{{\em Nucl. Phys. B}
  {\bfseries 678} (2004) 511--637},
  \href{http://arxiv.org/abs/hep-th/0306164}{{\ttfamily arXiv:hep-th/0306164}}.

\bibitem{Fuchs:2004dz}
J.~Fuchs, I.~Runkel, and C.~Schweigert, ``{TFT construction of RCFT
  correlators. 3. Simple currents},''
  \href{http://dx.doi.org/10.1016/j.nuclphysb.2004.05.014}{{\em Nucl. Phys. B}
  {\bfseries 694} (2004) 277--353},
  \href{http://arxiv.org/abs/hep-th/0403157}{{\ttfamily arXiv:hep-th/0403157}}.

\bibitem{Fuchs:2004xi}
J.~Fuchs, I.~Runkel, and C.~Schweigert, ``{TFT construction of RCFT correlators
  IV: Structure constants and correlation functions},''
  \href{http://dx.doi.org/10.1016/j.nuclphysb.2005.03.018}{{\em Nucl. Phys. B}
  {\bfseries 715} (2005) 539--638},
  \href{http://arxiv.org/abs/hep-th/0412290}{{\ttfamily arXiv:hep-th/0412290}}.

\bibitem{Frohlich:2004ef}
J.~Frohlich, J.~Fuchs, I.~Runkel, and C.~Schweigert, ``{Kramers-Wannier duality
  from conformal defects},''
  \href{http://dx.doi.org/10.1103/PhysRevLett.93.070601}{{\em Phys. Rev. Lett.}
  {\bfseries 93} (2004) 070601},
  \href{http://arxiv.org/abs/cond-mat/0404051}{{\ttfamily
  arXiv:cond-mat/0404051}}.

\bibitem{Frohlich:2006ch}
J.~Frohlich, J.~Fuchs, I.~Runkel, and C.~Schweigert, ``{Duality and defects in
  rational conformal field theory},''
  \href{http://dx.doi.org/10.1016/j.nuclphysb.2006.11.017}{{\em Nucl. Phys. B}
  {\bfseries 763} (2007) 354--430},
  \href{http://arxiv.org/abs/hep-th/0607247}{{\ttfamily arXiv:hep-th/0607247}}.

\bibitem{Kitaev:1997wr}
A.~Y. Kitaev, ``{Fault tolerant quantum computation by anyons},''
  \href{http://dx.doi.org/10.1016/S0003-4916(02)00018-0}{{\em Annals Phys.}
  {\bfseries 303} (2003) 2--30},
  \href{http://arxiv.org/abs/quant-ph/9707021}{{\ttfamily
  arXiv:quant-ph/9707021}}.

\bibitem{Kapustin:2010hk}
A.~Kapustin and N.~Saulina, ``{Topological boundary conditions in abelian
  Chern-Simons theory},''
  \href{http://dx.doi.org/10.1016/j.nuclphysb.2010.12.017}{{\em Nucl. Phys. B}
  {\bfseries 845} (2011) 393--435},
  \href{http://arxiv.org/abs/1008.0654}{{\ttfamily arXiv:1008.0654 [hep-th]}}.

\bibitem{Wen:2012hm}
X.-G. Wen, ``{Topological order: from long-range entangled quantum matter to an
  unification of light and electrons},''
  \href{http://dx.doi.org/10.1155/2013/198710}{{\em ISRN Cond. Matt. Phys.}
  {\bfseries 2013} (2013) 198710},
  \href{http://arxiv.org/abs/1210.1281}{{\ttfamily arXiv:1210.1281
  [cond-mat.str-el]}}.

\bibitem{Barkeshli:2014cna}
M.~Barkeshli, P.~Bonderson, M.~Cheng, and Z.~Wang, ``{Symmetry
  Fractionalization, Defects, and Gauging of Topological Phases},''
  \href{http://dx.doi.org/10.1103/PhysRevB.100.115147}{{\em Phys. Rev. B}
  {\bfseries 100} no.~11, (2019) 115147},
  \href{http://arxiv.org/abs/1410.4540}{{\ttfamily arXiv:1410.4540
  [cond-mat.str-el]}}.

\bibitem{Else:2017yqj}
D.~V. Else and C.~Nayak, ``{Cheshire charge in (3+1)-dimensional topological
  phases},'' \href{http://dx.doi.org/10.1103/PhysRevB.96.045136}{{\em Phys.
  Rev. B} {\bfseries 96} no.~4, (2017) 045136},
  \href{http://arxiv.org/abs/1702.02148}{{\ttfamily arXiv:1702.02148
  [cond-mat.str-el]}}.

\bibitem{Johnson-Freyd:2020twl}
T.~Johnson-Freyd, ``{(3+1)D topological orders with only a
  $\mathbb{Z}_2$-charged particle},''
  \href{http://arxiv.org/abs/2011.11165}{{\ttfamily arXiv:2011.11165
  [math.QA]}}.

\bibitem{Chen:2023qst}
X.~Chen, A.~Dua, M.~Hermele, D.~T. Stephen, N.~Tantivasadakarn, R.~Vanhove, and
  J.-Y. Zhao, ``{Sequential quantum circuits as maps between gapped phases},''
  \href{http://dx.doi.org/10.1103/PhysRevB.109.075116}{{\em Phys. Rev. B}
  {\bfseries 109} no.~7, (2024) 075116},
  \href{http://arxiv.org/abs/2307.01267}{{\ttfamily arXiv:2307.01267
  [cond-mat.str-el]}}.

\bibitem{Tantivasadakarn:2023zov}
N.~Tantivasadakarn and X.~Chen, ``{String operators for Cheshire strings in
  topological phases},''
  \href{http://dx.doi.org/10.1103/PhysRevB.109.165149}{{\em Phys. Rev. B}
  {\bfseries 109} no.~16, (2024) 165149},
  \href{http://arxiv.org/abs/2307.03180}{{\ttfamily arXiv:2307.03180
  [cond-mat.str-el]}}.

\bibitem{Tantivasadakarn_2024}
N.~Tantivasadakarn and X.~Chen, ``String operators for cheshire strings in
  topological phases,''
  \href{http://dx.doi.org/10.1103/physrevb.109.165149}{{\em Physical Review B}
  {\bfseries 109} no.~16, (Apr., 2024) }.
  \url{http://dx.doi.org/10.1103/physrevb.109.165149}.

\bibitem{ostrik2001module}
V.~Ostrik, ``Module categories, weak hopf algebras and modular invariants,''
  2001.

\bibitem{Choi:2022zal}
Y.~Choi, C.~C\'{o}rdova, P.-S. Hsin, H.~T. Lam, and S.-H. Shao,
  ``{Non-invertible Condensation, Duality, and Triality Defects in 3+1
  Dimensions},'' \href{http://arxiv.org/abs/2204.09025}{{\ttfamily
  arXiv:2204.09025 [hep-th]}}.

\bibitem{https://doi.org/10.48550/arxiv.math/0605185}
C.~J. Hillar and D.~Rhea, ``Automorphisms of finite abelian groups,'' 2006.
\newblock \url{https://arxiv.org/abs/math/0605185}.

\bibitem{Conrad_generatingsets}
K.~Conrad, ``Generating sets.''.

\bibitem{VAFA1986592}
C.~Vafa, ``Modular invariance and discrete torsion on orbifolds,''
  \href{http://dx.doi.org/https://doi.org/10.1016/0550-3213(86)90379-2}{{\em
  Nuclear Physics B} {\bfseries 273} no.~3, (1986) 592--606}.
  \url{https://www.sciencedirect.com/science/article/pii/0550321386903792}.

\bibitem{Vafa:1994rv}
C.~Vafa and E.~Witten, ``{On orbifolds with discrete torsion},''
  \href{http://dx.doi.org/10.1016/0393-0440(94)00048-9}{{\em J. Geom. Phys.}
  {\bfseries 15} (1995) 189--214},
  \href{http://arxiv.org/abs/hep-th/9409188}{{\ttfamily arXiv:hep-th/9409188}}.

\bibitem{Tantivasadakarn:2022ceu}
N.~Tantivasadakarn, R.~Verresen, and A.~Vishwanath, ``{Shortest Route to
  Non-Abelian Topological Order on a Quantum Processor},''
  \href{http://dx.doi.org/10.1103/PhysRevLett.131.060405}{{\em Phys. Rev.
  Lett.} {\bfseries 131} no.~6, (2023) 060405},
  \href{http://arxiv.org/abs/2209.03964}{{\ttfamily arXiv:2209.03964
  [quant-ph]}}.

\bibitem{Prem:2019etl}
A.~Prem and D.~J. Williamson, ``{Gauging permutation symmetries as a route to
  non-Abelian fractons},''
  \href{http://dx.doi.org/10.21468/SciPostPhys.7.5.068}{{\em SciPost Phys.}
  {\bfseries 7} no.~5, (2019) 068},
  \href{http://arxiv.org/abs/1905.06309}{{\ttfamily arXiv:1905.06309
  [cond-mat.str-el]}}.

\bibitem{deWildPropitius:1997wu}
M.~de~Wild~Propitius, ``{Confinement in partially broken Abelian Chern-Simons
  theories},'' \href{http://dx.doi.org/10.1016/S0370-2693(97)00985-4}{{\em
  Phys. Lett. B} {\bfseries 410} (1997) 188--194},
  \href{http://arxiv.org/abs/hep-th/9704063}{{\ttfamily arXiv:hep-th/9704063}}.

\bibitem{Coste_2000}
A.~Coste, T.~Gannon, and P.~Ruelle, ``Finite group modular data,''
  \href{http://dx.doi.org/10.1016/s0550-3213(00)00285-6}{{\em Nuclear Physics
  B} {\bfseries 581} no.~3, (Aug., 2000) 679–717}.
  \url{http://dx.doi.org/10.1016/S0550-3213(00)00285-6}.

\bibitem{Tambara:1998vmj}
D.~Tambara and S.~Yamagami, ``{Tensor Categories with Fusion Rules of
  Self-Duality for Finite Abelian Groups},''
  \href{http://dx.doi.org/10.1006/jabr.1998.7558}{{\em J. Algebra} {\bfseries
  209} no.~2, (1998) 692--707}.

\bibitem{Heidenreich:2021xpr}
B.~Heidenreich, J.~McNamara, M.~Montero, M.~Reece, T.~Rudelius, and
  I.~Valenzuela, ``{Non-invertible global symmetries and completeness of the
  spectrum},'' \href{http://dx.doi.org/10.1007/JHEP09(2021)203}{{\em JHEP}
  {\bfseries 09} (2021) 203}, \href{http://arxiv.org/abs/2104.07036}{{\ttfamily
  arXiv:2104.07036 [hep-th]}}.

\bibitem{Arias-Tamargo:2022nlf}
G.~Arias-Tamargo and D.~Rodriguez-Gomez, ``{Non-invertible symmetries from
  discrete gauging and completeness of the spectrum},''
  \href{http://dx.doi.org/10.1007/JHEP04(2023)093}{{\em JHEP} {\bfseries 04}
  (2023) 093}, \href{http://arxiv.org/abs/2204.07523}{{\ttfamily
  arXiv:2204.07523 [hep-th]}}.

\end{thebibliography}\endgroup

\end{document}